\documentclass[aps,superscriptaddress,twocolumn,floatfix,letterpaper,nofootinbib]{revtex4-1}

\usepackage[all,matrix,cmtip]{xy}
\usepackage{mathtools}
\usepackage{dsfont}

\usepackage{eufrak}
\newcommand\fM          {\EuFrak{M}}

\usepackage{euscript}
\newcommand\eA           {\EuScript{A}}
\newcommand\eB           {\EuScript{B}}
\newcommand\eC           {\EuScript{C}}

\newcommand\eM          {\EuScript{M}}

\newcommand\eT         {\EuScript{T}}
\newcommand\eV        {\EuScript{V}}

\newcommand\sA           {\mathsf{A}}
\newcommand\sB           {\mathsf{B}}
\newcommand\sC           {\mathsf{C}}
\newcommand\sD           {\mathsf{D}}

\newcommand\sGT         {\mathsf{GT}}

\newcommand\sI           {\mathsf{I}}

\newcommand\sK         {\mathsf{K}}

\newcommand\sM          {\mathsf{M}}
\newcommand\sN         {\mathsf{N}}

\newcommand\sR         {\mathsf{R}}

\newcommand{\bxt}[1]{\underset{#1}{\otimes}}
\newcommand{\ot}[1]{\underset{#1}{\otimes}}
\newcommand{\map}[1]{\overset{#1}{\rightarrow}}
\newcommand{\inj}[1]{\overset{#1}{\hookrightarrow}}

\newcommand{\Bulk}{\mbox{$\mathsf{Bulk}$}}
\newcommand{\bulk}{\mbox{$\mathsf{bulk}$}}
\newcommand{\w}{\mathrm{w}}
\newcommand{\id}{\mathrm{id}}

\newcommand{\Hom}{\mathrm{Hom}}

\newcommand{\Fun}{\mathrm{Fun}}
\renewcommand{\mod}{\mathrm{mod}}

\newcommand\seq[1]{\overset{\scriptscriptstyle #1}{\simeq}}

\newcommand\cTO{\mathds{TO}}
\newcommand\afTO{\eT\mathrm{O}_\mathrm{af}}
\newcommand\afcTO{\mathds{TO}_{\mathrm{af}}}
\newcommand\icTO{\eT\mathrm{O}_\mathrm{inv}}

\newcommand\Rep{\mathrm{Rep}}
\newcommand\sRep{\mathrm{sRep}}

\newcommand\cRep{\cR\mathrm{ep}}

\newcommand\cVec{\cV\mathrm{ec}}
\newcommand\scVec{\mathrm{s}\cV\mathrm{ec}}

\newcommand\eVec{\eV\mathrm{ec}}

\renewcommand\Ref[1]{Ref.~\onlinecite{#1}}

\newcommand\npo {\mbox{($n\hskip-2.5pt +\hskip-2.5pt 1$)}}
\newcommand\nmo {\mbox{($n\hskip-2.5pt -\hskip-2.5pt 1$)}}
\newcommand\npt {\mbox{($n\hskip-2.5pt +\hskip-2.5pt 2$)}}
\newcommand\nmt {\mbox{($n\hskip-2.5pt -\hskip-2.5pt 2$)}}

\newcommand\Cb{\mathbb{C}}
\newcommand\tto{\mathsf{I}}
\newcommand\dual{\widetilde}
\newcommand\rev{\mathrm{rev}}

\usepackage[normalem]{ulem}

\usepackage{amsthm}
\theoremstyle{theorem}

\newtheorem{Corollary}{Corollary}

\newtheorem{Proposition}{Proposition}
\newtheorem{Proposal}{Proposal}
\newtheorem{Conjecture}{Conjecture}
\newtheorem{Definition}{Definition}
\newtheorem{DefinitionPH}[Definition]{Definition$^\text{ph}$}
\theoremstyle{definition}
\newtheorem{Remark}{Remark}
\newtheorem{Example}{Example}

\begin{document}

\begin{titlepage}

\title{Algebraic higher symmetry and categorical symmetry\\ 
-- a holographic and entanglement view of symmetry
}

\author{Liang Kong} 
\affiliation{Shenzhen Institute for Quantum Science and Engineering, 
Southern University of Science and Technology, Shenzhen 518055, P.R. China}
\affiliation{Guangdong Provincial Key Laboratory of Quantum Science and Engineering, 
Southern University of Science and Technology, Shenzhen 518055, P.R. China}

\author{Tian Lan} 
\affiliation{Institute for Quantum Computing,
  University of Waterloo, Waterloo, Ontario N2L 3G1, Canada}

\author{Xiao-Gang Wen}
\affiliation{Department of Physics, Massachusetts Institute of
Technology, Cambridge, Massachusetts 02139, USA}

\author{Zhi-Hao Zhang} 
\affiliation{School of Mathematical Sciences, University of Science and Technology of China, 
Hefei 230026, P.R. China}
\affiliation{Shenzhen Institute for Quantum Science and Engineering, 
Southern University of Science and Technology, Shenzhen 518055, P.R. China}

\author{Hao Zheng} 
\affiliation{Shenzhen Institute for Quantum Science and Engineering, 
Southern University of Science and Technology, Shenzhen 518055, P.R. China}
\affiliation{Guangdong Provincial Key Laboratory of Quantum Science and Engineering, 
Southern University of Science and Technology, Shenzhen 518055, P.R. China}
\affiliation{Department of Mathematics, Peking University, Beijing 100871, China}

\begin{abstract} 

A global symmetry (0-symmetry) in an $n$-dimensional space acts on the whole
space.  A higher symmetry acts on closed submanifolds (i.e. loops and
membranes, {\it etc}), and those transformations form a higher group. In this
paper, we introduce the notion of \emph{algebraic higher symmetry}, which
generalizes higher symmetry and is beyond higher group.  We show that an
algebraic higher symmetry in a bosonic system in $n$-dimensional space is
characterized and classified by a \emph{local fusion $n$-category}.  We find
another way to describe algebraic higher symmetry by restricting to symmetric
sub Hilbert space where symmetry transformations all become trivial.  In this
case, algebraic higher symmetry can be fully characterized by a non-invertible
gravitational anomaly ({\it i.e.} an topological order in one higher
dimension). Thus we also refer to non-invertible gravitational anomaly as
\emph{categorical symmetry} to stress its connection to symmetry.  This
provides a holographic and entanglement view of symmetries.  For a system with
a categorical symmetry, its gapped state must spontaneously break part (not
all) of the symmetry, and the state with the full symmetry must be gapless.
Using such a holographic point of view, we obtain (1) the gauging of the
algebraic higher symmetry; (2) the classification of anomalies for an algebraic
higher symmetry; (3) the equivalence between classes of systems, with different
(potentially anomalous) algebraic higher symmetries or different sets of low
energy excitations, as long as they have the same categorical symmetry; (4) the
classification of gapped liquid phases for bosonic/fermionic systems with a
categorical symmetry, as gapped boundaries of a topological order in one higher
dimension (that corresponds to the categorical symmetry).  This classification
includes symmetry protected trivial (SPT) orders and symmetry enriched
topological (SET) orders with an algebraic higher symmetry.  

\end{abstract}


\maketitle

\end{titlepage}

{\small \setcounter{tocdepth}{1} \tableofcontents }

~

~

\section{Introduction} 

The notion of a symmetry plays a very important role in physics.  A quantum
system living on $n$-dimensional space\footnote{Here, a $n$-dimensional space
$M^n$ actually means a triangulation of $n$-dimensional manifold.  So $M^n$
should be viewed as a $n$-dimensional simplicial complex.  In this paper, we
mainly consider discrete lattice systems.} $M^n$ is defined by a vector space
$\cV$ formed by wave functions on $M^n$ and a Hamiltonian $H$. A symmetry in
such a system is a set of linear constraints on the allowed Hamiltonians.
Since the Hamiltonian is always a sum of local operators $H=\sum_{\v x} O_{\v
x}$, we can also more precisely describe a symmetry as a set of linear
constraints on the allowed local operators.  Those allowed local operators are
called \textbf{symmetric local operators} and they form an algebra of symmetric
local operators. A symmetric Hamiltonian is a sum of symmetric local operators.
The algebra of symmetric local operators contains all the information about the
symmetry and represents a very general way to describe the symmetry.  In this
paper, we will use this point of view to show that a symmetry in
$n$-dimensional space is described a \textbf{local fusion $n$-category}.

By a ``symmetry'', we usually mean a global symmetry, where we have a set of
unitary operators $W_\al$, labeled by $\al$, acting on the whole space $M^n$
(\ie a symmetry transformation) which give rise to the following linear
constraint on the local operators $W_\al O_{\v x}=O_{\v x}W_\al$.  If one digs
deeper, however, one finds that there are in fact several different kinds of
global symmetries. In quantum field theories, we have anomaly-free global
symmetries (gaugeable global symmetries) and anomalous global symmetries
(not-gaugeable  global symmetries or 't Hooft anomalies\cite{H8035}).  In
lattice systems, we have on-site symmetries (where the symmetry transformation
has a composition in terms of operators $W_\al(\v x)$ that acts only on lattice
site labeled by $\v x$: $W_\al=\otimes_{\v x} W_\al(\v x)$) and non-on-site
symmetries.\cite{CGL1314,W1313}  

These different kinds of global symmetries are closely related.  Consider a low
energy effective field theory of a lattice model.  The on-site symmetries in
the lattice model becomes the anomaly-free global symmetries in the effective
field theory, since the lattice on-site-symmetry is always gaugeable.  The
non-on-site symmetries in the lattice model become the anomalous global
symmetries in the effective field theory.\cite{W1313}  For the symmetries
related to spacial transformation, such as the lattice translation symmetry and
point group symmetry, sometimes they become anomalous symmetry in the effective
field theory, and sometimes they are anomaly-free.  In this paper, we consider
only internal symmetries instead of symmetries related to spacial
transformations.

There are also gauge symmetries in field theories and lattice theories. But
they are not symmetries in quantum systems,  and should not be called symmetry
at all.

Recently, in \Ref{GW14125148}, the notion of a global symmetry was generalized
to a $k$-form symmetry, which acts on all closed subspaces of codimension $k$
and becomes the identity operator if the closed subspaces are contractible.  It
was stressed that many results and intuitions for global symmetries (the 0-form
symmetries) can be extended to higher-form symmetries.  

In fact, closely related higher symmetries had been studied earlier (but under
various different names, such as logical operator, gauge-like symmetry, \etc),
where exactly solvable lattice Hamiltonians commuting with all closed string
and/or membrane operators were constructed to realize topological orders
\cite{K032,W0303,LW0316,Y10074601,B11072707,NOc0605316,NOc0702377}.  We call a
lattice symmetry generated by a $k$-codimensional operator as a $k$-symmetry,
where the codimension in this paper is defined with respect to the space
dimension.  Similar to a $k$-form symmetry, a $k$-symmetry acts on closed
subspaces of codimension $k$, but it does not become the identity operator when
the closed subspaces are contractible. A higher symmetry is a symmetry in a
lattice model.  A higher symmetry reduces to a higher form symmetry in the
ground state subspaces (\ie in low energy effective topological quantum field
theory).  Our local fusion higher category description of symmetry includes
those higher symmetries.

The emergence of higher symmetries was also studied before (again  under
different names, such as string-operators satisfying zero-law)\cite{HW0541},
where it was found that, unlike usual global symmetry (\ie 0-symmetry), the
emergent higher symmetries cannot be destroyed by any local perturbations.
Such a topological robustness was used to show that the emergent gapless $U(1)$
gauge bosons are robust against any local perturbations -- a topological
version of Goldstone theorem \cite{HW0541}. See
\Ref{KT13094721,TK151102929,BM170200868,W181202517,WW181211967,TW190802613} for
some recent discussions of lattice higher symmetries, their emergence,
anomalies and a classification of associated higher symmetry protected phases
on lattice. 

In this work, we study a new kind of symmetries that is beyond higher groups.
We refer to the new symmetries as \textbf{algebraic higher symmetries}, and
refer to higher groups as group-like higher symmetries.  Algebraic higher
symmetries include group-like higher symmetries as special cases.

Group-like higher symmetries and algebraic higher symmetries can both be
generated by $p$-dimensional operators $W_\al(S^p)$ labeled by an index $\al$
and $S^p$ (a $p$-dimensional closed submanifold), and $W_\al(S^p)$ only acts on
the degrees of freedom near $S^p$.  For a group-like higher symmetry, the
$k$-dimensional operators satisfy a group-like algebra
\begin{align}
 W_\al(S^p) W_\bt(S^p) = W_\ga(S^p),
\end{align}
while for an algebraic higher symmetry, they may satisfy a more
general multiplication algebra\cite{JW191213492}
\begin{align}
\label{WWNW}
 W_\al(S^p) W_\bt(S^p) = \sum_\ga N_{\al\bt}^\ga W_\ga(S^p).
\end{align}
In this case, the symmetry generator $W_\al(S^p)$ may be neither invertible nor
unitary. Such kind of algebraic symmetries was studied in 1+1D conformal field
theory via non-invertible defect lines (where invertible defect lines are known
to connect to symmetry)\cite{FSh0607247,DR10044725,CY180204445,TW191202817}.
We believe that local fusion higher categories classify the anomaly-free
algebraic higher symmetries, while  anomalous algebraic higher symmetries are
described by generic higher categories.

In Section~\ref{Lgauge}, we discuss an example, a lattice model described by a
Hamiltonian $H$, where the above algebraic higher symmetry does show up, i.e.
\begin{align}
 W_\al(S^p) H = H W_\al(S^p) .
\end{align}
Then, in Sections~\ref{lfcat} and \ref{symprd}, we discuss unbroken
anomaly-free algebraic higher symmetry from a point of view of trivial
symmetric product state and local fusion higher category.  In
Section~\ref{csym}, we show that an algebraic higher symmetry can be fully
described by a non-invertible gravitational anomaly\cite{JW190513279} (which is
the same as a topological order in one higher
dimension\cite{W1313,KW1458,KZ150201690,KZ170200673}), and this is a very
useful way to view symmetry. To stress its relation to symmetry, we also refer
to non-invertible gravitational anomaly (\ie topological order in one higher
dimension) as \textbf{categorical symmetry}\cite{JW191213492}.  In
Section~\ref{glpCS}, we obtain a classification of gapped liquid phases for
systems with a categorical symmetry.  It includes the classification of
symmetry protected trivial (SPT) phases and that of symmetry enriched
topological (SET) phases for algebraic higher symmetry.  In
Section~\ref{effCS}, we describe the emergence of categorical symmetries from
topological orders, when the excitations have a large separation of energy.  

The main point of this paper is about anomaly-free algebraic higher symmetries
that are generally described and classified by local fusion higher categories.
We also study topological orders with algebraic higher symmetries.  Our
approach is based on fusion higher category description of topological
orders,\cite{KW1458,KZ150201690,KZ170200673,LW180108530,GJ190509566,J200306663,KZ200514178}
which will be reviewed, clarified, and expanded in Section~\ref{toprev}.  A
brief summary of higher category description of topological orders can also be
found in the first few subsections of Section~\ref{main}.  This section tries
to summarize the results of this paper for physics readers.

We have a more mathematical version of this paper published as
\Ref{KZ200514178}.  The present paper contains more physical results and has
more physical discussions.

We remark that the precise definitions of fusion higher categories and local
fusion higher categories are difficult due to the lack of the universally
accepted and well-developed model for weak $n$-categories.  In this paper, we
try to give a physical definition via the notion of topological orders. Many
related concepts for topological order in arbitrary dimensions and for higher
categories are discussed this way in Section~\ref{toprev}.

We like to point out that the physical definition of topological orders given
in \Ref{CGW1038} is based on microscopic lattice models.  In fact, many
physical concepts are defined via microscopic lattice models, and we refer to
those kinds of definition as microscopic definitions.  There are also many
physical concepts which are defined via macroscopic measurements, such as
supperfluidity defined via vanishing viscosity and quantization of vorticity.
We refer to those kinds of definition as macroscopic definitions.  In this
respect, the definitions in mathematics are macroscopic definitions. So
mathematical definitions are closer to physical experiments.  Some notions in
symmetry and topological orders are defined microscopically, such as
topological excitations,\cite{K062,LW0510,LW1384} long range
entanglement,\cite{CGW1038} the characterization of algebraic higher symmetry
by \eqn{WWNW}, \etc.  Some other notions are defined microscopically, such as
topological degeneracy and the associated modular
transformations\cite{W8987,WN9077,W9039}, \etc.  So a lot of efforts of this
work is to convert microscopic definitions to macroscopic definitions, when
possible.  We will use $^\text{ph}$ to indicate the microscopic definitions.
Most results of this paper are presented via Propositions. Those results are
physical results based various physical arguments and beliefs.

Throughout this work, we use $n$d to denote the spacial dimension and $\npo$D
to denote the spacetime dimension, and the following convention of notations:
\begin{itemize}

\item $n$D topological orders: $\sA^n,\sB^n,\sC^n$ (mathsf font); 

\item fusion $n$-categories: $\cA^n,\cB^n,\cC^n$ (mathcal font); 

\item braided fusion $n$-categories: $\eA^n, \eB^n, \eC^n$ (euscript font).

\end{itemize}
Throughout this paper, superscripts always mean the spacetime dimension, or
level of higher category.  Also in this paper, we only consider finite
algebraic higher symmetry.  We mostly consider bosonic systems, except in
Sections \ref{cl1}, \ref{cl2}, and \ref{cl3}.  So when we say, for example, SPT
orders, we mean SPT orders in bosonic system.  In  Sections \ref{cl1},
\ref{cl2}, and \ref{cl3}, our results apply to both bosonic and fermionic
systems, and even anyonic systems,  via a more general notion of algebraic
higher symmetry.

\section{Summary of main results}
\label{main}

Since the main text of this paper is quite mathematical, in this section, we
summarize the main results in less rigorous physical terms, and introduce
concepts and notations along the way.

\subsection{Category of topological orders}

First, let us introduce some concepts and notations about topological order.
Let $\fM^{n+1}$, called the moduli space, be the space of Hamiltonians that
support  gapped liquid ground state \cite{ZW1490,SM1403}. An element in
$\pi_0(\fM^{n+1})$ is a gapped liquid phase, \ie a \textbf{topological order}
which is denoted by $\sM^{n+1}$.  So, a topological order is a gapped liquid
phase\cite{ZW1490,SM1403} (see Definition \ref{GLP} which is a microscopic
definition).  

\begin{figure}[t]
\begin{center}
\includegraphics[scale=0.7]{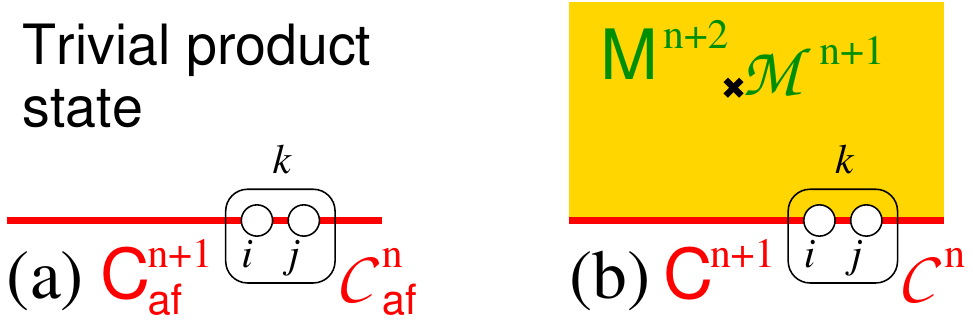} \end{center}
%1
\caption{ (a) an anomaly-free topological order $\sC^{n+1}_\text{af} \in
\afcTO^{n+1}$ in $\npo$-dimensional spacetime can be realized on lattice in the
same dimension, which can also be viewed as a boundary of a trivial product
state in one higher dimension.  The excitations in $\sC^{n+1}_\text{af}$ are
described by fusion $n$-category $\cC^n_\text{af}$.  (b) an anomalous
topological order $\sC^{n+1} \in \cTO^{n+1}$ in $\npo$-dimensional spacetime
can be realized as a boundary of an anomaly-free topological order $\sM^{n+2}$
in one higher dimension.  The excitations in $\sC^{n+1}$ are described by
fusion $n$-category $\cC^n$.  The excitations in $\sM^{n+2}$ are described by
fusion $\npo$-category $\cM^{n+1}$.  } \label{TOa} \end{figure}

\begin{figure}[t]
\begin{center}
\includegraphics[scale=0.7]{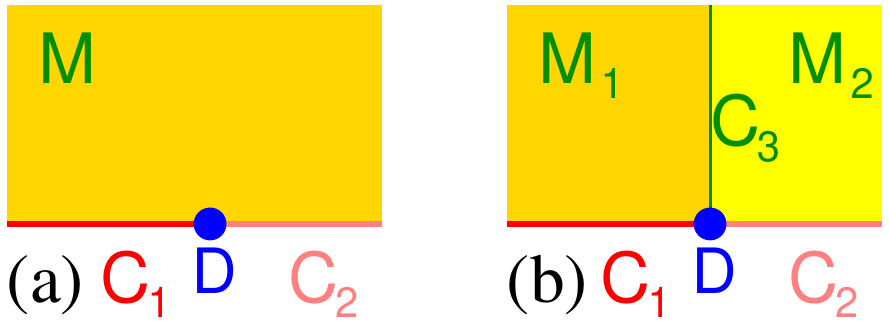} \end{center}
%2
\caption{ (a) an anomaly-free domain wall $\sD$ between two (potentially
anomalous) topological orders $\sC_1$ and $\sC_2$.  (b) an anomalous domain
wall $\sD$ between two (potentially anomalous) topological orders $\sC_1$ and
$\sC_2$.  $\sD$ is a boundary of domain wall $\sC_3$ which is a (potentially
anomalous) topological order.  } \label{DWa} \end{figure}

A topological order in $n$-dimensional space  is roughly described
macroscopically by the following data:\cite{KW1458,KZ150201690} 
\begin{enumerate}
\item the codimension-1, the codimension-2, \etc\ excitations above the gapped
liquid state (see Fig. \ref{TOa}); 

\item the domain walls between two high dimensional excitations; 

\item the domain walls connecting to other topological orders (see Fig.
\ref{DWa});

\item the monoid formed by the stacking topological orders.
\end{enumerate}
Roughly, the data in the first two items describes a fusion $n$-category
$\cM^n$, which is a partial description of topological order $\sM^{n+1}$.  If
we add the data in the third and fourth items to fusion $n$-category $\cM^n$,
we get a full description of the topological order $\sM^{n+1}$ (\ie a full
description of gapped liquid phase).

To incorporate all the above data in one framework, we can put all those
topological orders in $\npo$-dimensional spacetime together to form a
\textbf{category $\cTO^{n+1}$ of $\npo$D topological orders}\cite{KZ150201690}
(see Section~\ref{cata}). It consists of a collection of topological orders
(called the objects or 0-morphisms of the category), and 1-codimensional gapped
domain walls between two (not necessarily different) topological orders (called
1-morphisms of the category), and 2-codimensional domain walls between
1-codimensional domain walls (called 2-morphisms), so on and so forth.  The top
morphisms are $\npo$-morphisms, which are local operators (satisfying certain
symmetry constraints) acting on a spacetime point $(\v x,t)$.  The top
morphisms can also be viewed as instantons in spacetime.  The objects form a
monoid under the stacking operation.

To be more precise, we distinguish two different categories: $\cTO^{n+1}$ and
$\afcTO^{n+1}$. A topological order is called \textbf{anomaly-free} (\ie in
$\afcTO^{n+1}$) if it can be realized by lattice models in the same dimension
(see Section~\ref{cataf} and Fig. \ref{TOa}a), and is called \textbf{anomalous}
if otherwise (see Section \ref{cata} and Fig. \ref{TOa}b).\cite{W1313,KW1458}
An $\nmo$d domain wall between two anomaly-free $n$d topological orders is
called anomaly-free if it can be realized by a $\nmo$d lattice wall between two
$\npo$d lattice-model realization of two adjacent $\npo$d topological orders
(see Fig. \ref{DWa}a), and is called anomalous if otherwise (see Fig.
\ref{DWa}b).  Anomaly-free/anomalous higher codimensional domain walls can be
defined similarly. All potentially anomalous $\npo$D topological orders form a
category $\cTO^{n+1}$, in which 1-morphisms in $\cTO^{n+1}$ are defined by
potentially anomalous 1-codimensional gapped domain walls, higher morphisms are
higher codimensional gapped domain walls.  See \Ref{KZ150201690} for more
details.  Since objects in $\cTO^{n+1}$ are topological orders, we simply use
$\sA^{n+1},\sB^{n+1},\sC^{n+1}\in \cTO^{n+1}$ to denote $\npo$D topological
orders (see Section~\ref{gapliq}). The superscript $^{n+1}$ represents the
spacetime dimension and may be omitted if it is manifest from the context. We
denote the trivial $\npo$D topological order by $\tto^{n+1}$ (see Section
\ref{cattr}), and denote the stacking of two $\npo$D topological orders
$\sA^{n+1}$ and $\sB^{n+1}$ by $\sA^{n+1} \otimes \sB^{n+1}$. The data
$\tto^{n+1}$ and $\otimes$ endow the $\npo$-category $\cTO^{n+1}$ with a
structure of a symmetric monoidal $\npo$-category. All $\npo$D anomaly-free
topological orders, together with all anomaly-free domain walls of all
codimensions, form a symmetric monoidal $\npo$-category of anomaly-free $\npo$D
topological orders, denoted by $\afcTO^{n+1}$ (see Section~\ref{cataf} and
\Ref{KZ150201690} for more details).

We recall a few notions introduced in \Ref{KZ150201690}. We denote the trivial
$\nmo$D domain wall between $\sA^n$ and $\sA^n$ by $\id_\sA$, and the trivial
($n-$2)D domain wall between $\id_\sA$ and $\id_\sA$ by $\id_\sA^2$, so on and
so forth. 

As objects in a higher category, the notion of ``same topological order'' is
non trivial.  Physically, two anomaly-free topological orders, $\sM$ and
$\sM'$, are equivalent if they can deform into each other smoothly without
closing the energy gap (\ie without phase transition), \ie via a continuous
path.  However, there different paths that correspond to different ways that
$\sM$ and $\sM'$ are equivalent.  Those different classes of paths are
described by $\pi_1(\fM^{n+1})$.

Such a deformation corresponds to an \textbf{invertible domain wall} (which is
always gapped, see Definition \ref{invWall}) between the two topological
orders. Thus:
\begin{Definition}
Two anomaly-free topological orders $\sM$ and $\sM'$ are called
\textbf{equivalent} if they can be connected by an invertible domain wall
$\hat\ga$. We denote this isomorphism by $\sM \seq{\hat\ga} \sM'$ or $\hat\ga:
\sM \simeq \sM'$.  The invertible domain walls are classified by
$\pi_1(\fM^{n+1})$.
\end{Definition}
\noindent
The objects in $\afcTO^{n+1}$ are actually the equivalent classes of topological
orders, under the above equivalent relations (which correspond to isomorphisms
in category).  When we say two topological orders are the ``same'', they can be
equivalent in many different ways, described by different invertible domain
walls.

\subsection{Excitations in a topological order}

In an $n$d potentially anomalous topological order $\sC^{n+1}\in \cTO^{n+1}$
(\ie in $\npo$-dimensional spacetime), the point-like ($0$d), string-like (1d),
..., $\nmo$d excitations form a \textbf{fusion $n$-category}, which is denoted
as $\Hom(\sC^{n+1},\sC^{n+1})$, or simply $\cC^n$ (see Definition\ \ref{fc}).
By abusing the notation, we set 
\begin{align}
\Omega\sC^{n+1}:=\Hom(\sC^{n+1},\sC^{n+1})=\cC^n.  
\end{align}
We set the convention of the superscript:
$\Omega\sC^{n+1}=\Omega(\sC^{n+1})
$. Excitations of codimension-1 can be fused but not braided. If we exclude the
1-codimensional excitations, we obtain a braided fusion $\nmo$-category, which
is precisely the \textbf{looping} $\Om \cC^n$ of $\cC^n$ (see Section
\ref{looping}).  
The fusion $n$-category $\cC^n$ does not carry the full information about the
$n$d topological order, since $\cC^n$ only describes the excitations within the
$n$d topological order.  There are different topological orders (that differ by
stacking \textbf{invertible topological orders}\cite{KW1458,F1478,K1459}) which
have identical excitations.  To fully describe an $n$d topological order
$\sC^{n+1}$, we need not only the information about the excitations $\cC^n$,
but also the additional information on invertible topological orders. We can
also say that $n$d potentially anomalous topological orders (without any
symmetry) are classified, up to invertible topological orders, by fusion
$n$-categories\cite{KW1458,KZ150201690,J200306663,KZ200514178} (see  Proposition\
\ref{realizefcat} and \ref{FunVV}). 

Similar to topological order, it is tricky to determine if two fusion higher
categories are the ``same'' or not.  In general we can only say whether the two
fusion higher categories are equivalent or not.
\begin{Definition}
Two fusion higher categories, $\cM$ and $\cM'$ are \textbf{equivalent} if there
exist a functor $F:\cM\to \cM'$ and $G:\cM' \to \cM$ such that $F\circ G \simeq \id_\cM$ and $G\circ F \simeq \id_{\cM'}$, where $\simeq$ are natural isomorphisms. Such an equivalence $F$ is denoted by $\cM\seq{F}\cM'$ or $F: \cM\seq{}\cM'$.
\end{Definition}
Here, we like to clarify that, for simplicity we use the terms of functor,
natural isomorphism, algebra object, \etc,  while they should all be understood
as higher categorifications in higher categories.

\subsection{Holographic principle for topological order}

It was pointed out in \Ref{KW1458} that a potentially anomalous $\npo$D
topological order $\sC^{n+1}$ uniquely determines an \emph{anomaly-free}
topological order $\sM^{n+2}$ in one-higher dimension where $\sC^{n+1}$ can be
viewed as a boundary of $\sM^{n+2}$ (see Fig. \ref{TOa}b). This boundary-bulk
relation is the \emph{holographic principle of topological order}: ``anomaly''
= ``topological order in one-higher
dimension''.\cite{W1313,KW1458,KZ150201690,KZ170200673} Such a point of view on
anomaly is quite different from viewing anomaly as a non-invariance of path
integral measure, and the anomalies under the new point of view are in general
non-invertible (since the topological orders are in general non-invertible).
We denote this relation between two topological orders by 
\begin{align}
\sM^{n+2}=\Bulk(\sC^{n+1})
\end{align}
(see Proposition\ \ref{bulkCMp}).  Since
$\sM^{n+2}$ is anomaly-free, its bulk is trivial, i.e. 
\begin{align}
\Bulk(\sM^{n+2}) = \Bulk^2 (\sC^{n+1}) = \tto^{n+3}. 
\end{align}
In other words, $\Bulk$ is a ``categorified'' differential. 

In fact, we have a stronger version of the holographic principle (see
\eqn{bulkCM} and Fig. \ref{dEq}b): \frmbox{the excitations in the topological
order $\sC^{n+1}$, described by the fusion $n$-category $\Omega\sC^{n+1}$, can
already uniquely determine the bulk \emph{anomaly-free} topological order
$\sM^{n+2}$.  We denote the map from fusion higher categories $\Omega\sC^{n+1}$
to topological orders $\sM^{n+2}$ as
\begin{align}
 \bulk(\Omega\sC^{n+1}) = \sM^{n+2}.
\end{align}
}
The bulk topological order which in turn determines a fusion $n$-category $\Om
\sM^{n+2}$ describing excitations in $\sM^{n+2}$.  After dropping  all
1-codimensional excitations, we obtain a braided fusion $\nmo$-category
$\Omega^2\sM^{n+2}$. For simplicity, throughout this work, we also use the
following convention, for example, 
\begin{align}
\cC^n & := \Omega\sC^{n+1}, & \eC^{n-1} & := \Om \cC^n = \Omega^2\sC^{n+1}; 
\nonumber\\
\cM^{n+1} & := \Omega\sM^{n+2}, & \eM^n & := \Om \cM^{n+1} = \Omega^2\sM^{n+2}. 
\end{align}
The boundary-bulk relation $\bulk(\Omega\sC^{n+1}) = \bulk(\cC^n) = \sM^{n+2}$ reduces to the
main results in \Ref{KZ150201690,KZ170200673} (see Section~\ref{BBrel})
\begin{align}
\eM^n = Z_1(\cC^n) \quad \mbox{or} \quad 
\Omega^2\sM^{n+2} = Z_1(\Omega\sC^{n+1}),  
\end{align}
where $Z_1$ is the monoidal center (or $E_1$-center, or Drinfeld center for
fusion 1-categories).  For a more detailed description of topological orders in
arbitrary dimensions, see Section~\ref{toprev}, as well as
\Ref{KW1458,KZ150201690,KZ170200673,LW180108530,GJ190509566,J200306663,KZ200514178}.

\subsection{Algebraic higher symmetry}
\label{IIahs}

Now, we are ready to describe algebraic higher symmetry.
First, let us describe a very general view of symmetry.  
\begin{DefinitionPH}
A \textbf{symmetry} is simply ``a way'' to select a set of local operators
$\{O\}$, called \textbf{symmetric local operators}, that form a linear vector
space:
\begin{align}
 O_1 + O_2 \in \{O\},\ \ \forall \ \ O_1,O_2 \in  \{O\} ,
\end{align}
and form a linear algebra
\begin{align}
 O_1  O_2 \in \{O\},\ \ \forall \ \ O_1,O_2 \in  \{O\} .
\end{align}
\end{DefinitionPH}
\noindent
The symmetric Hamiltonians are simply sums of those selected local operators.

The standard way to select the symmetric local operators is via symmetry
transformations that form a group $G$:
\begin{align}
 \{O_G\ | \ W_g O_G = O_G W_g,\ g\in G\},
\end{align}
where the symmetry transformation $W_g$ acts on the whole space.  The
Hamiltonians formed by the sums of local operators in $\{O_G\}$ is said to have
a $0$-symmetry $G$.

For a 0-symmetry given by a group $G$ in spatial $n$-dimension, if the ground
state of a symmetric Hamiltonian is a symmetric product state, then point-like
excitations are described by the representations of $G$, which are called
\textbf{charged particles}. We denote the category of these representations by
$1\cRep G\equiv \cRep G$.  These excitations can be fused and braided, and can
be condensed to form higher dimensional excitations, called
\textbf{condensation descendants}.  All these excitations form a (symmetric)
fusion $n$-category, denoted by $n\cRep G$.  Due to Tannaka duality\cite{T3801}
between $\Rep(G)$ and $G$, the fusion and braiding properties (\ie the
conservation law) of the point-like excitations can fully determine the
symmetry group $G$.  When $n=2$, we believe that the constructed $n\cRep G$
outlined above is the same as that in \Ref{GKm0602510}.

In fact,  fusion $n$-category $n\cRep G$ can also determine a set of local
operators in $n$-dimensional space, denoted as $\{ O_{n\cRep G}\}$.  The set
$\{ O_{n\cRep G}\}$ describes all possible local interactions among the
excitations described by $n\cRep G$ that preserve all the fusion and braiding
properties of the excitations.  For example, $\{ O_{n\cRep G}\}$ contain all
the operators that create particle-anti-particle pairs.  It also contain all
the operators that create a small loop of string-like excitations, small ball
of membrane-like excitations, \etc.  There are also potential interactions
between those excitations.  We believe, all those operators generate the whole
set $\{ O_{n\cRep G}\}$.  However, $\{ O_{n\cRep G}\}$ does not contain
operators that create single particle that carries non-trivial representation
(\ie single charged particle).  Such operators will break the symmetry.  

In the above, we have described two ways (\ie two symmetries) which select two
sets of local operators, $\{ O_G\}$ and $\{ O_{n\cRep G}\}$.  We believe that
\frmbox{there is one-to-one correspondence between the local operators in the
two sets, $\{O_G\}$ and $\{ O_{n\cRep G} \}$, such that the two corresponding
local operators share the same properties (such as identical operator algebra
relations).  In other words, the linear algebras formed by $\{O_G\}$ and $\{
O_{n\cRep G} \}$ are isomorphic.} 
In general
\begin{DefinitionPH}
consider two symmetries (\ie two ways) that select two sets of local operators
$\{O\}$ and $\{O'\}$.  The two symmetries are said to be
\textbf{holographically equivalent (holo-equivalent)} if the linear algebras
formed by $\{O\}$ and $\{O'\}$ are isomorphic,
\end{DefinitionPH}
\noindent
The reason we use the term \emph{holographically} is due to the Propositions
\ref{MbulkR} and \ref{RRbulk}.  Note that two holo-equivalent symmetries may be
generated by transformations that are not related by a unitary transformation.
So ``holo-equivalent'' is more general then ``equivalent'' for symmetries.

Thus, the symmetry described by the transformations $G$ and the symmetry
described by the fusion $n$-category $n\cRep G$ are holo-equivalent. This
correspondence represents a categorical view of symmetry, which is heavily used
in \Ref{LW160205936,LW160205946}.

The 0-symmetry transformations $W_g$ that acts on the whole space can be
generalized so that the generalized symmetry transformations $W_i$ can act on
any loops, any closed membranes, \etc. We call the new symmetry \emph{algebraic
higher symmetry}, which can be beyond higher groups.  The algebraic higher
symmetry described by the transformations $W_\al$ select a set of 
local operators 
\begin{align}
\label{HW}
\{O_W \ | \ W_\al O_W = O_W W_\al\} ,
\end{align}
where $\al$ labels different symmetry transformations.  The label $\al$ may
include various closed subspaces of the space manifold, where the symmetry
acts.  In Section~\ref{Lgauge} and \ref{symprd}, we discuss some examples of
algebraic higher symmetries via the symmetry transformations $W_\al$.  But a
mathematical definition  (\ie a macroscopic definition not involving lattice)
of algebraic higher symmetries in terms of symmetry transformations $W_\al$ is
not easy to formulate.  

In the following, we will use the categorical view of symmetry to obtain a
mathematical definition of algebraic higher symmetry.  First, we have a
mathematical definition of \textbf{anomaly-free} property of  algebraic higher
symmetry: 
\begin{DefinitionPH}
\label{afahs}
An $n$d algebraic higher symmetry is \textbf{anomaly-free} if there exists a
symmetric gapped Hamiltonian in the same dimension whose ground state is a
non-degenerate product state.  Or in other words, the gapped ground state is
non-degenerate for any closed space manifolds.  Such non-degenerate ground
state is called \textbf{trivial symmetric state}.  The excitations on top of
such a ground state are called \textbf{charge objects}, which carry
``representations'' of the algebraic higher symmetry.
\end{DefinitionPH}
\noindent
We note that the excitations (the charge objects) may be point-like,
string-like, membrane-like, \etc.  In particular, for an algebraic $k$-symmetry
that acts on closed subspace of codimension-$k$, its charge objects has
dimension-$k$.

Motivated by the Tannaka duality of $0$-symmetry described by a group, we
propose that an anomaly-free algebraic higher symmetry in $n$d boson systems is
completely characterized by the excitations on top of its trivial symmetric
state.  In this paper, we use this property to define algebraic higher
symmetry. 

\begin{figure}[t]
\begin{center}
\includegraphics[scale=0.7]{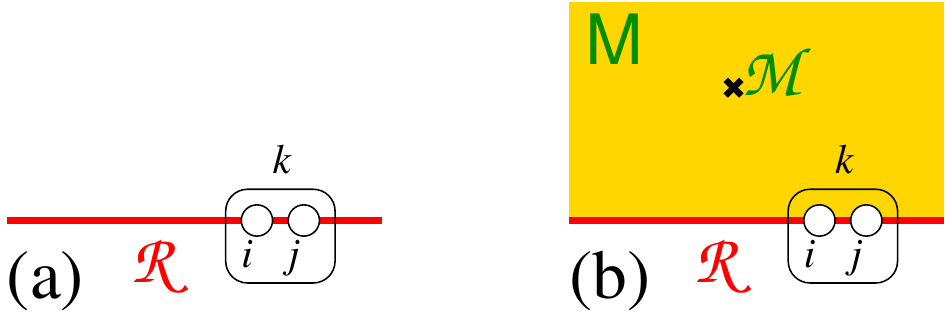} \end{center}
%3
\caption{ (a) An algebraic higher symmetry in $n$d bosonic systems is fully
characterized by its charge objects (the excitations in trivial symmetric
state), which form a local fusion $n$-category $\cR$.  The symmetry selects a
set of local operators $\{ O_\cR\}$ which are said to have the algebraic higher
symmetry $\cR$.  (b) A categorical symmetry for bosonic systems in
$n$-dimensional space is characterized by an anomaly-free topological order
$\sM$ in one higher dimension.  The categorical symmetry $\sM$ also select a
set of local operators, which is given by all the boundary local interactions,
$\{O_\sM\}$, of the bulk topological order $\sM$.  An algebraic
higher symmetry $\cR$ is holo-equivalent to a categorical symmetry given by
$\sM=\bulk(\cR)$.  The holo-equivalence means that the algebraic higher
symmetry $\cR$ and the categorical symmetry $\sM=\bulk(\cR)$ select equivalent
sets of local operators, \ie there is one-to-one correspondence between
$\{O_\cR\}$ and $\{O_\sM\}$, such that the two corresponding local
operators have the same operator algebra relations.  In this sense, the systems with an
algebraic higher symmetry $\cR$ also have the categorical symmetry
$\sM=\bulk(\cR)$.  } \label{RM}
\end{figure}

Those excitations on a trivial symmetric state form a very special fusion
$n$-category ${\cal R}$ (called the representation category of the symmetry).
To see in which way the fusion $n$-category is special, we note that the
symmetry decribed by $\cR$ can be explicitly broken.  This explicit symmetry
breaking process will change $\cR$ to another fusion $n$-category $n\cVec$,
where $n\cVec$ describes point-like, string-like, \etc\ excitations in a
product state without any symmetry (see Section~\ref{looping}).  So $\cR$ is a
special fusion $n$-category that is equipped with a top-faithful monoidal
functor $ {\cal R} \map{\bt} n\cVec $, where the functor $\bt$ describes the
explicit symmetry breaking process.  Such a fusion $n$-category $\cR$ is said
to be local.  \frmbox{The anomaly-free bosonic \textbf{algebraic higher
symmetries} are classified by local fusion $n$-categories $\cR$, \ie by the
data $\cR \map{\bt} n\cVec$ (see Fig.  \ref{RM}a and Section~\ref{lfc}).}  We
can use this classification as a formal definition of algebraic higher
symmetry.  For simplicity, in this paper, we usually drop $\bt$ and use the
representation category $\cR$ to describe an algebraic higher symmetry.  For
example, a finite 0-symmetry $G$ in $n$-dimensional space has a representation
category $n\cRep G$ and can also be referred as a $n\cRep G$ symmetry.

As a symmetry, the algebraic higher symmetry characterized by $\cR$, also
select a set of symmetric local operators $\{ O_\cR \}$, which describe all
possible local interactions between excitations described by $\cR$.  If the set
of local operators selected by the transformations $W_\al$ (see \eqn{HW})  has
a one-to-one correspondence with the set of local operators selected by the
local fusion $n$-category $\cR$, \ie if $\{O_W\} \simeq \{ O_\cR \}$, then
$\cR$ describes the algebraic higher symmetry defined by the transformations
$W_\al$.

It is possible that two local fusion $n$-categories, $\cR$ and $\cR'$, select
the equivalent local operator algebras. 
\begin{DefinitionPH}
If $\{ O_\cR \}$ and $\{ O_{\cR'} \}$ form isomorphic linear algebras (\ie
there is a one-to-one correspondence between $\{ O_\cR \}$ and $\{ O_{\cR'} \}$
such that the corresponding operators have the same operator algebra
relations),  then the two symmetries are called \textbf{holo-equivalent} .
\end{DefinitionPH}
\noindent
Later we will show that $n\cRep G$ and $n\cVec_G$ are both local fusion
$n$-categories if $G$ is a finite group.  Their corresponding algebraic higher
symmetries are holo-equivalent.

In this paper, we mainly discuss anomaly-free algebraic higher symmetry.  For
simplicity, by an algebraic higher symmetry we mean an anomaly-free algebraic
higher symmetry unless indicated otherwise.

We further generalize the notion of algebraic higher symmetry, by introducing a
notion of $\cV$-local fusion $n$-category (see Def. \ref{vlocal}), which has a
top-faithful surjective monoidal functor $ {\cal R} \map{\bt} \cV$,
where $\cV$ is a fusion $n$-category.  When $\cV=n\cVec$, $\cR$ describes
algebraic higher symmetry in $n$d bosonic systems.  When $\cV=n\scVec$, where
$n\scVec$ is the fusion $n$-category of super $n$-vector spaces, $\cR$
describes algebraic higher symmetry in $n$d fermionic systems.  \frmbox{The
anomaly-free fermionic algebraic higher symmetries are classified by the data
$\cR \map{\bt} n\scVec$, where $\cR$ is a fusion $n$-category.} For some
discussions on fermionic topological orders (with $\cR=n\scVec$) see
\Ref{GWW1017,LW150704673,BK160501640}, and on fermionic SPT/SET orders (with
$\cR\map{\bt}n\scVec$) see
\Ref{GW1441,KTT1429,WS14011142,GK150505856,LW160205946,FH160406527,WG170310937,KT170108264,LW180901112,GW181211959}.

More general choices of $\cV$ can describe systems formed by anyons or other
higher dimensional topological excitations.  So the notion of a generalized
algebraic higher symmetry allows us to study the symmetry of bosonic and
fermionic systems at equal footing.  It is interesting to see that the boson,
fermion, and anyon statistics can be encoded in a generalization of algebraic
higher symmetry.  

\subsection{Dual symmetry} \label{dsymm}

An algebraic higher symmetry can be understood via a more general notion:
\emph{categorical symmetry}.  Before explaining  categorical symmetry, let us
explain a simpler notion of dual symmetry.  It was pointed out in
\Ref{JW191213492} that an $n$d system with 0-symmetry $G$ also has a dual
algebraic $\nmo$-symmetry denoted by $\dual G^{(n-1)}$. 

We may use the holographic view to understand the appearance of the dual
symmetry. We note that the symmetric sub-Hilbert space of a $G$-symmetric
system in $n$-dimensional space can be viewed as a boundary of a
one-higher-dimensional $G$-gauge theory\cite{KK11045047,JW191213492} denoted by
$\sGT^{n+2}_G$.  The fusion (the conservation) of the bulk point-like gauge
charges in the $G$-gauge theory gives rise to the 0-symmetry $G$.  The bulk
$\sGT^{n+2}_G$ also has $\nmo$d gauge flux.  The fusion (the conservation) of
the bulk gauge flux in the $G$-gauge theory gives rise to the algebraic
$\nmo$-symmetry $\dual G^{(n-1)}$ (see Section~\ref{Lgauge}).  We stress that
both the 0-symmetry $G$ and the dual algebraic $\nmo$-symmetry $\dual
G^{(n-1)}$ are present at all the boundaries if we view the boundaries as
lattice boundary Hamiltonians or lattice boundary conditions\cite{JW191213492}
(for details, see next subsection).  However, for a gapped boundary, viewed as
a quantum ground state, one of the 0-symmetry and algebraic $\nmo$-symmetry, or
some of their combinations must be spontaneously broken
\cite{L190309028,JW191213492}.  

If we condense all gauge flux, we obtain a boundary with the 0-symmetry $G$ and
the spontaneously broken algebraic $\nmo$-symmetry $\dual G^{(n-1)}$.  The
boundary excitations are described by $n\cRep G$.  This boundary corresponds
to the usual $G$-symmetric product state whose excitations are also described
by $n\cRep G$.  

If we condensed all gauge charges, we obtain a boundary with the
dual algebraic $\nmo$-symmetry $\dual G^{(n-1)}$ and the spontaneously broken
0-symmetry $G$.  The boundary excitations are described by a local fusion
$n$-category $n\cVec_G$.  This is the usual spontaneous $G$-symmetry breaking
state.  The non-trivial fusion (the conservation) of the symmetry breaking
domain walls is also described by $n\cVec_G$, which gives rise to the dual
algebraic $\nmo$-symmetry $\dual G^{(n-1)}$.  Thus the dual symmetry
$\dual G^{(n-1)}$ can also be represented by its representations category,
which is just the fusion $n$-category, $n\cVec_G$, of $G$-graded vector spaces.  
For such a boundary, the  dual algebraic $\nmo$-symmetry
$\dual G^{(n-1)}$ is not spontaneously broken.

If the boundary Hamiltonians have both the  0-symmetry $G$ and the dual
algebraic $\nmo$-symmetry $\dual G^{(n-1)}$, we should see a boundary phase
where both the 0-symmetry $G$ and the dual algebraic $\nmo$-symmetry
$\dual G^{(n-1)}$ are not spontaneously broken.  Indeed, such a boundary
phase does exist, and it must be gapless.  This is because to get a gapped
boundary, we must condense enough bulk excitations at the boundary, which break
one of the 0-symmetry and algebraic $\nmo$-symmetry, or some of their
combinations.  If we do not condense any bulk excitations, the boundary can
only be gapless.\cite{KZ190504924,KZ191201760}.

We see that it is better to view a system with $G$-symmetry as a boundary of
the $G$-gauge theory in one-higher-dimension. This holographic point of view
allows us to see the accompanying dual symmetry (i.e. the algebraic
$\nmo$-symmetry $\dual G^{(n-1)}$) clearly.  Using a categorical language, the
point-like excitations carrying group representations (the charge objects) in
an $n$d $G$-symmetric product state generate a local fusion $n$-category
$n\cRep G$.  The same local fusion $n$-category $n\cRep G$ also describes the
excitations on a boundary of $G$-gauge theory $\sGT^{n+2}_G$, i.e.
$\sGT^{n+2}_G = \bulk(n\cRep G)$ (see Section~\ref{BBrel}).  This links the
0-symmetry $G$ to the $G$-gauge theory $\sGT^{n+2}_G$ in one higher dimension.
The boundary with excitations $n\cRep G$ can be obtained from $\sGT^{n+2}_G$ by
condensing the gauge flux.

$\sGT^{n+2}_G$ has another boundary whose excitations are described by another fusion $n$-category $n\cVec_G$. This boundary is obtained by condensing gauge charges. In this case, the gauge-flux excitations are not condensed, and their non-trivial fusion gives rise to the dual algebraic $\nmo$-symmetry $\dual G^{(n-1)}$.  In fact, $n\cVec_G$ is the representation category
that describes the charge objects of the dual symmetry $\dual G^{(n-1)}$. In summary, we have 
\begin{align}
\sGT^{n+2}_G = \bulk(n\cVec_G) = \bulk(n\cRep G).
\end{align}

We see that both 0-symmetry $G$ and its dual algebraic $\nmo$-symmetry $\dual
G^{(n-1)}$ share the same $G$-gauge theory $\sGT^{n+2}_G$ in one higher
dimension.  Thus, we can view $\sGT^{n+2}_G$ as a combined symmetry, denoted by
$G \vee \dual G^{(n-1)}$. The combined symmetry is referred as categorical
symmetry.  It is in this sense we say that the categorical symmetry $G \vee
\dual G^{(n-1)}$ is bigger then the symmetry $G$ and the dual symmetry $\dual
G^{(n-1)}$.  We like to mention that the combined symmetry is similar to the
``materialized symmetry'' in \Ref{K032}.  However, there is a difference: the
categorical symmetry $G \vee \dual G^{(n-1)}$ is a symmetry on $n$d boundary,
while the materialized symmetry is for $\npo$d bulk.

It is possible to realize above model-independent discussion by concrete
lattice models.  We expect that Levin-Wen type of lattice models can be
generalized to higher dimensions.  Similar to the 2+1D case
\cite{LW0510,KK11045047}, an $n+$2D model is built on a chosen fusion
$n$-category $\cC$ and a gapped boundary is built on a chosen $\cC$-module.
Then the $G$-gauge theory $\sGT^{n+2}_G$ can be realized by such a lattice
model by choosing $\cC=n\cRep G$. One of its gapped boundary $n\cRep G$ can be
realized by the boundary lattice model built on the obvious $n\cRep G$-module
$n\cRep G$.  The other gapped boundary $n\cVec_G$ can be realized by the
boundary lattice model built on the $n\cRep G$-module $n\cVec$, where the
module structure on $n\cVec$ is induced from the fiber functor $n\cRep G \to
n\cVec$, and $n\cVec_G$ is the category of $n\cRep G$-module endo-functors on
$n\cVec$. Mathematically, it is just a manifestation of Morita equivalence
between $n\cRep G$ and $n\cVec_G$.

We would like to mention that a structure similar to categorical symmetry was
found previously in AdS/CFT correspondence,\cite{M9831,Wh9802150,HO181005338}
where a global symmetry $G$ at the high-energy boundary is related to a gauge
theory of group $G$ in the low-energy bulk. In this paper, we stress that the
categorical symmetry encoded by the bulk $G$-gauge theory not only contains the
$G$ symmetry at the boundary, it also contains a dual algebraic higher symmetry
$\dual G^{(n-1)}$ at the boundary.  We developed a categorical theory for this
holographic point of view for both 0-symmetry and algebraic higher symmetry.
This allows us to gauge the algebraic higher symmetry, classify the anomalies
for a given algebraic higher symmetry, identify which algebraic higher
symmetries are holo-equivalent, identify duality relations for low energy
effective theories, and classify SET/SPT orders with a given algebraic higher
symmetry.

\subsection{Categorical symmetry -- a holographic of view of symmetry}
\label{IIcats}

The above is just the simplest example of categorical symmetry.  We can
generalize the above discussion, and show that, when restricted to the
symmetric sub-Hilbert space, an $n$d system with an algebraic higher symmetry
$\cR$ can be viewed as a boundary of an anomaly-free topological order
$\sM=\bulk(\cR)$ (see \eqn{bulkCM}).  This allows us to see that our system
actually has a categorical symmetry, characterized by topological order
$\sM$.  

Let us first define what is a categorical symmetry.  
In short, 
\begin{align}
&\ \ \ \
\text{a categorical symmetry}
 \\
& = \text{a non-invertible gravitational anomaly}
\nonumber\\
& =
\text{a topological order in one higher dimension}.
\nonumber 
\end{align}
To give a more detailed definition, we note that a symmetry is explicitly
defined via the algebra of the symmetric local operators that its selects.  Let
us define categorical symmetry this way.
\begin{DefinitionPH}
For an $\npo$d
anomaly-free topological order $\sM$, the corresponding \textbf{categorical
symmetry} is given by
\begin{itemize}
  \item a special boundary of $\sM$ such that all the excitations in $\sM$ are
either condensed or have nearly zero energy gap. All the bulk excitations
have an energy gap larger than a positive fixed value $\Del_\text{bulk}$.
Those nearly zero-energy boundary excitations
define a low energy boundary Hilbert space;
  \item the symmetric local operators $\{ O_\sM \}$ are the local
operators acting within the low energy boundary Hilbert space.
\end{itemize}
\end{DefinitionPH}
\noindent
We note that a bulk topological order $\sM$ can have many different special
boundaries that satisfy the above conditions.  We conjecture that different
choices of the special boundaries give rise to different sets of symmetric
local operators, $\{ O_\sM \}$ and $\{ O_\sM' \}$, that generate equivalent
operator algebra.  In other words, $\{ O_\sM \}$ and $\{ O_\sM' \}$ are holo-equivalent.

Although we define categorical symmetry via a topological order in one higher
dimension, in fact, as pointed out in \Ref{JW191213492}, at least some
categorical symmetries can be defined via the \emph{patch symmetry
transformations} without going to one higher dimension. So we believe that the
categorical symmetry is really a property of $n$d systems.

\begin{figure}[t]
\begin{center}
\includegraphics[scale=0.7]{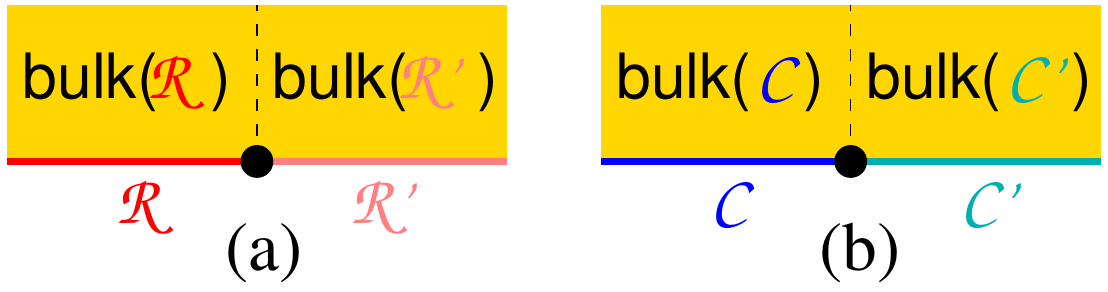} \end{center}
%4
\caption{ (a) Two algebraic higher symmetries $\cR$ and $\cR'$ are
holo-equivalent if they have the same categorical symmetry $\bulk(\cR)\simeq
\bulk(\cR')$.  (b) Two sets of low energy excitations $\cC$ and $\cC'$ are
holo-equivalent if they have the same categorical symmetry $\bulk(\cC)\simeq
\bulk(\cC')$.  Here holo-equivalent means the states with symmetry $\cR$ or
$\cR'$ (or formed by $\cC$ or $\cC'$) have an one-to-one correspondence.  }
\label{dEq}
\end{figure}

For an $n$d categorical symmetry described by an $\npo$d topological order
$\sM$, consider one of its special boundary, such that all the excitations in
$\sM$ are either condensed or have small but non-zero energy gap on the
boundary.  Here small means much smaller than the bulk gap $\Del_\text{bulk}$.
In this case, the  special boundary can be viewed as a gapped boundary, whose
non-condensing excitations are described by a fusion $n$-category $\cR$ that
satisfy $\bulk(\cR)=\sM$.  The fusion $n$-category $\cR$ defines an algebra
higher symmetry which is holo-equivalent to the categorical symmetry $\sM$.  In
other words, $\cR$ selects a set of symmetric local operators $\{ O_\cR\}$ and
$\sM$ selects a set of symmetric local operators $\{ O_\sM\}$.  The two sets of
local operators generate equivalent algebra.
We find that (see Fig. \ref{RM}b and Section~\ref{BBrel}).  
\begin{Proposition}
\label{MbulkR}
an algebraic higher symmetry $\cR$ and a categorical symmetry $\sM$ are
holo-equivalent, \ie $\{ O_\cR\}$ and $\{ O_\sM\}$ are isomorphic linear alegras,
if and only if $\sM\simeq \bulk(\cR)$.
\end{Proposition}
\noindent
Using the notion of categorical symmetry, we can easily tell when two algebraic
higher symmetries are holo-equivalent 
\begin{Proposition}
\label{RRbulk}
two algebraic higher symmetries, $\cR$ and $\cR'$, are holo-equivalent if and only
if $\bulk(\cR)\simeq \bulk(\cR')$ (see Fig. \ref{dEq}a).
\end{Proposition}

We note that the dimension-0, dimension-1, \etc\ excitations described by $\eM$
in the bulk topological order $\sM$, can be viewed as the dimension-0,
dimension-1, \etc\ excitations on the boundary, if they do not condense.  The
fusion rule of those bulk excitations corresponds to the conservation law,
which leads to the categorical symmetry of the boundary (where the boundary is
viewed as a system).  However, when the boundary is viewed as a ground state,
some of the bulk excitations may condense on the boundary, and the categorical
symmetry associated with those condensing excitations are spontaneously broken.
So the boundary, when viewed as a system (\ie as a Hamiltomian), has the full
categorical symmetry. But when viewed as a state, the boundary may
spontaneously break part of the categorical symmetry due to condensation of
bulk excitations.  From this point of view, the categorical symmetry has some
special properties:\cite{L190309028,JW191213492}
For a system with a non-trivial categorical symmetry ${\sM}$, 
\begin{enumerate}
\item
its gapped ground state must
spontaneously break the categorical symmetry partially
(\ie some excitations in $\eM=\Om^2\sM$ condense);
\item
it is impossible to spontaneously break the categorical symmetry completely in
a gapped state, and possibly, nor in a gapless state (\ie  it is impossible to condense all excitations in $\eM$);  
\item
the symmetric ground
state with the full categorical symmetry must be gapless (\ie if none of the excitations in $\eM$ condense, the boundary must be gapless).
\end{enumerate}

To see an example of categorical symmetry, \Ref{JW191213492} shows that a
Hamiltonian on $n$-dimensional lattice with 0-symmetry $\Z_2$ also has a
$\nmo$-symmetry $\Z_2^{(n-1)}$.  So the system actually has a larger $\Z_2\vee
\Z_2^{(n-1)}$ categorical symmetry.  Such a  $\Z_2\vee \Z_2^{(n-1)}$
categorical symmetry is nothing but the $\Z_2$ topological order
$\sGT^{n+2}_{\Z_2}$ (or $\Z_2$ gauge theory) in one higher dimension (\ie in
$\npo$-dimensional space).  The $\Z_2$ symmetry corresponds to the mod-2
conservation of the point-like $\Z_2$ gauge charge.  The $\Z_2^{(n-1)}$
$\nmo$-symmetry corresponds to the mod-2 conservation of the $\nmo$-dimensional
$\Z_2$ gauge flux.

The above system can have a gapped phase where $\Z_2^{(n-1)}$ is
spontaneously broken and $\Z_2$ is not broken, which is the usual $\Z_2$
symmetric phase.\cite{JW191213492} Using the categorical language, we may say
that this phase has an (un-broken) algebraic higher symmetry characterized by
the local fusion $n$-category $\cR=n\cRep\Z_2$ (which is nothing but the usual
$\Z_2$ 0-symmetry).  

The system can also have a gapped phase where $\Z_2$ is spontaneously broken
and $\Z_2^{(n-1)}$ is not broken, which is the usual $\Z_2$ symmetry broken
phase.\cite{JW191213492}  This phase has an (un-broken) algebraic higher
symmetry characterized by the local fusion $n$-category $\cR=n\cVec_{\Z_2}$,
which describes the conservation of symmetry-breaking domain walls.  

The quantum critical point of the $\Z_2$ symmetry breaking transition has the
full categorical symmetry $\Z_2\vee \Z_2^{(n-1)}$.  In particular, in
$1$-dimensional space ($n=1$), the $\Z_2\vee \Z_2^{(0)}$ categorical
symmetry leads to the emergent $\Z_2\times \Z_2$ symmetry for right-movers and
left-movers \cite{JW191213492}.

\subsection{Emergence of algebraic higher symmetry and
categorical symmetry}

In a practical $n$d condensed matter system, we often have an on-site
0-symmetry described by a symmetry group $G$. Then the system also has a $G\vee
\dual G^{(n-1)}$ categorical symmetry.  But how to have a more general higher
symmetry or algebraic higher symmetry $\cR$, as well as their associated
categorical symmetry $\sM=\bulk(\cR)$ in a practical condensed matter system?
Certainly, we can try to realize algebraic higher symmetry by fine tuning.
Here we will describe a situation to have an algebraic higher symmetry without
fine tuning. In fact, algebraic higher symmetry and categorical symmetry can
emerge at low energies.

We will first discuss the emergence of a categorical symmetry $\sM$.  Once we
have an emergent categorical symmetry $\sM$ (which may or may not be
spontaneously broken), then we can determine the emergent algebraic higher
symmetry $\cR$ (which may or may not be spontaneously broken) by solving the
equation $\bulk(\cR) = \sM$.  Such a equation may have many solutions for
$\cR$, but different solutions are all holo-equivalent.

Let us consider a topological order (with or without symmetry) on an $n$d
lattice, whose excitations are described by a fusion $n$-category $\cC$. $\cC$
may contain topological excitations not associated with symmetry. $\cC$ may
also contain charge objects if we have symmetry.  
Assuming the excitations have a large separation of energy scale, such that all
the low energy excitations (point-like, string-like, \etc) are described a
subcategory $\cC^\text{low}$ of $\cC$.  All other excitations not in
$\cC^\text{low}$ have large energy gaps which is assumed to be infinity.  Thus
at low energies, we only see the excitations in $\cC^\text{low}$. Here we treat
all excitations in $\cC^\text{low}$ at equal footing, and do not distinguish
which excitations are charge objects from a symmetry and which excitations are
topological excitations.  In other words, we pretend all the excitations in
$\cC^\text{low}$ to be topological excitations and pretend the system to have a
(potentially anomalous) topological order without symmetry, whose excitations
are described by $\cC^\text{low}$.  

We see that once we know the low energy excitations $\cC^\text{low}$ (which may
contain possible charge objects from symmetry), the higher energy lattice
regularization becomes irrelevant.  Thus we can directly consider a field
theory with low energy excitations $\cC^\text{low}$.  We ask what is the low
energy emergent categorical symmetry?
The answer is very simple: \frmbox{the low energy effective categorical
symmetry for a $n$d \emph{field theory} with low energy excitations
$\cC^\text{low}$ is given by a topological order $\sM^\text{low}
=\bulk(\cC^\text{low})$ (see \eqn{bulkCM}) in one higher dimension.} Here by
\emph{field theory}, we mean a theory whose UV regularization is not specified.
When we say a field theory have a property, we mean that there exist a UV
regularization of the field theory, such that the regularized theory has the
property.  It is possible that the same field theory with a different
regularization may not have the property.  In particular, when we say two field
theories are connected by phase transitions, we mean that for any UV
regularization of the first field theory, we can find a UV regularization for
the second field theory, such that the regularized theories are connected  by
phase transitions.

The emergent categorical symmetry $\sM^\text{low}$ is very useful (see Section
\ref{effCS}):  \frmbox{ The categorical symmetry $\sM^\text{low}$ represents
the full information that controls all the low energy properties of the
system.}  For example, given a set of low energy excitations $\cC^\text{low}$,
we like to ask, when the low energy excitations condense, what kind new phases
are possible?  What kind of critical points are possible at the phase
transitions?  Do we have any principle to address those issues?  The answer is
yes, and the answer is the emergent categorical symmetry.  This because all the
possible low energy systems (described by all possible interactions of
excitations in $\cC^\text{low}$) share the same emergent categorical symmetry
$\sM^\text{low}$.  We may view the emergent categorical symmetry
$\sM^\text{low}$ as an ``topological invariant'' of the low energy systems.  We
believe that all other topological invariants of the low energy systems are
contained in the emergent categorical symmetry $\sM^\text{low}$.

In this paper, we obtain many results assuming exact algebraic higher symmetry
and categorical symmetry.  Those result remain valid for systems with emergent
categorical symmetry $\sM^\text{low} =\bulk(\cC^\text{low})$.
This allows us to
apply the results of this paper to some practical situations.  In the next
subsection, we consider two applications along this line.

\subsection{Categorical symmetry and duality}

A symmetry is useful since it can constrain the properties of a system, such as
possible phases and phase transitions, the critical properties at the phase
transitions, \etc.  From the above discussion, we see that the constraint from
a symmetry actually comes from the corresponding categorical symmetry.  This
is because the possible physical properties of a system with an algebraic
higher symmetry $\cR$ are the same as the possible physical properties of a
boundary of the topological order $\sM=\bulk(\cR)$ in one higher dimension.
In particular, as we have mentioned before, if two symmetries $\cR$ and
$\cR'$ have the equivalent categorical symmetry $\bulk(\cR)\simeq \bulk(\cR')$,
then the two symmetries provide the same constraint on the physical properties
(see Fig. \ref{RM} and \ref{dEq}a), at least within the symmetric sub-Hilbert
space.  In this case, the two symmetries are holo-equivalent (see
Sections \ref{csym} and \ref{effCS}).

Here, we like state a stronger result: \frmbox{if two algebraic higher
symmetries $\cR$ and $\cR'$ have the equivalent monoidal center $Z_1(\cR)
\simeq Z_1(\cR')$, then the two symmetries provide the same constraint on the
physical properties, and the two symmetries are holo-equivalent.} In other
words, the sets of local operators selected by the two symmetries, $\{ O_\cR\}$
and $\{ O_{\cR'}\}$, have an one-to-one correspondence (for example via a
duality transformation, see \Ref{JW191213492}) and generate the same algebra.
The Hamiltonians as sums of those symmetric local operators also have an
one-to-one correspondence, and the corresponding Hamiltonians have the same
spectrum.

The above result is motivated by the following consideration: Let
$\sM^\text{inv}$ be an invertible topological order in $\npo$-dimensional
space, and $\cC_0$ be the fusion $n$-category describing the excitations in one
gapped boundary of $\sM^\text{inv}$.  Then $\cR$ and $\cR'\equiv \cR\otimes
\cC_0$ will have the same monoidal center $Z_1(\cR')= Z_1(\cR)$, but different
bulks: $\bulk(\cR')= \sM\otimes \sM^\text{inv} \neq \bulk(\cR)=\sM$.
Therefore, requiring $Z_1(\cR')= Z_1(\cR)$ does not imply $\bulk(\cR')\simeq
\bulk(\cR)$ and does not imply the holo-equivalence.  However, if $\cR$ is
local and describe an algebraic higher symmetry, then $\cR'= \cR\otimes \cC_0$
is not local and does not describe an algebraic higher symmetry.  In other
words, the excitations in $\cC_0$ are topological, which comes from the
invertible topological order $\sM^\text{inv}$ in one higher dimension.
Symmetry breaking cannot make them trivial.  This is why we think that there is
no top-faithful functor $\bt$ that map $\cR'= \cR\otimes \cC_0$ into $n\cVec$.
Thus we believe that 
\begin{Proposition}
\label{RZ1}
if $\cR$ and $\cR'$ are both local (\ie both describe algebraic higher
symmetries), then $Z_1(\cR')\simeq Z_1(\cR)$ implies $\bulk(\cR')\simeq\bulk(\cR)$.
\end{Proposition}

As a result, all possible phases in a system with $\cR$ symmetry have a
one-to-one correspondence with all possible phases in a system with $\cR'$
symmetry.  In fact, we have a stronger result, all possible states on a system
with $\cR$ symmetry have an one-to-one correspondence with all possible states
on a system with $\cR'$ symmetry.  Those states include gapped states and
gapless states \etc. In \Ref{JW191213492}, some lattice exact duality mappings
were discussed for some very simple examples to explicitly demonstrate such a
result.  This duality relation can be an important application of categorical
symmetry.

For example, an $n$d system with $G$ 0-symmetry can be mapped to an $n$d system
with the dual $\dual G^{(n-1)}$ $\nmo$-symmetry, and vice versa.  The $G$
0-symmetry and the $\dual G^{(n-1)}$ $\nmo$-symmetry are holo-equivalent
symmetries.  Using the categorical notation, we say the $n\cRep G$ symmetry and
the $n\cVec_G$ symmetry are holo-equivalent symmetries, since $Z_1(n\cRep
G)\simeq Z_1(n\cVec_G)$.

The above duality result can be generalized even further (see Fig. \ref{dEq}b and Section~\ref{effCS}):
\frmbox{Consider two $n$d field theories with low energy excitations described
by two fusion $n$-categories $\cC$ and $\cC'$ respectively.   The two field
theories are dual to each other (\ie are \textbf{holo-equivalent}), if they have
equivalent categorical symmetries $\bulk(\cC)\simeq \bulk(\cC')$ (see
\eqn{bulkCM}), provided that all other excitations remain to have high
energies.} We like to remark that the two field theories may have different
symmetries described by different charge objects, forming two different
subcategories in $\cC$ and $\cC'$.  The two field theories may also have
different low energy topological excitations.  In other words, we do not care
which excitations are topological excitations and which excitations are charge
objects of the symmetries.

When two systems have the same
categorical symmetry $\sM$, both systems can be simulated by the boundaries
of the same bulk topological order $\sM$ (since the categorical symmetry is
the bulk topological order).  Hence the two systems are holo-equivalent. This
means that the possible states of the system $\cC$ (including condensed states,
gapless states, \etc) have an one-to-one correspondence with the possible
states of the system $\cC'$ (see Section~\ref{effCS}).  Those states are just the
possible boundary states of the same $\sM$.

\subsection{Gauging the algebraic higher symmetry and the corresponding
$\cR$-gauge theory}

\begin{figure}[t]
\begin{center}
\includegraphics[scale=0.7]{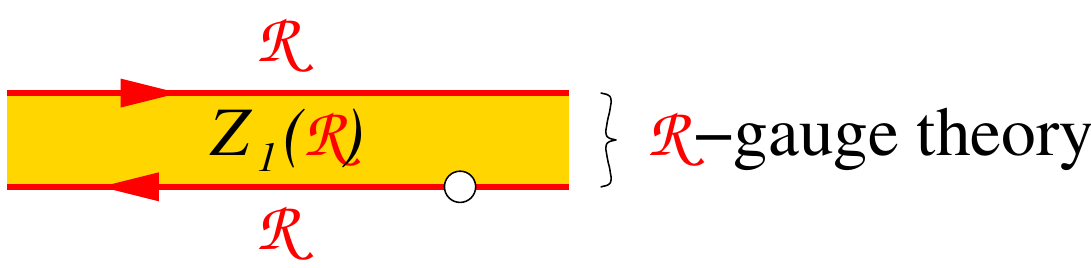} \end{center}
%5
\caption{Gauging the $\cR$-symmetry: stacking two local fusion $n$-category
$\cR$ over their common bulk $Z_1(\cR)$ gives rise to an fusion $n$-category
$\cR\ot{Z_1(\cR)}\cR^\rev$, describing the excitations in $n$d anomaly-free
topological order $\sGT_\cR^{n+1}$, which is the $\cR$-gauge theory.  The
boundary Hamiltonians of $\cR$-gauge theory share the same low energy
properties with the Hamiltonians with algebraic higher symmetry $\Om\cR$, if
$\cR=\Si\Om\cR$.  We like to remark that $\cR$'s on the two boundaries differ
by a parity transformation as indicated by the arrows and superscript $^\rev$.
} \label{RRgauge} 
\end{figure}

Given an $n$d product state with an on-site 0-symmetry $G$ (\ie an anomaly-free
0-symmetry), we can gauge the symmetry to obtain a state with topological order
and no symmetry.  The resulting topological order is nothing but the $G$-gauge
theory $\sGT^{n+1}_G$.  The excitations in $\sGT^{n+1}_G$ are described by a
fusion $n$-category $\Omega\sGT_G^{n+1}$.  In fact $\Omega\sGT_G^{n+1}=\Si
Z_1\big((n-1)\cRep(G)\big)$, where $\Si$ is the \textbf{delooping} (see Section
\ref{looping}).  

Similarly, given an $n$d product state with an anomaly-free higher symmetry, we
can gauge the higher symmetry to obtain a state with topological order and no
symmetry.  The resulting topological order is described by a higher gauge
theory.

Now  given an $n$d product state with an anomaly-free algebraic higher symmetry
$\cR$, can we gauge the algebraic higher symmetry to obtain a state with
topological order and no symmetry?  If we can, then the corresponding
topological order is a gauge theory for the algebraic higher symmetry $\cR$.
We denote such a gauge theory by $\sGT^{n+1}_\cR$, the excitations in which
form a fusion $n$-category $\Omega\sGT_\cR$.

In this paper, we propose a way to gauge  algebraic higher symmetry, which
gives us a construction of $\cR$-gauge theory (by constructing the
corresponding topological order $\sGT^{n+1}_\cR$).  Our approach is based on
the holographic view of  the $\cR$-symmetry, which is very different from the
usual gauging based on spacetime dependent symmetry transformations.

Under the holographic point of view, an algebraic higher symmetry $\cR$ gives
rise to a 1-higher-dimensional topological order $\sM$ such that
$\Omega\sM=Z_1(\cR)$ (see Fig. \ref{RM})b.  Then the topological order obtained
by gauging the $\cR$-symmetry in a symmetric product state can be obtained by
simply stacking two $\cR$ boundaries through their common bulk $Z_1(\cR)$ (see
Fig.  \ref{RRgauge}).  This is an algebraic way to gauge an symmetry, which
work for 0-symmetries, higher symmetries, and algebraic higher symmetries (see
Section~\ref{Rgauge}).  \frmbox{The excitations in an $n$d $\cR$-gauge theory
$\sGT_\cR^{n+1}$ are described by an multi-fusion $n$-category
\begin{align}
\Om\sGT_\cR^{n+1} =Z_0(\cR)= \cR\bxt{Z_1(\cR)}\cR^\rev ,
\end{align}
where $Z_0(\cR):=\Fun(\cR,\cR)$ is the $E_0$-center.  }

\subsection{Dual of an algebraic higher symmetry}

\begin{figure}[t]
\begin{center}
\includegraphics[scale=0.7]{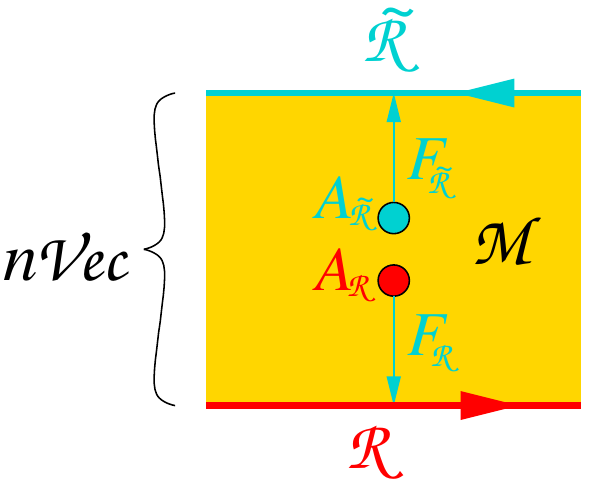} \end{center}
%6
\caption{ 
Stacking $\cR$ and $\dual\cR$ through their common bulk gives rise to a trivial
product state $\cR\ot{\eM}\dual\cR^\rev=n\cVec$, if the two algebraic higher symmetries,
$\cR$ and $\dual\cR$, are dual to each other. Condensing the condensable algebra
$A_\cR$ (resp. $A_{\dual\cR}$) produce the $\cR$ (resp. $\dual\cR$) boundary.  
The functor $F_\cR: \eM\to \cR$, induced by moving the bulk excitations to the boundary, maps $A_\cR$ to the trivial excitation on the $\cR$ boundary.  Similar for $F_{\dual\cR}$.  }
\label{RdR}
\end{figure}

Using a similar holographic approach, we can also define the \textbf{dual
symmetry} $\dual \cR$ for an arbitrary algebraic higher symmetry $\cR$ as
follows: 
$\cR$ and $\dual\cR$ are dual to each other, if they have the same
bulk $Z_1(\cR)=Z_1(\dual\cR)=\eM$ and if the stacking of $\cR$ and $\dual\cR$
through their bulk gives rise to a trivial topological order (\ie a product
state, see Fig. \ref{RdR}, Def.  \ref{RdRdef}, and Proposition~\ref{RBlocal}). 
\begin{align}
\cR\ot{\eM}\dual\cR^\rev = n\cVec.
\end{align}
Such a definition reproduces our previous result: the dual of $G$ 0-symmetry
is the $\t G^{(n-1)}$ $\nmo$-symmetry, \ie $n\cRep(G)$ and $n\cVec_G$ are dual
to each other.\cite{JW191213492}

We would like to mention that a gapped boundary of $\eM$, such as $\cR$, is
induced by condensing some excitations in $\eM$ at the boundary.  The
collection of those condensing excitations form a so called condensable algebra
$A_\cR$.  The condensable algebra $A_\cR$ uniquely determines the gapped
boundary.  A different condensable algebra $A_{\dual\cR}$ unique determines a
different gapped boundary $\dual\cR$.  Roughly speaking, moving bulk excitation
in $\eM$ to the $\cR$ boundary induces a map from $\eM$ to $\cR$, the
mathematical description of which is a functor $F_\cR: \eM\to \cR$.  Under such
a map, the condensable algebra $A_\cR$ is mapped to the trivial excitations in
$\cR$ (\ie condensed,  see Fig.  \ref{RdR}).

\subsection{Anomalous algebraic higher symmetry}

Can an algebraic higher symmetry have anomaly? How to describe its anomaly?
First, an anomalous symmetry is characterized by two things: symmetry and
anomaly.  So an anomalous algebraic higher symmetry is characterized by a pair
$(\cR,\al)$, where $\cR$ is for symmetry and $\al$ for anomaly.  

For a 0-symmetry in $n$-dimensional space, an anomalous symmetry is
characterized by a pair $(G,\om_{n+2})$ where $\om_{n+2}\in H^{n+2}(G,U(1))$ is
an $\npt$-cocycle.  A more physical way to understand the anomalous symmetry
$(G,\om_{n+2})$ is to view it as the boundary symmetry of a 1-dimension-higher
SPT state,\cite{W1313} which is also characterized by the pair $(G,\om_{n+2})$.
We can gauge the $G$-symmetry in the SPT state to get a ``twisted'' $G$-gauge
theory (the Dijkgraaf-Witten theory\cite{DW9093}), denoted by
$\sGT^{n+2}_{G,\om_{n+2}}$.  In fact, $\sGT^{n+2}_{G,\om_{n+2}}$ is the
categorical symmetry that is holo-equivalent to the anomalous symmetry
$(G,\om_{n+2})$.  Thus, we can also describe the anomalous symmetry
$(G,\om_{n+2})$ via its  holo-equivalent categorical symmetry
$\sGT^{n+2}_{G,\om_{n+2}}$, which is a ``twisted'' $G$-gauge theory in one
higher dimension.\cite{W1313,JW191213492} This is the point of view that we
will use in this paper.

In fact, under the holographic point of view, a ``twisted'' $G$ gauge theory in
one higher dimension defines an anomalous $0$-symmetry.  The boundaries of the
``twisted'' $G$ gauge theory give rise to all possible phases (including
symmetry breaking phases), as well as all other properties, of systems with the
anomalous $G$ 0-symmetry.  

\begin{figure}[t]
\begin{center}
\includegraphics[scale=0.7]{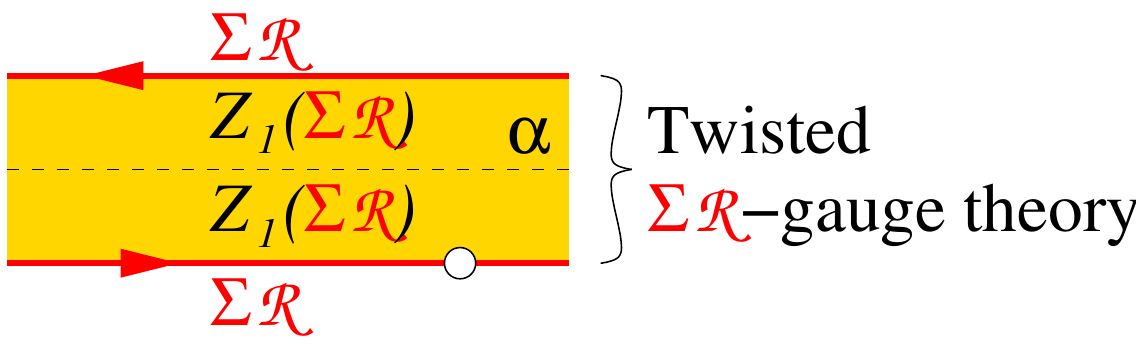} \end{center}
%7
\caption{The topological order described by a twisted $\Si\cR$ gauge theory has
excitations described by
$\Si\cR\ot{Z_1(\Si\cR)}\al\ot{Z_1(\Si\cR)}\Si\cR^\rev$.  The twist is done
by an automorphism $\al$ in $Z_1(\Si\cR)$, that keeps the
condensable algebraic $A_{\dual{\Si\cR}}$ of the dual symmetry $\dual{\Si\cR}$
unchanged $\al(A_{\dual{\Si\cR}}) \simeq A_{\dual{\Si\cR}}$. The boundary
Hamiltonians of twisted $\Si\cR$-gauge theory correspond to the Hamiltonians
with anomalous algebraic higher symmetry $\cR$.  } \label{RalRgauge}
\end{figure}

Similarly, an $n$d bosonic system with an anomalous higher symmetry described
by a higher group $\cB(G,\pi_2,\cdots)$ (using the notation in
\Ref{ZW180809394}) has a holo-equivalent categorical symmetry characterized
by a ``twisted'' higher gauge theory in one-higher dimension.  The different
anomalies for the higher group $\cB (G,\pi_2,\cdots)$ are (partially)
characterized by cocycles in $H^{n+2}[\cB (G,\pi_2,\cdots);\R/\Z]$.  The
``twisted'' higher gauge theory in one higher dimension defines the anomaly of
the anomalous higher symmetry.

The categorical symmetry of an $n$d Hamiltonian with an
anomaly-free algebraic higher symmetry $\cR$ is given by $\sM=\bulk(\cR)$.  We
like to ask whether $\sM=\bulk(\cR)$ describes the $\cR$-gauge theory in one
higher dimension.  The answer is no, simply because $\cR$ describes a symmetry
in $n$d, not in one higher dimension $\npo$d.  The gauge theory in one higher
dimension cannot be a $\cR$-gauge theory since $\cR$ lives in one lower
dimension.

When we discuss $G$-gauge theory in all the dimensions, we have used the fact
that the same $G$-symmetry can be defined in all the dimensions.  For an
algebraic symmetry $\cR$ in $n$d, what is the corresponding symmetry in one
higher dimension $\npo$d? This is a highly non-trivial question.  It turns out
that an algebraic symmetry, in general, cannot be promoted to one higher
dimensions.  Only a special class of algebraic symmetries, described by
symmetric local fusion $n$-categories, can be promoted to one higher dimension.
This is because a symmetric fusion $n$-category $\cR$ in $n$d can be viewed as
a braided fusion $n$-category, describing 2-codimensional and higher
excitations in one higher dimension (\ie in $\npo$d).  We then can do a
delooping to obtain a fusion $\npo$-category $\Si\cR$ (see Section
\ref{looping}). If $\cR$ is a symmetric local $n$-category, $\Si\cR$ is again a
symmetric local $\npo$-category.  So, $\Si\cR$ describes the $\cR$-symmetry in
one higher dimension.  Since $\Si\cR$ is also symmetric and local, we can
promote further to obtain the corresponding $\cR$ symmetry in all higher
dimensions $\Si^2\cR$, $\Si^3\cR$, \etc,.  So in this subsection, we assume
$\cR$ to be symmetric local fusion $n$-category.

Now, we can state the non-trivial result \frmbox{ the categorical symmetry
for an anomaly-free algebraic symmetry $\cR$, $\sM=\bulk(\cR)$, is the same as
the $\Si\cR$-gauge theory $\sGT^{n+2}_{\Si\cR}$ in one higher dimension:
\begin{align}
 \bulk(\cR) = \sGT^{n+2}_{\Si\cR},
\end{align}
provided that $\cR$ is symmetric.
}
The excitations in the  $\Si\cR$-gauge theory are given by
\begin{align}
\label{OmGTSiR}
\Om \sGT^{n+2}_{\Si\cR}  = \Si\cR \ot{Z_1(\Si\cR)}\Si\cR^\rev, 
\end{align}
which defines the \textbf{gauging} of the algebraic higher symmetry $\Si\cR$.
\Eqn{OmGTSiR} describes the excitations in $\bulk(\cR)$ given by $\Om
\bulk(\cR) = \Si Z_1(\cR)$. The fact that $\Si Z_1(\cR) = \Si\cR
\ot{Z_1(\Si\cR)}\Si\cR^\rev$ follows\cite{J200306663} from the following
identity\cite{KZ200308898}: 
\begin{align}
\Si Z_1(\cR) = Z_0(\Si\cR) = \Si\cR \ot{Z_1(\Si\cR)}\Si\cR^\rev
. 
\end{align}

Similarly, an anomalous algebraic higher symmetry $(\cR,\al)$ is defined
via its categorical symmetry which corresponds to a twisted $\Si\cR$-gauge
theory in one higher dimension.  The twist is produced by an automorphism
$\al$ of $Z_1(\Si\cR)$ (the dash-line in Fig.  \ref{RalRgauge}).  Such
a twisted $\Si\cR$ gauge theory, denoted by $\sGT_{\Si\cR,\al}^{n+2}$, is
characterized by its excitations described by (see Fig. \ref{RalRgauge})
\begin{align}
\Om \sGT_{\Si\cR,\al}^{n+2} = 
\Si\cR \ot{Z_1(\Si\cR)} \al \ot{Z_1(\Si\cR)}\Si\cR^\rev .
\end{align}
The automorphism $\al$ is not arbitrary. 
It must satisfy $\al(A_{\dual{\Si\cR}})\simeq
A_{\dual{\Si\cR}}$, where $A_{\dual{\Si\cR}}$ is the condensable algebra for
the dual symmetry of $\Si\cR$ (see Section~\ref{Rgauge}).  This result allows us
to classify types of anomalies that an algebraic higher symmetry can have.
Since invertible domain wall include invertible topological orders, the above
anomalies include invertible gravitational anomalies, symmetry ('t Hooft)
anomalies (which are always invertible), and invertible mixed
symmetry-gravitational anomalies.  To summarize, \frmbox{Anomalous algebraic
higher symmetries $\cR$ are classified by the automorphisms $\al$ of
$Z_1(\Si\cR)$ such that $ \al(A_{\dual{\Si\cR}})\simeq A_{\dual{\Si\cR}}$.  Its
categorical symmetry $\sM$ satisfies $\Om\sM=\Si\cR \ot{Z_1(\Si\cR)} \al
\ot{Z_1(\Si\cR)}\Si\cR^\rev $.  }

As an application of the above result, we consider 1d bosonic system with an
anomalous $\Z_2^3$ symmetry.  The possible anomalies are classified by
$H^3(\Z_2^3, U(1))$, which correspond to 2d $\Z_2^3$-SPT orders.  The
categorical symmetry of the 1d anomalous $\Z_2^3$ symmetry is given by the 2d
topological order obtained by gauging the corresponding 2d $\Z_2^3$-SPT states.
It was found that a particular anomalous $\Z_2^3$ symmetry, described by a so
called type-III cocycle in $H^3(\Z_2^3, U(1))$, has a categorical symmetry
described by the 2d $D_4$ gauge theory $\sGT^3_{D_4}$.\cite{GNm0603191,WW1454}
Certainly, the 1d anomaly-free $D_4$ symmetry also has a categorical symmetry
described by the 2d $D_4$ gauge theory $\sGT^3_{D_4}$.  Therefore, this
particular 1d anomalous $\Z_2^3$ symmetry is holo-equivalent to 1d $D_4$
symmetry.  In general, we conjecture that \frmbox{two anomalous algebraic
higher symmetries, $(\cR,\al)$ and $(\cR',\al')$, are holo-equivalent if they
satisfy
\begin{align}
&\ \ \ \
\Si\cR \ot{Z_1(\Si\cR)} \al \ot{Z_1(\Si\cR)}\Si\cR^\rev 
\nonumber\\
&\hspace{.5cm} \simeq \Si\cR' \ot{Z_1(\Si\cR')} \al' \ot{Z_1(\Si\cR')}{\Si\cR'}^\rev ,
\end{align}
which implies that $(\cR,\al)$ and $(\cR',\al')$ have the same categorical
symmetry.  }

One may ask: can a categorical symmetry $\sM$ be always viewed as an anomalous
algebraic higher symmetry?  If the categorical symmetry satisfies $\Om\sM =
\Si\cR \ot{Z_1(\Si\cR)} \al \ot{Z(\Si\cR)}\Si\cR^\rev $, then
the categorical symmetry can indeed be viewed as an anomalous $\cR$-symmetry.
We believe there exist categorical symmetries that do not have the above form,
and those categorical symmetries cannot be viewed as an anomalous algebraic
higher symmetry.  Categorical symmetry play a similar role as anomalous
symmetry, but it is more general.  In fact, after introducing categorical
symmetry, we do not need to use anomalous symmetry any more.  The effect of
anomalous symmetry can all be covered by categorical symmetry.

\subsection{Classification of gapped liquid phases for systems with a
categorical symmetry} \label{catsymmB}

\begin{figure}[t]
\begin{center}
\includegraphics[scale=0.7]{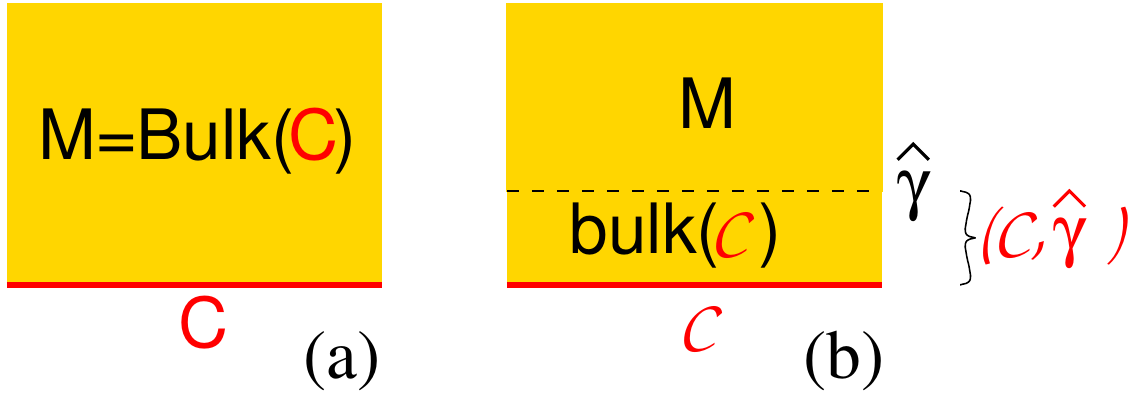} \end{center}
%8
\caption{ (a) Gapped liquid phases in systems with categorical symmetry $\sM$
are classified by anomalous topological orders $\sC$ such that
$\sM=\Bulk(\sC)$.  Those gapped liquid phases are partially classified by
fusion $n$-categories $\cC$ such that $\eM=\Om^2\sM=Z_1(\cC)$.  (b) Gapped
liquid phases in systems with categorical symmetry $\sM$ are classified %(in a many-to-one fashion)
by the pairs $(\cC,\hat\ga)$, where $\cC$ satisfies
$\bulk(\cC)=\sM$ and $\hat\ga$ is an invertible domain wall in $\sM$.  }
\label{CwM}
\end{figure}

We can also use the holographic point of view to classify SET/SPT orders with a
given algebraic higher symmetry $\cR$.  But first, let us classify possible
gapped liquid phases in $n$d systems with a categorical symmetry  $\sM$
(assuming $n\geq 1$).  Using boundary-bulk relation, we find that 
(see Section~\ref{pSETSPT}) 
\frmbox{for
$n$d lattice systems with a categorical symmetry  $\sM$, all their gapped
liquid phases (which partially break the categorical symmetry spontaneously)
are classified by potential anomalous $n$d topological orders $\sC$ (\ie $n$d
boundary topological orders) that satisfy (see Fig.  \ref{CwM}a)
\begin{align}
\sM=\Bulk(\sC) .
\end{align}
}
We note that the categorical symmetry $\sM$ is an anomaly-free
topological order in one higher dimension.  

Since the fusion $n$-categories $\cC$ (describing the excitations) partially
describes an $n$d topological order (up to invertible topological
orders\cite{KW1458,F1478,K1459} and SPT orders), we get a partial
classification if we only use $\cC$ (see Section~\ref{glpCS}): \frmbox{ the
gapped liquid phases (up to invertible topological orders and SPT orders) in
$n$d lattice systems with categorical symmetry $\sM$ are classified by fusion
$n$-categories $\cC$ that satisfy $\bulk(\cC) =\sM$.}

Motivated by the results in \Ref{LW160205936,LW160205946}, we conjecture that,
comparing $\sC$ and $\cC$, the missing information is the invertible
topological orders and SPT orders.  Thus we can use a pair $(\cC, \hat\ga)$ to
describe $\sC$, where $\hat\ga$ corresponds to invertible topological orders or
SPT orders. More precisely (see Fig. \ref{CwM}b and Section~\ref{glpCS}), \frmbox{gapped liquid phases
in $n$d lattice systems with a fixed categorical symmetry $\sM$ are classified
by pairs $(\cC,\hat\ga)$, where $\cC$ is a fusion $n$-category that satisfies
$\bulk(\cC)=\sM$ and $\hat\ga$ is an invertible domain wall in $\sM$. } 

The invertible domain walls formed by the trivial excitations in the bulk
topological order $\sM$ are invertible topological orders.  The invertible
domain walls formed by the topological excitations in the bulk topological
order $\sM$ are SPT orders, since the fusion of the topological excitations
encode symmetry.  There are also invertible domain walls that exchange
excitations. Those invertible domain walls correspond to the automorphisms of
the fusion higher category $\Om \sM = \cM$ that describes the excitations in
the bulk topological order $\sM$.

\subsection{Classification of SET orders and SPT orders with an algebraic
higher symmetry} \label{setspt}

For systems with a categorical symmetry $\sM$, in the above, we classify
anomaly-free gapped liquid phases $\sC$ which partially break the categorical
symmetry.  Here we assume the unbroken symmetry is an algebraic higher symmetry
$\cR$, such that $\sM=\bulk(\cR)$. 
In this case, the classification of gapped liquid phases $\sC$ in last
subsection includes the classification of anomaly-free gapped liquid phases
with a given algebraic higher symmetry $\cR$.  But which  gapped liquid phases
do not break the symmetry $\cR$ and which spontaneously break the symmetry
$\cR$?  To identify the gapped liquid phases $\sC$ that do not break the
symmetry $\cR$, first $\sC$ should include $\cR$ as its excitations, \ie $\cR$
is a subcategory of $\cC=\Om\sC$, or more precisely, there is a top-fully
faithful functor $\iota: \cR \inj{} \cC$ (see Proposition~\ref{RC} and Def. \ref{topfully}).  Second, $\cR$
and $\cC$ have the same bulk topological order $\bulk(\cC)= \bulk(\cR)$ (\ie
the same categorical symmetry).  

This understanding gives us a classification of anomaly-free gapped liquid
phases with an anomaly-free algebraic higher symmetry $\cR$.  First, let us
define the notion ``anomaly-free gapped liquid phases with an anomaly-free
algebraic higher symmetry $\cR$'' more carefully.  We have defined anomaly-free
algebraic higher symmetry in Def. \ref{afahs}.  ``A phases with an symmetry''
means that the symmetry is not spontaneously broken.  But how do we determine
if an  algebraic higher symmetry $\cR$ is spontaneously broken or not?  There
is a macroscopic way to do so (see Section~\ref{pSETSPT}): \frmbox{A gapped
phase has a symmetry $\cR$ (\ie $\cR$ is not spontaneously broken) if the
excitations of phase contain $\cR$.}  A gapped liquid phase is anomaly-free if
it can be realized as the ground of lattice Hamiltonian in the same dimension.
Again, there is a way to describe this property macroscopically (see Section
\ref{ahsTopS}): \frmbox{if the excitations (described by fusion $n$-category
$\cC$) in the gapped liquid phase with an algebraic higher symmetry $\cR$
satisfy $\bulk(\cC)\simeq \bulk(\cR)$, then the  gapped liquid phase is
anomaly-free.} Here $\bulk(\cC)\simeq \bulk(\cR)$ means that the two
topological orders $\bulk(\cC)$ and $\bulk(\cR)$ are equivalent, \ie they can
be connected by an invertible (also called transparent) domain wall $\hat\ga$.
Note that the  invertible  domain wall is not unique.  Thus the two topological
orders $\bulk(\cC)$ and $\bulk(\cR)$ can be equivalent in many different ways.
We denote a way of equivalence by $\bulk(\cC)\seq{\hat\ga} \bulk(\cR)$ or
$\hat\ga: \bulk(\cC)\seq{} \bulk(\cR)$.

Now, we are ready to state some classification results.  First, let us describe
a simple partial classification: \frmbox{Given an algebraic higher symmetry
described by a local fusion $n$-category $\cR$, anomaly-free symmetric gapped
liquid phases (up to invertible topological orders and SPT orders) are
classified by fusion $n$-categories $\cC$ that (1) admit a top-fully faithful
functor $\iota: \cR \inj{} \cC$, and (2) satisfy $\bulk(\cC)\simeq  \bulk(\cR)$
(see Section~\ref{pSETSPT}).}

To get a more complete classification that includes invertible toppological
orders and SPT orders for the algebraic higher symmetry $\cR$, we need to
include the equivalence (\ie the invertible domain wall) $\hat\ga:
\bulk(\cR)\seq{}\bulk(\cC)$ and use the data $( \cR \inj{\iota} \cC ,\hat\ga) $
to classify anomaly-free symmetric gapped liquid phases.  However, not every
equivalence $\hat\ga$ should be included.  We know that the categorical
symmetry described by $\bulk(\cR)$ or by $\bulk(\cC)$ includes the algebraic
higher symmetry $\cR$.  The equivalence $\hat\ga$ should not change $\cR$ that
is contained in $\bulk(\cR)$ and in $\bulk(\cC)$ \cite{KZ200308898}.  For
details and the main classification results, see Section~\ref{ahsTop}.  Here, we
just quote a classification of $\cR$-SPT orders, obtained by setting $\cC=\cR$
\frmbox{$n$d SPT orders and invertible topological orders with an anomaly-free
algebraic higher symmetry $\cR$ are classified by the invertible domain wall
$\hat\al$ in $Z_1(\cR)$, such that its induced automorphisms $\al$ of
$Z_1(\cR)$ satisfy  $ \al(A_{\dual\cR}) \simeq A_{\dual\cR}$, where
$A_{\dual\cR}$ is the condensable algebra in $Z_1(\cR)$ that determines a
boundary $\dual\cR$ corresponding to the dual of the symmetry $\cR$ (see Fig.
\ref{RdR}).} This result generalizes the previous classification of SPT orders
for higher
symmetries.\cite{KT13094721,TK151102929,BM170200868,W181202517,WW181211967}

\section{A higher category theory of topological orders in higher
dimensions}

\label{toprev}

In this section, we present a review, a clarification, and an expansion of a
higher category theory for topological orders in higher dimensions, based on
\Ref{KW1458,KZ150201690,KZ170200673,LW180108530,GJ190509566,J200306663,KZ200514178}.
Many notions of higher category and topological order will be introduced and
explained for physics and mathematics audience. Those notions will be used to
understand algebraic higher symmetry and categorical symmetry, as well as to
classify topological orders and SPT orders with those symmetries, in the rest
of this paper.  Readers who are familiar with higher category and topological
order can skip this section.

\subsection{Topological orders as gapped liquid phases}
\label{gapliq}

\begin{figure}[t]
\begin{center}
\includegraphics[scale=0.4]{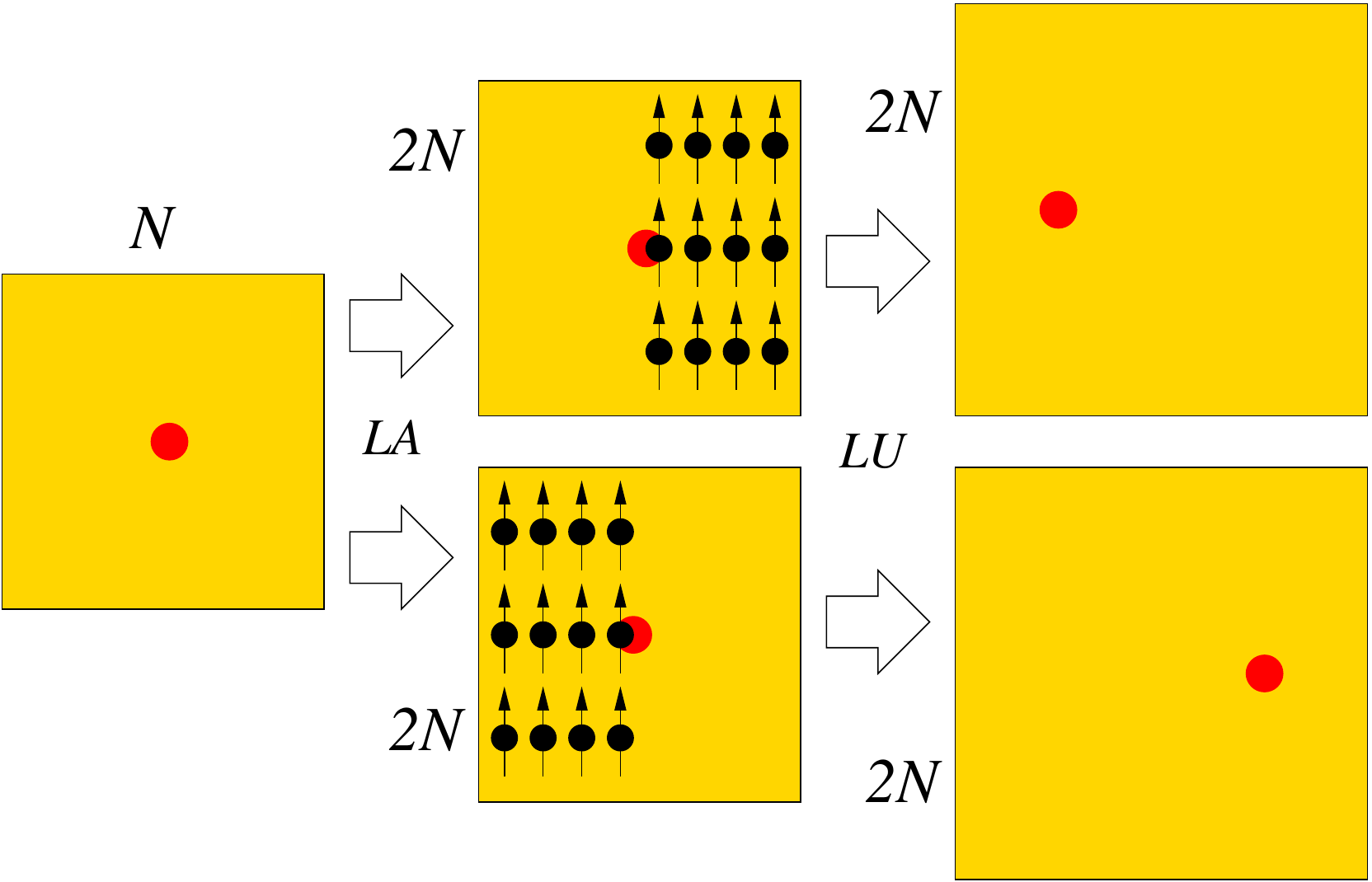} \end{center}
%9
\caption{ 
There are two kind of equivalent relations: (1) finite-depth local unitary (LU)
transformations, and (2) local addition (LA) of product states.
}
\label{N2N}
\end{figure}

In this section, we give a microscopic description of topological order.
Topological orders\cite{W8987,WN9077,W9039} are gapped liquid phases without
symmetry.  The notion of gapped liquid phases is introduced in
\Ref{CGW1038,ZW1490,SM1403}:
\begin{DefinitionPH}
\label{GLP}
A \textbf{gapped liquid phase} is an equivalence class of gapped states, under
the following two equivalence relations (see Fig. \ref{N2N}):\\
(1) two gapped states connected by a finite-depth local unitary transformation
are equivalent;\\
(2) two gapped states differ by a stacking of product state
are equivalent, where the degrees of freedom of the product state may have a non-uniform but bounded density.\\
If there is no symmetry, the local unitary transformation has no symmetry
constraint, and the corresponding gapped liquid phases of local bosonic or
fermionic systems are \textbf{topological orders}\cite{W8987,WN9077,W9039}.  In
the presence of \textbf{finite internal symmetry}, the local unitary
transformation is required to commute with the symmetry transformations, and
the corresponding gapped liquid phases include \textbf{spontaneous symmetry
breaking orders}, \textbf{symmetry protected trivial (SPT)
orders},\cite{GW0931,CGL1314,GW1441} \textbf{symmetry enriched topological
(SET) orders}
\cite{KW0906,CGW1038,LV1334,MR1315,HW1351,X13078131,HW13084673,CY14126589,CQ160608482,HL160607816}.
\end{DefinitionPH} 
\noindent
In this paper, we only consider bosonic systems with finite internal
symmetries.  We do not consider spacetime symmetries (such as time reversal
and translation symmetries), nor continuous symmetry (such as $U(1)$ symmetry).
So in this paper, when we refer \emph{symmetry}, we only mean \emph{finite
internal symmetry}.

We would like to remark the above definition has an important feature. A gapped
liquid phase with some non-invertible topological excitations on top of it is
not a gapped liquid phase according to the definition.  (The notion of
non-invertible topological excitations is defined in the next section.) We note
that the Hamiltonian here may not have translation symmetry.  Thus it is hard
to tell if the ground of a Hamiltonian has excitation in it or not.  Using our
above definition, a gapped liquid state is a ground state of a Hamiltonian that
has no non-invertible topological excitations.  However, a gapped liquid state
may contain invertible topological excitations.  In fact, two gapped liquid
states differ by invertible topological excitations are very similar, and both
can be viewed as proper ground states.

To see the above point, let us start with a gapped state of $N$ sites with a
topological excitation in the middle.  We may double the system size by
stacking a product state of $N$ sites to the left half of the system, or to the
right half of the system.  Both operations are equivalence relations, and the
resulting states of $2N$ sites should be equivalent, \ie be connected by a
finite-depth  local unitary transformation.  However, in the presence of the
non-invertible topological excitation, the excitation appears at left $1/4$
or right $1/4$ of the enlarged systems (see Fig. \ref{N2N}).  Such two enlarged
systems are not connected by finite-depth  local unitary transformations, which
can only move the non-invertible topological excitation by a finite distance.
Thus a gapped liquid state with some non-invertible topological excitations can
non-longer be viewed as a gapped liquid state.

However, a gapped liquid state with some invertible topological excitations can
still be viewed as a gapped liquid state.  This is because  finite-depth  local
unitary transformations can move invertible topological excitations by a large
distance across the whole system.  Thus the gapped liquid phases defined above
may contain some invertible topological excitations.

\Ref{KW1458,KZ150201690} outline a description of topological orders (\ie
without symmetry) in any dimensions, via braided fusion higher categories.
Here, we would like to review and expand the discussions in \Ref{KW1458,KZ150201690}.
We would like to remark that the needed higher category theory is still not fully
developed. So our discussion here is just an outline.  We hope that it can be a
blue print for further development.  However, our discussions become rigorous
at low dimensions (such as 1d and 2d).

\subsection{Trivial, local, and topological excitations}
\label{topexc}

\begin{figure}[t]
\begin{center}
\includegraphics[scale=0.7]{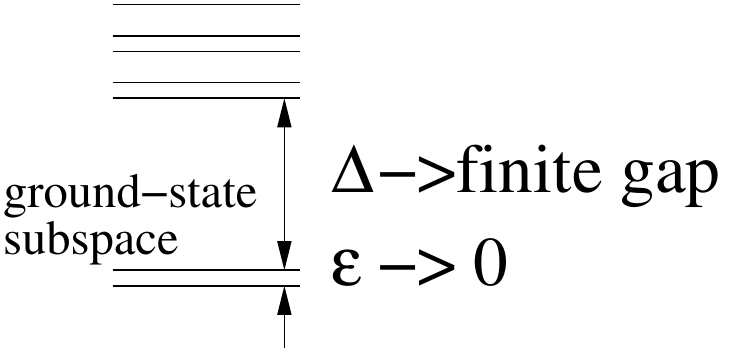} \end{center}
%10
\caption{ 
The ground state subspace below the energy gap $\Del$.  The energy splitting
$\eps$ in the subspace approaches to $0$ in thermodynamic limit, which
the energy gap $\Del$ remain non-zero.
}
\label{gap}
\end{figure}

The reason that gapped liquid phases (which include topological orders) can be
described by higher categories is that higher category is the natural language
to describe excitations within a gapped liquid phase, as well as domain walls
between different  gapped liquid phases.  To understand this connection, let us
define excitation more carefully.  We find that there two different ways to
define types of excitations, which result in different kinds of higher category
theories.\cite{KW1458,KZ150201690} However, only the first definition of types
and its associated higher category theory is more
developed.\cite{GJ190509566,J200306663,KZ200514178} We will concentrate on this one.

We consider a gapped liquid state, which is
the ground state of a local Hamiltonian $H$.  As discussed in last section, a
gapped liquid state does not contain any non-invertible topological
excitations.
To define excitations in $H$, for example, to define
string-like excitations, we can add several trap Hamiltonians $\del
H_\mathrm{str}(S^1_a)$, labeled by $a$, to $H$ such that $H + \sum_a \del
H_\mathrm{str}(S^1_a)$ has an energy gap.  $\del H_\mathrm{str}(S^1_a)$ is only
non-zero along the string $S^1_a$.  Here we require $\del
H_\mathrm{str}(S_a^1)$ to be local along the string $S^1_a$.  $\del
H_\mathrm{str}(S_a^1)$ can be any operator, as long as its acts on the degrees
of freedom near the string $S_a^1$.  The ground state subspace
$\cV_\text{fus}(S^1_1,S^1_2,\cdots)$ of $H + \sum_a \del H_\mathrm{str}(S^1_a)$
(where is
also called the fusion space) corresponds to string-like excitations located at
$S^1_1$, $S^1_2$, \etc (see Fig. \ref{gap}).  

\begin{figure}[t]
\begin{center}
\includegraphics[scale=0.7]{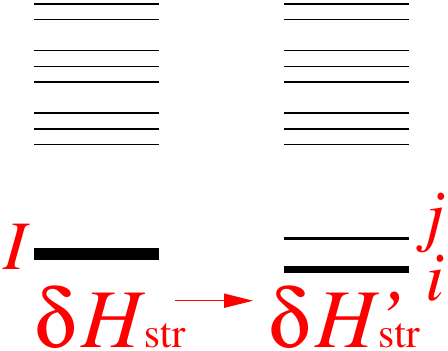} \end{center}
%11
\caption{ 
The ground state subspace of a composite excitation can be split by some change
of trap Hamiltonian $\del H_\text{str} \to \del H'_\text{str}$.  The ground
state subspace of a composite excitation $i$ can be viewed as a direct sum of
the ground state subspaces of excitations $j$ and $k$.  In this sense, we can
denote $I=i\oplus j$.
}
\label{specsplit}
\end{figure}

If the ground state subspace is stable (\ie unchanged) against any small change
of $\del H_\mathrm{str}(S^1_a)$, then we say the correspond string on $S^1_a$
is a \textbf{simple} excitation (or a simple morphism in mathematics).  If the
ground state subspace has accidental degeneracy (\ie can be split by some small
change of $\del H_\mathrm{str}(S^1_a)$, see Fig. \ref{specsplit}), then we say
the correspond string on $S^1_a$ is a \textbf{composite} excitation (or a
composite morphism in mathematics).  A composite excitation $I$ is a direct sum
of several simple excitations
\begin{align}
 I = i\oplus j \oplus \cdots .
\end{align} 
In other words, $I$ can be viewed as an accidental degeneracy of excitations
$i,j,\cdots$.  We see that different string-like excitations can be labeled by
different trap Hamiltonians $\del H_\mathrm{str}(S^1_a)$ (\ie different
non-local operators on $S^1_a$'s). To summarize
\begin{DefinitionPH}
\label{exc}
Excitations are something that can be trapped.  In other words, excitations are
described by the ground state subspace of the Hamiltonian with traps.
\end{DefinitionPH} 
\noindent

But the above definition give us too many different strings, and many of
different strings actually have similar properties.  So we would like to introduce a
equivalence relation to simplify the types of strings.  We define two strings
labeled by $\del H_\mathrm{str}(S^1)$ and $\del \t H_\mathrm{str}(\t S^1)$ as
equivalent, if we can deform $\del H_\mathrm{str}(S^1)$ into $\del \t
H_\mathrm{str}(\t S^1)$ without closing the energy gap.  The equivalence
classes of the strings define the types of the strings.  We would like to point out
that if $S^1$ is an open segment, the corresponding string is equivalent to
the trivial string $\one_s$ described by $\del H_\mathrm{str}(S^1)=0$, since
we can shrink the string along $S^1$ to a point without closing the gap.
\begin{DefinitionPH}
\label{type}
Two excitations are \textbf{equivalent} (\ie are of the same \textbf{type}) if
they can be connected by local-unitary transformations and by stacking of
product states.
\end{DefinitionPH} 
\noindent
We would like to remark that if the two excitations are defined on a closed
sub-manifold, then we can define their equivalence by deforming their trap
Hamiltonians into each other in the space of local trap Hamiltonians without
closing the energy gap.  The above definition is more general, since the
local-unitary transformations and stacking of product states can be applied to
a part of the sub-manifold that support the excitations, and we can examin if
the two excitations turn into each other on the part of the sub-manifold.
\begin{Definition}
\label{trivialtype}
An excitation is \textbf{trivial} if it is equivalent to the \textbf{null}
excitation defined by a vanishing trap Hamiltonian.
\end{Definition} 
\begin{Definition}
\label{invertibletype}
An excitation is \textbf{invertible} if there exists another  excitation such
that the fusion of the two excitations is equivalent to a trivial excitation.
\end{Definition} 

The above equivalence relation can also be phrased in a way similar
to Def. \ref{GLP}:
\begin{Proposition}
A \textbf{type} of excitations is an equivalence class of gapped ground 
states with
added trap Hamiltonian acting on a $m$-dimensional subspace $S^m$, under the
following two equivalence relations:\\ 
(1) two gapped states connected by a
finite-depth local unitary transformation acting on the subspace $S^m$ are
equivalent;\\ 
(2) two gapped states differ by a stacking of product states
located on the subspace $S^m$ are equivalent.
\end{Proposition} 
\noindent
We see that, when $m>0$, the excitations defined above are gapped liquid state
on the sub space $S^m$, and there is no lower dimensional non-invertible
excitations on $S^m$.

We also would like to introduce the notion of non-local equivalence and
non-local type: 
\begin{DefinitionPH}
\label{nltype}
Two excitations are \textbf{non-locally equivalent} (\ie are of the same
\textbf{nl-type}) if they can be connected by non-local-unitary transformations
and by stacking of product states.
\end{DefinitionPH} 
\noindent
\begin{DefinitionPH}
\label{localtype}
An excitation is \textbf{local} if it has the same nl-type as the null
excitation.
\end{DefinitionPH} 
\noindent
We see that a trivial excitation is always a local excitation.  But a local
excitation may not be a trivial excitation.
\begin{DefinitionPH}
\label{toptype}
An excitation is \textbf{topological} if it is non-local.
\end{DefinitionPH} 
\noindent
Again, the above non-local equivalence relation can also be phrased in a way
similar to Def. \ref{GLP}:
\begin{Proposition}
A \textbf{nl-type} of excitations is an equivalence class of gapped states with
added trap Hamiltonian acting on a $m$-dimensional subspace $S^m$, under the
following two equivalence relations:\\ (1) two gapped states connected by a
non-local unitary transformation acting on the subspace $S^m$ are
equivalent;\\ (2) two gapped states differ by a stacking of product states
located on the subspace $S^m$ are equivalent.
\end{Proposition} 
\noindent
We also believe that
\begin{Proposition}
\label{nlmp}
Two excitations have the same nl-type if and only if they can be connected by
gapped or gapless domain walls.  We note that the morphisms in higher category
only correspond to gapped  domain walls.
\end{Proposition}
\noindent
We would like to remark that for point-like excitations the notion of \emph{type} and
\emph{nl-type} coincide.

Those different concepts of excitations were discussed in \Ref{KW1458}, where
the \emph{nl-type} was called \emph{elementary type}, and \emph{topological}
excitation was called \emph{elementary topological} excitation.  The
\emph{local} excitation was called \emph{descendant}  excitation in
\Ref{KW1458}.

\subsection{Examples of excitations}

To illustrate the above concepts that we just introduced, let us consider a
2d $\Z_2$ topological order\cite{RS9173,W9164} for bosons
described by 2+1D $\Z_2$ gauge theory.  
\begin{Example}
\label{excZ2top}
The $\Z_2$ topological order has four \textbf{types} of point-like excitations,
labeled by $\one,e,m,f$, where $e$ is the $\Z_2$ charge, $m$ is the $\Z_2$
vortex, and $f$ is a fermion -- the bound state of $e$ and $m$.  $\one$ is a
\textbf{trivial} point-like excitation. The $\Z_2$ topological order also has
four \textbf{nl-types} of point-like excitations, labeled by $\one,e,m,f$.
$\one$ is a \textbf{local} point-like excitation, and $e,m,f$ are
\textbf{topological} point-like excitations.  

The $\Z_2$  topological order has only one \textbf{nl-type} of string-like
excitations, which is a \textbf{local} string-like excitation.  The $\Z_2$
topological order has six \textbf{types} of string-like excitations, generated by $\one_s,e_s,m_s,f_s$,
with two additional types given by
$f_s\otimes m_s=e_s\otimes f_s$ and $m_s\otimes f_s=f_s\otimes e_s$:
\begin{align}
 e_s\otimes e_s &= 2e_s, \nonumber\\
  m_s\otimes m_s &= 2m_s, \nonumber\\
 e_s\otimes m_s &= f_s\otimes m_s = e_s\otimes f_s,\nonumber\\ 
 m_s\otimes e_s &= m_s\otimes f_s = f_s\otimes e_s.
\end{align}
The $e_s$-type of
string-like excitation is formed by the $e$-particles, condensing into a 1d
phase of spontaneous $\Z_2$ symmetry breaking state.  We note that the
$e$-particles have a mod-2 conservation, and an emergent $\Z_2$ symmetry.
Similarly, the $m_s$-type of string-like excitation is formed by the
$m$-particles, condensing into a 1d phase of spontaneous $\Z_2$ symmetry
breaking state.  The $f_s$-type of string-like excitation is formed by the
$f$-particles, condensing into a 1d topological superconducting phase (\ie
the 1d invertible topological order of fermions where the string ends have
Majorana zero modes\cite{K0131}).  $\one_s$ is the \textbf{trivial} string-like
excitation.  Although $\one_s,e_s,m_s,f_s$ are four different \textbf{types} of
string-like excitations, they are all \textbf{local} string-like excitations,
\ie belong to the trivial \textbf{nl-type} of string-like excitations. We also
comment that $f_s$ is an invertible string-like excitations, or the $e,m$-
exchange transparent
domain wall; if we move $e$ throuch $f_s$, it becomes $m$ and vice versa. 
\end{Example}

Next we consider a 3d trivial product state of bosons.  
\begin{Example}
\label{excTri}
Such a state has trivial \textbf{nl-type} of point-like, string-like, and
membrane-like excitations, \ie all excitations are \textbf{local}.  It also has
trivial \textbf{type} of point-like and string-like, but it has non-trivial
\textbf{types} of membrane-like excitations.  In fact those non-trivial
\textbf{types} of membrane-like excitations corresponding to 2d topological
orders.  Thus there are infinite \textbf{types} of membrane-like excitations
corresponding to infinite different 2d topological orders, even though the
3d state has trivial topological order and is a trivial product state of
bosons.  All those membrane-like excitations are \textbf{local} but not
\textbf{trivial}.
\end{Example}

\begin{Remark}
\label{CCrem}
We remark that the our above description of 3d trivial topological order is
different from that in \Ref{GJ190509566,J200306663}.
\Ref{GJ190509566,J200306663} only include membrane excitations that correspond
to 2d topological orders with gappable boundary.  There are still infinite type
of membranes of this kind.  In our description, the membrane excitations
include both 2d topological orders with gappable boundary and 2d topological
orders whose boundary cannot be gapped.
\end{Remark}

We see that to have a complete \emph{macroscopic} description of trivial
product state of bosons in $n$-dimensional space without symmetry, we need to
classify $(n-1)$d topological orders of bosons, which corespond to \emph{types}
codimension-1 excitations.  We also see that to have a complete
\emph{macroscopic} description of $n$d topological order of bosons without
symmetry, we need to classify $(n-1)$d SET orders of bosons/fermions with
symmetries (\ie the emergent symmetry).

\subsection{Trivial topological order (the product states) and its excitations}
\label{cattr}

In the last section, we see that the \emph{types} of dimension-$k$ excitations
in a trivial product state in $n$-dimensional space correspond to topological
orders (gapped liquid phases) in $k$-dimensional space.  Thus the study of the
trivial topological order and its excitations of various dimensions allows us
to understand topological orders in spatial dimensions less then $n$.  This
motivated us to develop a comprehensive theory of trivial topological order.

All trivial topological orders are product states and all product states belong
to one phase, if there is no symmetry.  We denote the trivial topological order
in $n$-dimensional space as $\tto^{n+1}$ ($n+1$ is the spacetime dimension).
$\tto^{n+1}$ is also referred as an object.  Once we have the trivial
topological order $\tto^{n+1}$, we also have accidental degeneracy of several
$\tto^{n+1}$'s (\ie  several product states).  We denote a gapped liquid phase
formed by $m$ degenerate product states as $\underbrace{ \tto^{n+1} \oplus
\cdots \oplus \tto^{n+1}}_{m \text{ copies}} = m \tto^{n+1} $.  So, after the
completion, the collection of trivial topological orders has objects $m
\tto^{n+1}$.  We refer $\tto^{n+1}$ as simple object, while $m\tto^{n+1}$
($m>1$) as composite object.  We see that the composite object does not
correspond to a stable phase, since the accidental $m$-fold degeneracy can be
easily split by local perturbations in the Hamiltonian.

The collection of trivial topological orders in $\npo$D spacetime, $\{ m
\tto^{n+1}\}$, is a set.  However, the objects in the set have many relations.
Two objects can be connected by a gapped codimension-1 domain wall $a:
m_1\tto^{n+1} \to m_2\tto^{n+1}$,  which is called an 1-morphism.  For example,
an 1-morphism $a: 2\tto^{n+1} \to 3\tto^{n+1}$ can be represented as
\begin{align}
 a = 
\bpm
 0 (_{\t\tto^{n+1}_1}|_{\tto^{n+1}_1}), &  0 (_{\t\tto^{n+1}_1}|_{\tto^{n+1}_2}) \\ 
 1 (_{\t\tto^{n+1}_2}|_{\tto^{n+1}_1}), &  0 (_{\t\tto^{n+1}_2}|_{\tto^{n+1}_2}) \\ 
 0 (_{\t\tto^{n+1}_3}|_{\tto^{n+1}_1}), &  0 (_{\t\tto^{n+1}_3}|_{\tto^{n+1}_2})  \\ 
\epm,  
\end{align}
Physically, it means that there is a gapped domain wall between the first
product state in $2\tto^{n+1} = \tto^{n+1}_1\oplus \tto^{n+1}_2$ and the second
product state in $3\tto^{n+1} = \t\tto^{n+1}_1\oplus \t\tto^{n+1}_2\oplus
\t\tto^{n+1}_3$, and such a gapped domain wall is not degenerate.  We denote
such a gapped domain wall as $(_{\t\tto^{n+1}_2}|_{\tto^{n+1}_1})$.  All other
domain walls between different product states have higher energy density or
gapless.  In this paper, we do not consider gapless domain walls and we always
assume gapless domain walls have infinite  energy density.

We can have another 1-morphism $b: 2\tto^{n+1} \to 3\tto^{n+1}$ 
\begin{align}
 b = 
\bpm
 0 (_{\t\tto^{n+1}_1}|_{\tto^{n+1}_1}), &  0 (_{\t\tto^{n+1}_1}|_{\tto^{n+1}_2}) \\ 
 2 (_{\t\tto^{n+1}_2}|_{\tto^{n+1}_1}), &  0 (_{\t\tto^{n+1}_2}|_{\tto^{n+1}_2}) \\ 
 0 (_{\t\tto^{n+1}_3}|_{\tto^{n+1}_1}), &  0 (_{\t\tto^{n+1}_3}|_{\tto^{n+1}_2})  \\ 
\epm,  
\end{align}
Physically, it means that there is a gapped domain wall between the first
product state in $2\tto^{n+1} $ and the second product state in $3\tto^{n+1} $,
and such a gapped domain wall are 2-fold degenerate.  So we express the gapped
domain wall as $(_{\t\tto^{n+1}_2}|_{\tto^{n+1}_1}) \oplus
(_{\t\tto^{n+1}_2}|_{\tto^{n+1}_1}) =2(_{\t\tto^{n+1}_2}|_{\tto^{n+1}_1})$.
The most general 1-morphism $c: 2\tto^{n+1} \to 3\tto^{n+1}$ has a form
\begin{align}
 c = \bpm
\oplus_{\v k} m_{11}^{\v k} (_{\t\tto^{n+1}_1}{\v k}_{\tto^{n+1}_1}), & \oplus_{\v k} m_{12}^{\v k} (_{\t\tto^{n+1}_1}{\v k}_{\tto^{n+1}_2}) \\ 
\oplus_{\v k} m_{21}^{\v k} (_{\t\tto^{n+1}_2}{\v k}_{\tto^{n+1}_1}), & \oplus_{\v k} m_{22}^{\v k} (_{\t\tto^{n+1}_2}{\v k}_{\tto^{n+1}_2}) \\ 
\oplus_{\v k} m_{31}^{\v k} (_{\t\tto^{n+1}_3}{\v k}_{\tto^{n+1}_1}), & \oplus_{\v k} m_{32}^{\v k} (_{\t\tto^{n+1}_3}{\v k}_{\tto^{n+1}_2})  \\ 
\epm,  
\end{align}
where $m_{ij}^{\v k}\in\N$.  Here, for example, $(_{\t\tto^{n+1}_2}{\v
k}_{\tto^{n+1}_1})$ denote a gapped domain wall between the first product state
in $2\tto^{n+1} $ and the second product state in $3\tto^{n+1} $, and ${\v k}$
labels different types of accidentally degeneracte gapped domain wall between
the two product states.  $m_{ij}^{\v k}$ is the accidental degeneracy of the
domain walls of the same type ${\v k}$.  We see that an 1-morphism is like a
matrix that can also be added.

In particular, a 1-morphism $\v k: \tto^{n+1} \to \tto^{n+1}$, denoted by
$(_{\tto^{n+1}}{\v k}_{\tto^{n+1}})$, is the codimension-1 excitation discussed
in the last section, where ${\v k}$ labels the different \emph{types} of the
excitations, as defined in Def. \ref{type}.  Such an excitation corresponds to
a topological order in $\nmo$-dimensional space.  We use
$\Hom(\tto^{n+1},\tto^{n+1})$ denote the collection of all morphisms from
$\tto^{n+1}$ to $\tto^{n+1}$, which happen to the collection of all topological
orders in $\nmo$-dimensional space.  We would like to remark that
$\Hom(\tto^{n+1},\tto^{n+1})$ is also complete in the sense that it not only
contain stable topological orders, its also contain accidental degeneracy of
topological orders.  In other words, if $\v a,\v b \in
\Hom(\tto^{n+1},\tto^{n+1})$, then the accidental degeneracy of $\v a$ and $\v
b$, $\v a\oplus \v b$, is also in $\Hom(\tto^{n+1},\tto^{n+1})$.  Thus just
like the collection of trivial topological orders $\{ m \tto^{n+1}\}$,
$\Hom(\tto^{n+1},\tto^{n+1})$ is also closed under the ``degeneracy'' operation
$\oplus$.  

We would like to point out that there is a 1-morphism in
$\Hom(\tto^{n+1},\tto^{n+1})$ that corresponds to a codimension-1 trivial
topological order (\ie a product state in $\nmo$-dimensional space or
$n$-dimensional spacetime).  We denote such a 1-morphism as $\tto^n \in
\Hom(\tto^{n+1},\tto^{n+1})$.

\begin{figure}[t]
\begin{center}
\includegraphics[scale=0.4]{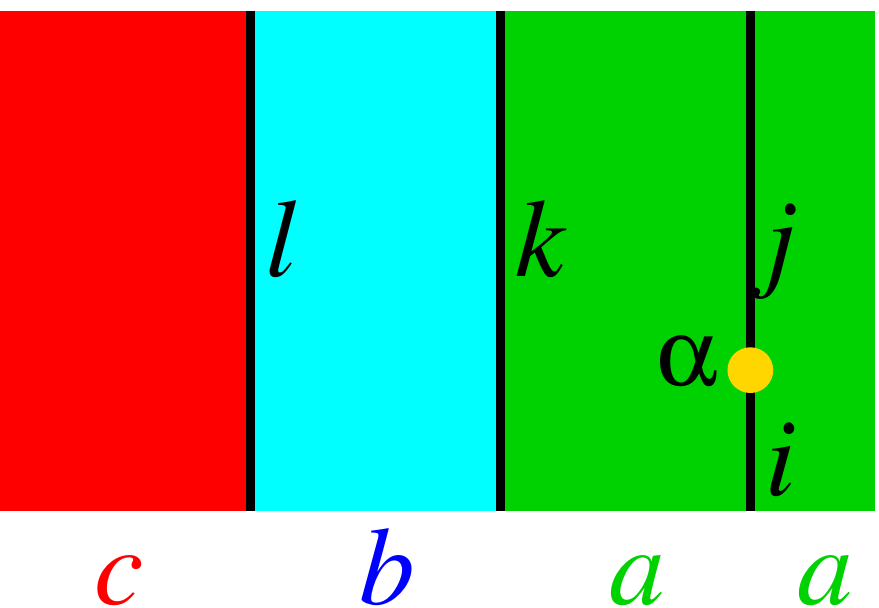} \end{center}
%12
\caption{The 2-dimensional excitations $a,b,c$ are objects. The 1-dimensional
domain walls $\v i,\v j,\v k,\v l$ are 1-morphisms.  $\v \al$ is a 2-morphism (domain wall)
connecting two 1-morphisms $\v i$ and $\v j$.  The fusion of domain walls
$\v k,\v l$ between excitations $\v a,\v b,\v c$ via the ``glue'' $\v b$
is given by $\v l\ot{\v b} \v k$. 
}
\label{composition}
\end{figure}

Two codimension-1 topological orders $\v a,\v b$ may also be connected by a
gapped domain wall of codimension-2: $\v k: \v a\to \v b$ (see Fig.
\ref{composition}).  We call $\v k$ a 2-morphism.  To be precise, here,
the ``domain wall'' really means \emph{types} of domain walls.  We regard two
domain walls as equivalent if they differ only by local unitary transformations
and local addition of product states on the wall.  The collection of
2-morphisms from $\v a$ to $\v b$ is denoted as $\Hom(\v a,\v b)$.  We see that
the collection of 2-morphisms from $\tto^n$ to $\tto^n$, $\Hom(\tto^n,\tto^n)$,
is the collection of condimension-2 excitations, which are also topological
orders in $\nmt$-dimensional space.  Such a collection also contain a product
state (trivial topological order) in $\nmt$-dimensional space, denoted as
$\tto^{n-1} \in \Hom(\tto^n,\tto^n)$.  Also, the collection of 2-morphisms from
$\v a$ to $\v a$, $\Hom(\v a,\v a)$, is the collection of condimension-2
excitations on the codimension-1 excitation $\v a$.

We would like to remark
that it is possible that $\Hom(\v a,\v b)=0$,  which means that
there is no gapped domain wall between $\v
a$ and $\v b$. Here $0$ denotes the ``zero'' category, which can be roughly
thought as the linearized and categorified version of the empty set.

The above discussion can be continued. This allows us to define 3-morphisms,
4-morphisms, \etc.  The $n$-morphisms correspond to codimension-$n$ or
dimension-0 (\ie point-like) excitations.  The point-like excitations are world
lines in spacetime.  The domain wall on world-lines are $\npo$ mophisms.  In
general, a point-like excitation $\v p$ (an $n$-morphism) may have degenerate
ground states (of the Hamiltonian with traps).  We denote the vector space of
the degenerate ground states as $\cV_\text{fus}(\v p,\cdots)$, where $\cdots$ represent
other excitations which are fixed.  Then a $\npo$-morphism $\v o$ from one
point-like excitations $\v p_1$ to the other $\v p_2$ (where the two
excitations are near each other) is a  linear operator acting near $\v
p_1$ and $\v p_2$ from $\cV_\text{fus}(\v p_1,\cdots)$ and $\cV_\text{fus}(\v p_2,\cdots)$: $
\cV_\text{fus}(\v p_1,\cdots)\overset{\v o}{\to} \cV_\text{fus}(\v p_2,\cdots)$. We denote such a $\npo$-morphism
as $\v o: \v p_1\to \v p_2$.

Just like the objects (also called 0-morphisms), the morphisms also can be
divided into two classes: the \emph{simple} morphisms (which correspond to
stable excitations whose ground state cannot be split by any local
perturbations near the excitations) and \emph{composite} morphisms (which
correspond to unstable excitations with accidental degeneracy).  

\begin{figure}[t]
\begin{center}
\includegraphics[scale=0.6]{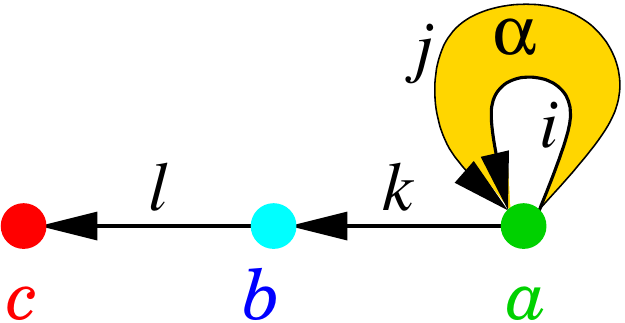} \end{center}
%13
\caption{A dual representation of Fig. \ref{composition}.  A higher category
is a collection of vertices (objects), arrows (1-morphisms), oriented surfaces
(2-morphisms), \etc, connected in a certain way.  In other words, a higher
category is a simplicial complex.  } \label{PandA}
\end{figure}

The objects (\ie the 0-morphisms), as well as $m$-morphisms can also fuse or
compose.  Let $\v a,\v b,\v c$ be three $(m-1)$-morphisms, and $\v k \in
\Hom(\v a,\v b)$ and $\v l \in \Hom(\v b,\v c)$ are two $m$-morphisms.  Then, a
composition of $\v k$ and $\v l$ is given by a $m$-morphism from $\v a$ to $\v
c$: $\v l\ot{\v b} \v k \in \Hom(\v a,\v c)$.  The subscript $\v b$
indicates that $\v k$ and $\v l$ are fused together via the ``glue'' $\v b$ (see
Fig. \ref{composition}).  The picture Fig. \ref{composition} also has a dual
representation Fig. \ref{PandA}

In the above, we discussed excitations of various dimensions in a trival
topological order.  We may reverse the thinking and use all the  excitations to
characterize the trival topological order, or more generally, a non-trivial
topological order.  This is equivalent to using higher categories to
characterize topological orders or trivial orders.  However, in order to use
excitations to describe topological orders or trivial orders, the first issue
one faces is weather to use \emph{type} or use \emph{nl-type} of excitations to
construct higher categories.  The notions of \emph{type} and \emph{nl-type}
were discussed in \Ref{KW1458,KZ150201690}.  In physics, when we refer
topological excitations, we usually mean the \emph{nl-types} of excitations,
which seems to suggest using \emph{nl-type} to construct higher category.
However, in mathematics, it is more natural to use \emph{types} of excitations
to build the higher categories that describe topological
orders.\cite{GJ190509566} In some sense, \emph{nl-types} are like the basis
vectors in a vector space. The completion under ``+'' give rise to all the
\emph{types} which form the whole vector space. In higher category theory, such
a completion corresponds to condensing the nl-types of excitations to construct
all the types of excitations.  
\begin{DefinitionPH}
\label{dexc}
\textbf{Descendent excitations} are excitations of dimension 1,2,3, \etc, which
are obtained by condensing lower dimensional excitations.
\end{DefinitionPH} 

The process of adding all the \emph{types} of excitations in a category is
called \emph{condensation completion} which is discussed in \Ref{GJ190509566}.
(Note that the condensation completion in \Ref{GJ190509566} only includes
defects that correspond to gapped liquid phases that have gappable boundaries,
see Remark \ref{CCrem}.)  In this paper, we do a more general condensation
completion that includes all the \textbf{descendent excitations} that
correspond to all possible gapped liquid phases. In other words, we use
\emph{types} of all excitations to build the higher categories. 

In $n$-dimensional space, the trivial topological order has dimension-$(n-1)$
excitations, dimension-$\nmt$ excitations, \etc.  Those excitations can fuse
(the $\otimes$ operation) and can have accidental degeneracy (the $\oplus$
operation). The excitations can also have domaon walls between them (the
morphisms). The collection of excitations, plus those extra structures form a
\emph{fusion $n$-category}, which is denoted as $\Hom(\tto^{n+1},\tto^{n+1})$.

The precise definition of a fusion $n$-category is difficult to write down due to the lack of a universally accepted and well developed model of weak $n$-categories. But this is not the only problem. Recently, by ignoring this problem, Johnson-Freyd managed to solve other important problems and provided a workable definition in \Ref{J200306663}. Due to its complexity, we choose to not to give Johnson-Freyd's definition, but to provide a rough and physical intuition underlying the definition instead. 
\begin{DefinitionPH}
\label{fc}
A \textbf{fusion $n$-category} is an $n$-category, which is 
\begin{itemize}
\item $\Cb$-linear: the $n$-morphisms are required to form complex vector spaces, 
\item additive (with $\oplus$ operation); 
\item monoidal: with fusion $\otimes$ operation, which is compatible with the $\Cb$-linear and additive structures; 
\item semi-simple (all composite object $\v x$ has a unique decomposition $\v x=\v a\oplus \v b
\cdots$) and the tensor unit is simple; 
\item condensation complete: the $0$-morphisms (the objects), $1$-morphisms, $2$-morphisms, \etc\ include all the decedent excitations; 
\end{itemize}
and satisfies certain fully dualizable condition amounts to the invariance of
the physical reality by deforming and folding of the associated topological
order. 
\end{DefinitionPH}

\begin{Remark} \label{rem:fusion-n-cat}
Because our  descendant excitations include topological orders whose boundary
cannot be gapped, our definition of fusion $n$-category is different
from that proposed by \Ref{GJ190509566,J200306663} where the descendent
excitations only include topological orders with gappable boundary (see also Remark
\ref{CCrem}).
\end{Remark}

Since the excitations in a trivial topological order is surrounded by product
states, we can add more product states to form a higher dimensional trivial
topological order, and view the same excitations as excitations in a  higher
dimension trivial topological order.  In fact, we can view any excitations in a
trivial topological order as excitations in an infinite dimensional trivial
topological order.  So we can always braid the excitations in a trivial
topological order by viewing the excitations as in an infinite dimensional
trivial topological order.  Since the excitations in an infinite dimensional
trivial topological order have trivial braiding and exchange properties, those
excitations form a symmetric fusion higher category with trivial exchange
properties.  Thus
\begin{Proposition} 
\label{sfctri}
The fusion $n$-category $\Hom(\tto^{n+1},\tto^{n+1})$, that describes the
excitations in a trivial topological order $\tto^{n+1}$ in $n$-dimensional
space, can always be promoted into a braided fusion $n$-category. In fact, such
a braided fusion $n$-category is a symmetric fusion $n$-category with trivial
exchange properties.
\end{Proposition}

\subsection{The category of anomaly-free topological orders}
\label{cataf}

\begin{table*}[t]
  \caption{Correspondence between concepts in higher category 
and concepts in topological order.\cite{KW1458,KZ150201690}}
  \label{tab:dic}
  \centering
  \begin{tabular}{|p{2.5in}|p{4.5in}|}
    \hline
    \textbf{~~Concepts in higher category} ~~~ &  ~~\textbf{Concepts in physics}\\
    \hline
Symmetric monoidal $\npo$-category
$\afcTO^{n+1}$  
& 
The collection of
all $n$d anomaly-free topological orders,  which can all be
realized by bosonic lattice model in the same dimension.
\\
    \hline
    A simple object (0-morphism) of $\afcTO^{n+1}$  & A topological order
(with non-degenerate ground state on $S^n$). \\
    \hline
    Simple 1-morphisms of $\afcTO^{n+1}$ connecting different objects& 
The types of domain wall between different topological orders\\
    \hline
    Simple 1-morphisms of $\afcTO^{n+1}$ connecting the same object& The types of codimension-1
topological excitations within a single topological order. They can fuse
(compose).\\
    \hline
    Simple 2-morphisms of $\afcTO^{n+1}$ & The types of codimension-2 topological excitations. They can
    fuse as well as braid (both induced from composition).\\
    \hline
    Simple $\nmo$-morphisms of $\afcTO^{n+1}$ & The types of string-like topological excitations\\
    \hline
    Simple $n$-morphisms of $\afcTO^{n+1}$ & The types of point-like topological excitations\\
    \hline
    $\npo$-morphisms of $\afcTO^{n+1}$ & The operators (instantons in spacetime)\\
    \hline
    Composite  morphisms & The types of topological excitations with accidental degeneracy\\
    \hline
    Trivial morphisms & The types of excitations that can be created by local operators (the \emph{trivial} excitations)\\
    \hline
  \end{tabular}
\end{table*}

\begin{DefinitionPH}
\label{aftop}
An \textbf{anomaly-free} topological order is a gapped liquid phase that can be
realized by a local bosonic lattice models in the same dimension.  
\end{DefinitionPH}
\noindent
In a trivial topological order $\tto^{n+2}$ $\npo$-dimensional space, a
type of codimension-1 excitation correspond to an anomaly-free topological
order in $n$-dimensional space. 
This because
we can remove the product state around a codimension-1 excitation and view it
as an anomaly-free topological order.  Thus $\Hom(\tto^{n+2},\tto^{n+2})$ is
the set of anomaly-free $n$d topological orders.  Those  $n$d topological
orders have excitations on them and have domain walls between them, which
correspond to morphisms.  They can also fuse $\otimes$ and have accidental
degeneracies $\oplus$. Besides, we include all descendant excitations
(condensation completion).  Adding those structures to the set
$\Hom(\tto^{n+2},\tto^{n+2})$, we make it into a fusion $\npo$-category
(see Table \ref{tab:dic}), which leads to the definition
of the first version of category of anomaly-free topological orders:
\begin{Definition}
\label{afcat}
The \textbf{category of anomaly-free topological orders} in $n$-dimensional
space, is the fusion $\npo$-category given by $\Hom(\tto^{n+2},\tto^{n+2})$
where $\tto^{n+1}$ is the trivial topological order (\ie a product state) in
$\npo$-dimensional space.  We denote the category of anomaly-free topological
orders as $\afTO^{n+1}$.
\end{Definition}
\noindent

In the above, we have defined anomaly-free topological orders via a microscopic
approach, since we used the notion of product states and condensing low
dimensional excitations to construct descendant excitations.  Can we define
anomaly-free topological orders using only the macroscopic notions?  Here we
would like to point out that the anomaly-free topological orders have a
defining macroscopic property called the principle of remote
detectability:\cite{L13017355,KW1458}
\begin{Proposition}
\label{remote}
A topological order is anomaly-free if and only if any excitations of
non-trivial \textbf{nl-type} can be detected remotely (such as via braiding) by
some other excitations.
\end{Proposition}
\noindent
Here the \textbf{nl-type} also has a defining macroscopic property
\begin{Proposition}
\label{nltype1}
Two excitations have the same \textbf{nl-type} if and only if they can be
connected by gapped or gapless domain walls.
\end{Proposition}
\noindent
The gapped domain walls are the morphisms that we have discussed, while the
gapless domain walls are not included in our discussion.  Also the notion of
``detecting remotely'' is not defined carefully.  This reveals the difficulty
to define anomaly-free topological order macroscopically beyond 2-dimensional
space. The above just points out a possible direction.

The category $\afTO^{n+1}$ include both topological phases (called stable
topological orders) and correspond to simple objects in $\afTO^{n+1}$) and
first-order phase transitions between two  stable topological orders (called
unstable topological orders).  The stable topological order corresponds to
simple object and the unstable topological order corresponds to composite
object in $\afTO^{n+1}$.  For example, the first-order phase transition point
between two stable topological orders, the simple objects $\sA$ and $\sB$,
correspond to the composite object $\sA\oplus \sB$, which can be viewed as the
accidental degeneracy of the two stable topological orders $\sA$ and $\sB$.  As
a fusion category, the objects in  $\afTO^{n+1}$ has this $\oplus$ operation.

The fusion higher category $\afTO^{n+1}$ has a special property, reflecting the following physics fact.
The stacking of two stable anomaly-free topological orders $\sM_1^{n+1}$ and
$\sM_2^{n+1}$ always given us a third stable anomaly-free topological order
$\sM_3^{n+1}$, and the result does not dependent on the order. Also note that
the stacking is nothing but the fusion along $\tto^{n+2}$.
This means that
\begin{Proposition}
The simple objects, $\sM_1^{n+1}$, $ \sM_2^{n+1}$, $ \sM_3^{n+1}$, in
$\afTO^{n+1}$ form a commutative monoid under the fusion $\ot{\tto^{n+2}}$:
\begin{align}
 \sM_1^{n+1} \ot{\tto^{n+2}} \sM_2^{n+1} = 
 \sM_2^{n+1} \ot{\tto^{n+2}} \sM_1^{n+1} = 
 \sM_3^{n+1}. 
\end{align}
\end{Proposition}
\noindent
$\ot{\tto^{n+2}}$ is also abbreviated as $\otimes$.  Since a commutative monoid
may have a submonoid which is actually an Abelian group,  anomaly-free
topological orders have a subset of \textbf{invertible topological
orders},\cite{KW1458,F1478,K1459} which form an Abelian group under the
stacking $\otimes$.  All the invertible topological orders in $n$-dimensional
space, plus their accidental degeneracies, also form a fusion $\npo$-category
denoted as $\icTO^{n+1}$.

From the above discussion, we see that we can ignore all the unstable
topological orders, and restrict ourselves to only stable topological orders
(which is more natural from a physics point of view).  After dropping all the
composite objects from $\afTO^{n+1}$, we obtain a monoidal $\npo$-category
$\afcTO^{n+1}$.  The objects in $\afcTO^{n+1}$ still support the stacking
$\otimes$ operation, but do not support the accidental-degeneracy (or
first-order phase transition) $\oplus$ operation.  Thus $\afcTO^{n+1}$ is
monoidal but not fusion.

\subsection{The category of anomalous topological orders}
\label{cata}

In the last section, we see that $\Hom(\tto^{n+2},\tto^{n+2})$ gives rise to
the collection of anomaly-free (stable and unstable) topological orders in
$n$-dimensional space.  Similarly, $\Hom(\tto^{n+2},\sM^{n+2})$ gives rise to
the collection of gapped boundaries of a $\npo$d topological order described by
$\sM^{n+2}$.  Those gapped domain walls have domain walls (morphisms) between
them, as well as accidental degeneracy $\oplus$ operation.  But they do not
have stacking $\otimes$ operation.  But we can fuse the morphisms in
$\Hom(\tto^{n+2},\tto^{n+2})$ to $\Hom(\tto^{n+2},\sM^{n+2})$ from right, and
fuse the morphisms in $\Hom(\sM^{n+2},\sM^{n+2})$ to
$\Hom(\tto^{n+2},\sM^{n+2})$ from left.  Both $\Hom(\tto^{n+2},\tto^{n+2})$ and
$\Hom(\sM^{n+2},\sM^{n+2})$ are fusion $\npo$-categories.  Thus
$\Hom(\tto^{n+2},\sM^{n+2})$ is a right module of fusion $\npo$-category
$\Hom(\tto^{n+2},\tto^{n+2})$ and a left module of fusion $\npo$-category
$\Hom(\sM^{n+2},\sM^{n+2})$.  

As a collection of gapped boundaries of a $\npo$d anomaly-free topological
order $\sM^{n+2}$, $\Hom(\tto^{n+2},\sM^{n+2})$ support $\oplus$ operation but
does not $\otimes$ operation.  In order to allow the staking $\otimes$
operation, we consider a collection of $\Hom(\tto^{n+2},\sM^{n+2})$ for all
different $\sM^{n+2}$, \ie we consider all the gapped boundaries of all $\npo$d
anomaly-free topological orders.  Such an enlarged collection support the
stacking $\otimes$ operation, by stacking both boundary and bulk.  However, in
the enlarged collection $\otimes$ operation becomes messy, since there are two
kinds of accidental degeneracies: the accidental degeneracies of the boundary
and the accidental degeneracies of the bulk, suggesting that there are two
kinds of $\oplus$ operations.  To keep things simple, we will drop all the
unstable gapped boundaries and all the unstable bulk topological order, \ie we
restrict $\sM^{n+2}$ to be simple objects and consider only simple 1-morphisms
in $\Hom(\tto^{n+2},\sM^{n+2})$.
In this way, we to obtain
\begin{Definition}
\label{acat}
The \textbf{category of potential anomalous topological orders} in
$n$-dimensional space, is the monoidal $\npo$-category given by the union of
all $\Hom(\tto^{n+2},\sM^{n+2})$'s after dropping all the composition
1-morphisms, where $\tto^{n+2}$ is the unit object in $\afcTO^{n+2}$ and
$\sM^{n+2}$ is a simple object in $\afcTO^{n+2}$.  It is a right module over
fusion $\npo$-category $\Hom(\tto^{n+2},\tto^{n+2})$, a left module over fusion
$\npo$-category $\cM^{n+1}=\Hom(\sM^{n+2},\sM^{n+2})$ and thus a bimodule.
Here $\sM^{n+2}$ is a stable anomaly-free topological order in
$\npo$-dimensional space, which determine the \textbf{anomaly}.  We denote the
category of anomalous topological orders as $\cTO^{n+1}$.  Such a category
describes all the gapped boundaries of all the anomaly-free topological orders
\end{Definition}

\subsection{Invertible domain wall between topological orders}

We have seen that the collection of domain walls (plus the excitations on the
walls and their $\otimes$, $\oplus$ operations) in a stable $n$d topological
orders $\sC$ is given by $\Hom(\sC,\sC)$.  In fact, $\cC=\Hom(\sC,\sC)$ is a
fusion $n$-category.  The objects in $\cC$ (the domain walls) support the
fusion $\otimes$ operation.  Roughly, an object $a$ (a domain wall) is
\textbf{invertible} if there exist another object $b$ such that $a\otimes b
\simeq b\otimes a \simeq \tto$, where $\tto$ is the trivial object (the unit of
$\otimes$ operation).

Let us give a more careful definition.  An invertible 0d domain wall $\hat\ga$
between two 1d topological orders $\sA^1$ and $\sB^1$ is well-defined (see
Section~4.3 in \Ref{KZ150201690}), and is denoted by $\sA^1\seq{\hat\ga} \sB^1$
or $\hat\ga: \sA^1\simeq \sB^1$.  Higher dimensional invertible domain walls
are defined by induction: 
\begin{Definition}
\label{invWall}
A gapped domain wall $\sM^{n-1}$ between two gapped walls $\sA^n$ and $\sB^n$
is called \textbf{invertible} if there is a gapped domain wall $\sN^{n-1}$
between $\sB^n$ and $\sA^n$ such that there exist an invertible gapped domain
wall between $\sM^{n-1}\ot{\sB^n}\sN^{n-1}$ and $\id_{\sA^n}$, i.e.
$\sM^{n-1}\ot{\sB^n}\sN^{n-1}\simeq \id_{\sA^n}$, and one between
$\sN^{n-1}\ot{\sA_n} \sM^{n-1}$ and $\id_{\sB^n}$, i.e. $\sN^{n-1}\ot{\sA_n}
\sM^{n-1}\simeq \id_{\sB^n}$. 
\end{Definition}
\noindent
The invertible domain wall will be used extensively later.

\subsection{Looping and delooping}
\label{looping}

Looping and delooping operations reveal the layered structures in higher
categories.  From an $n$-category we can construct a fusion $\nmo$-category via
a process called looping (see Fig. \ref{PandA} where the morphisms are viewed
as paths and loops):
\begin{Definition}
Given an $n$-category $\cC$, we choose an object $\v a$ (the ``base point'').
We can construct a fusion $\nmo$-category, denoted as $\Om_{\v a}\cC$, whose
objects are given by the morphisms $\v k: \v a\to \v a$.  In other words
$\Om_{\v a} \cC = \Hom(\v a,\v a)$.  If $\cC$ is a fusion $n$-category, we
usually choose the base point to be the unit of fusion $\one_\cC$: $\Om \cC =
\Hom(\one_\cC,\one_\cC)$, and $\Om \cC$ becomes a braided fusion $n$-category.
\end{Definition}

To apply the looping to a physical situation, let us consider a single simple
object $\sC^{n+1}$ in $\cTO^{n+1}$, which corresponds to an $n$d gapped
boundary of an anomaly-free topological order $\sM^{n+2}=\Bulk(\sC^{n+1})$ in
$\npo$-dimensional space (see \eqn{BulkCM}).  $\sC^{n+1}$ is also an $n$d
anomalous topological order.  For special case $\sM^{n+2}=\tto^{n+2}$,
$\sC^{n+1}$ becomes an $n$d anomaly-free topological order.
$\Hom(\sC^{n+1},\sC^{n+1})$ is the collection of $(n-1)$d excitations on the
boundary.  Here we include the morphisms, as well as the $\otimes$ and $\oplus$
operations to view $\Hom(\sC^{n+1},\sC^{n+1})$ as a fusion $n$-category,
denoted by $\cC^n = \Hom(\sC^{n+1},\sC^{n+1})$.  Thus $\cC^n$ describe all the
codimension-1, codimension-2, \etc\ excitations on the $n$d boundary
$\sC^{n+1}$

The unit object under $\otimes$ in $\cC^n$ is denoted as
$\one_{\cC^n}=\id_{\sC^{n+1}}$, which is the trivial codimension-1 excitations
in $\sC^{n+1}$.  Then the looping $\Om\cC^n = \Hom(\one_{\cC^n},\one_{\cC^n})$
is a fusion $\nmo$-category, which describes the codimension-2 excitations on
the $n$d boundary $\sC^{n+1}$.  Those excitations can also braid and $\Om\cC^n$
is in fact a braided fusion $\nmo$-category.  

We see that the looping of a fusion category $\cC^n$ is obtained by dropping
the objects (the codimension-1 excitations) and include only the morphisms of
the trivial object (the codimension-2 excitations).  The looping operation can
be continued, and the commutativity increases.  For example $\Om\Om\cC^n\equiv
\Om^2 \cC^n$ is a symmetric fusion $\nmt$-category, \etc.

There is reverse process of looping, called delooping (see Fig. \ref{PandA}).
From a fusion $n$-category, we can construct an $\npo$-category via delooping:
\begin{Definition}
Given an fusion $n$-category $\cC$, we can construct a $\npo$-category,
denoted as $\Si_* \cC$, which has only one object $*$ and whose morphisms
are given by the objects of $\cC$. In other words, $\Hom(*,*) =\cC$.  We
can complete $\Si_* \cC$ by adding the composite objects $*\oplus *
\cdots$, to obtain an additive $\npo$-category with $\oplus$ operation.  We can also
do a condensation completion, by adding objects (the gapped liquid
phases) formed by the codimenson-1 excitations (\ie the 1-morphisms of $\Si_*
\cC$), the codimenson-2 excitations (\ie the 2-morphisms of $\Si_* \cC$), \etc.
The resulting $\npo$-category is called the delooping of $\cC$ and is denoted by $\Si \cC$.
\end{Definition}

\begin{Remark} \label{rem:delooping}
Our definition of delooping is compatible with Definition$^{\mathrm{ph}}$\,\ref{fc} (see also Remark \ref{CCrem} and \ref{rem:fusion-n-cat}), and is different from that in \Ref{GJ190509566,J200306663}. 
\end{Remark}

As an application to physics, let us consider a braided fusion $\nmo$-category
$\eC^{n-1}$ that describes the codimension-2 and higher excitations in
$n$-dimensional space. Then the delooping $\Si_{\tto^n} \eC^{n-1}$ is the fusion
$n$-category with only one object $\tto^n$, which correspond to the trivial
codimension-1 excitation in the  $n$-dimensional space.  We can do a
condensation completion by adding $\tto^n\oplus\tto^n\cdots$, as well as all
the descendant codimension-1 excitations, obtained from condensing
codimension-2 and higher excitations.  The resulting fusion $n$-category is
$\Si \eC^{n-1}$.  If we can add a braiding structure to $\Si \eC^{n-1}$, making
it a braided fusion $n$-category, then the delooping plus condensation
completion can be continued.  $\Si\Si\eC^{n-1}=\Si^2\eC^{n-1}$ is a fusion
$\npo$-category.

We note that excitations in a trivial topological order $\tto^{n+1}$ in
$n$-dimensional space are described by a fusion $n$-category
$\Hom(\tto^{n+1},\tto^{n+1})$.  It contains $(n-1)$d, $\nmt$d, $\cdots$, $0$d
excitations.  If we drop the $(n-1)$d excitations, the remaining $\nmt$d,
$\cdots$, $0$d excitations correspond to excitations in trivial
topological order $\tto^{n}$ in $\nmo$-dimensional space, and are described by
$\Hom(\tto^{n},\tto^{n})$.  This way we find
\begin{align}
\Om \Hom(\tto^{n+1},\tto^{n+1}) = \Hom(\tto^{n},\tto^{n}).
\end{align}

All the excitations in trivial topological order are descendent excitations.
Thus if we add one layer of descendent excitations in one higher dimension, we
obtain excitations of a trivial topological order in one higher dimension.
Therefore we have
\begin{align}
\Si  \Hom(\tto^{n},\tto^{n}) = \Hom(\tto^{n+1},\tto^{n+1}).
\end{align}
We note that the codimension-1 excitations in a trivial topological order is
embeded in a product state in 1 higher dimension.  We can also view the same
excitation as embeded in a product state in 2 higher dimension. In this case,
the excitation becomes codimension-2 and can braid.  Thus
$\Hom(\tto^{n+1},\tto^{n+1})$ can also be viewed as a braided fusion
$n$-category, and we can perform delooping.  In fact, the braiding is trivial,
and $\Hom(\tto^{n+1},\tto^{n+1})$ can be viewed as a symmetric fusion
$n$-category.

Since $\Hom(\tto^{n+2},\tto^{n+2})=\afTO^{n+1}$ is the fusion higher category
of anomaly-free topological orders in $n$-dimensional space, we find that
\begin{align}
 \Om \afTO^{n+1} = \afTO^n,\ \ \Si \afTO^n = \afTO^{n+1}.
\end{align}
\noindent

We note that in $0$-dimensional space, the category of anomaly-free topological
order has only one simple object $\tto^1$, which corresponds to a single
quantum state $|\psi\>$.  The set of 1-morphisms $\Hom(\tto^1,\tto^1)=\C$ is
the set of 1-by-1 complex matrices.  We see that the category of anomaly-free
topological orders in $1$-dimensional space is given by $\cVec$ -- the category
of complex vector spaces:
\begin{align}
\afTO^1 = \Si\C=\cVec.
\end{align}
The higher category of $n$-vector spaces is given by the iterated delooping
\begin{align}
 n\cVec = \Si^{n-1} \cVec.
\end{align}

\begin{Remark}
Our definition of $n\cVec$ is different from that in \Ref{GJ190509566} for $n>2$ (recall Remark\,\ref{rem:delooping}). We suspect that the difficulty of defining $n\cVec$ mathematically might be due to the complexity of higher topological orders. 
\end{Remark}

We see that the category of anomaly-free topological orders
is given by
\begin{align}
 \afTO^{n+1} =  \Si^n \cVec = (n+1)\cVec.
\end{align}
\noindent
We would like to remark that the fusion $\npo$-category $ \afTO^{n+1}
=\Hom(\tto^{n+2},\tto^{n+2}) = (n+1)\cVec$ can also be promoted into a braided
fusion $\npo$-category, which is actually, a symmetric fusion $\npo$-category
with trivial exchange property (see Proposition~\ref{sfctri}).  After we promote
$\npo\cVec$ to a braided fusion $\npo$-category, we can denote it as
$\npo\eVec$. In other words, $\npo\eVec$ is the braided fusion
$\npo$-category obtained from the fusion $\npo$-category $\npo\cVec$ by
adding the trivial braiding structure (which is always doable).

Consider an anomaly-free topological order $\sM^{n+1} \in \afTO^{n+1}$. Its
excitations are described by a fusion $n$-category
$\cM^n=\Hom(\sM^{n+1},\sM^{n+1})$.  The objects in $\cM^n$ are codimension-1
excitations, which cannot braid and cannot be remotely detected by any
excitations.  Thus, according to Proposition~\ref{remote}, those codimension-1
excitations must all have the trivial nl-type.  We believe that 
\begin{Proposition}
all the excitations with the trivial nl-type are descendant excitations, coming
from the condensations of lower dimensional excitations.
\end{Proposition}
\noindent
Thus, the codimension-1 excitations in an anomaly-free topological order are
all descendant excitations.  Droping those codimension-1 excitations gives us
the looping $\Om \cM^n$. The delooping of $\Om \cM^n$ addes back those
descendant codimension-1 excitations. We find
\begin{Proposition}
The excitations in an anomaly-free topological order described by fusion
$n$-category $\cM^n$ satisfy the following relation
\begin{align}
 \cM^n = \Si\Om \cM^n .
\end{align}
\end{Proposition}
\noindent
The reverse may not be true: a fusion $n$-category $\cM^n$ satisfying $\cM^n =
\Si\Om \cM^n$ may not describe the excitations in an anomaly-free topological
order.

We see that the category of anomaly-free topological orders
 $\afTO^{n+1}=(n+1)\cVec$ is a fusion $\npo$-category.  Since the
 condensation completion always include excitations induced by condensing the
 trivial excitations and $n\cVec$ is formed only by those excitations induced by
 condensing the trivial excitations, we find that
 \begin{Corollary}
 a fusion $n$-category $\cC^n$ is a bimodule of
 $n\cVec$:
 \begin{align}
  \cC^n\otimes n\cVec = n\cVec \otimes \cC^n.
 \end{align}
 \end{Corollary}
 \noindent
 We also have
 \begin{Corollary}
 The $n$-category $\cM^n$ that describes the excitations in an anomaly-free
 topological order $\sM^{n+1} \in \afTO^{n+1}$, $\cM^n
 = \Hom(\sM^{n+1},\sM^{n+1})$, is a fusion $n$-category.
 \end{Corollary}
 \noindent
 This is because the $n$-category $\cM^n$ contains all the descendent
 excitations.

\subsection{Boundary-bulk relation}
\label{BBrel}

Consider an anomaly-free stable topological order $\sM^{n+2}$ in
$\npo$-dimensional space, $\sM^{n+2} \in \afcTO^{n+2}$, and its gapped boundaries.  The $\npo$d bulk
topological order and the $n$d gapped boundaries have a very direct relation.
\Ref{KW1458,KZ150201690,KZ170200673} proposed a holographic principle for this
boundary-bulk relation: \emph{boundary uniquely determines bulk}. The
boundary-bulk relation has several versions, differ in mathematical details.

In the first version, we consider a linear $\npo$-category $\cB^{n+1}$ and a
fusion $\npo$-category $\cM^{n+1}$ that acts on $\cB^{n+1}$ from left.
$\cB^{n+1}$ is also a right module over the fusion $\npo$-category
$\Hom(\tto^{n+2},\tto^{n+2})$.  The pair $(\cB^{n+1},\cM^{n+1})$ describes a
category of all gapped boundaries of an $\npo$d anomaly-free topological order
in $\afcTO^{n+2}$.  We believe that 
\begin{Proposition}
there is only one anomaly-free topological order $\sM^{n+2}$ in $\afcTO^{n+2}$,
which gives rise to the category of the gapped boundaries
\begin{align}
&\ \ \ \ (\cB^{n+1},\cM^{n+2})
\nonumber\\
& = \Big(\Hom(\tto^{n+2},\sM^{n+2}), \Hom(\sM^{n+2},\sM^{n+2}) \Big)
.
\end{align}
\end{Proposition}
\noindent
We would like to point that the pair $(\cB^{n+1},\cM^{n+2})$ not only uniquely
determines the bulk topological order $\sM^{n+2}$, it also gives extra
information about how the bulk is connected to the boundary.  If we
ignore such information, we believe that the linear $\npo$-category of the
gapped boundaries can already  uniquely determines the bulk topological order:
\begin{Proposition}
There is only one anomaly-free topological order $\sM^{n+2}$ in $\afcTO^{n+2}$,
which gives rise to the linear $\npo$-category for the gapped boundaries
\begin{align}
\cB^{n+1} = \Hom(\tto^{n+2},\sM^{n+2})
.
\end{align}
\end{Proposition}

In the second version, we consider a particular gapped boundary $\sC^{n+1} \in
\cB^{n+1}$.  Now $\Hom(\sM^{n+2},\sM^{n+2})$ does not act within $\sC^{n+1}$
since the fusion with excitations in $\Hom(\sM^{n+2},\sM^{n+2})$ (\ie the $n$d
excitations in $\sM^{n+2}$) may change $\sC^{n+1}$ to some other boundaries
$\t{\sC}^{n+1}$.  However, the $(n-1)$d excitations in $\sM^{n+2}$ act within
$\sC^{n+1}$. The $n$-category of all the $(n-1)$d excitations is given by
$\Hom(\id_{\sM^{n+2}},\id_{\sM^{n+2}})$ where $\id_{\sM^{n+2}} \in \Hom(\sM^{n+2},\sM^{n+2})$ is the unit morphism (that corresponds to the trivial $n$d
excitation in the bulk topological order $\sM^{n+2}$).  In fact, $\Hom(\id_{\sM^{n+2}},\id_{\sM^{n+2}})$ is a braided fusion category, which is actually
defined as  the looping $\Om \Hom(\sM^{n+2},\sM^{n+2})$.  Therefore, there is
a braided fusion $n$-category $\eM^n = \Om \Hom(\sM^{n+2},\sM^{n+2})$, that
acts on $\sC^{n+1}$.  A gapped boundary is described by a pair $(\sC^{n+1},\eM^n)$.  We believe that the pair $(\sC^{n+1},\eM^n)$ uniquely
determines the bulk topological order and how the bulk topological order is
connected to the boundary.  If we ignore the information about how the bulk is
connected to the boundary, we believe that $\sC^{n+1} \in \Hom(\tto^{n+2},\sM^{n+2})$ uniquely determines the bulk topological order:
\begin{Proposition}
\label{BulkCMp}
There is only one anomaly-free topological order $\sM^{n+2}$ in $\afcTO^{n+2}$,
which gives rise to the boundary
\begin{align}
\sC^{n+1} &\in \Hom(\tto^{n+2},\sM^{n+2})
.
\end{align}
We denote such boundary-bulk relation as
\begin{align}
\label{BulkCM}
 \Bulk (\sC^{n+1}) = \sM^{n+2} .
\end{align}
\end{Proposition}
\noindent
The above is the accurate meaning of \emph{boundary uniquely determines 
bulk}.
 
$\cC^n$ can determine the boundary topological order $\sC^{n+1}$ up to an invertible topological order.  Since we believe that all
invertible topological orders are anomaly-free, the excitations
$\cC^n=\Hom(\sC^{n+1},\sC^{n+1})$ in the  boundary topological order $\sC^{n+1}$ can
already determine the bulk topological order $\sM^{n+2}$. We obtain
\begin{Proposition}
\label{bulkCMp}
For any fusion $n$-category $\cC^n$,
there is only one anomaly-free topological order $\sM^{n+2}$ in $\afcTO^{n+2}$ admitting a boundary $\sC^{n+1}\in \Hom(\tto^{n+2},\sM^{n+2})$ 
such that
\begin{align}
\cC^n &= \Om\sC^{n+1}.
\end{align}
We denote such boundary-bulk relation as
\begin{align}
\label{bulkCM}
 \bulk(\cC^{n}) = \sM^{n+2}
\end{align}
\end{Proposition}
\noindent
Here, we have assumed the following. 
\begin{Proposition}
\label{realizefcat}
A fusion $n$-category $\cC^n$ can always be realized by the
excitations in a potentially anomalous topological order $\sC^{n+1}$ such that
$\cC^n = \Om\sC^{n+1}$.
\end{Proposition}
\noindent
The above result was shown for $n=1$ case.  Given a fusion category $\cC$, we
can explicitly construct a 2d string-net liquid state,\cite{LW0510} that has a
boundary realizing the fusion category $\cC$.\cite{LW1384} For $n>1$, the
general construction is sketched in Proposition~\ref{FunVV}.

In the third version, we only consider the excitations on a particular gapped
boundary $\sC^{n+1} \in \cB^{n+1}$, 
and instead of determining bulk topological orders,
we ask only whether boundary excitations can determine bulk
excitations.
The boundary excitations are described
by a fusion $n$-category $\cC^n = \Hom(\sC^{n+1},\sC^{n+1})$.  Again
$\Hom(\sM^{n+2},\sM^{n+2})$ ($n$d excitations in $\sM^{n+2}$) does not act
within $\cC^n$.  However, the $(n-1)$d excitations in $\sM^{n+2}$ act within
$\cC^n$.  The braided fusion $n$-category of all the $(n-1)$d excitations is
given by $\eM^n=\Om \Hom(\sM^{n+2},\sM^{n+2})$, which acts on $\cC^n$.  In
other words, $\cC^n$ is a left module over $\eM^n$. It is also a right module
over
$n\cVec=\Hom(\tto^{n+1},\tto^{n+1})$.  A gapped boundary, up to an invertible
topological order, is described by a pair $(\cC^n,\eM^n)$.  In fact, the data
$(\cC^n,\eM^n)$ determines the boundary excitations, which in turn determine
the gapped boundary, up to an invertible topological order. 
We believe that the data $(\cC^n,\eM^n)$ can determine the
category of the bulk excitations (\ie determine the bulk topological order up
to an invertible topological order). In fact, we believe that $\cC^n$ alone can
uniquely determine the category of the bulk excitations. 
\begin{Proposition}
There is only one anomaly-free topological order $\sM^{n+2}$ in $\afcTO^{n+2}$,
up to invertible topological orders, which gives rise to the category of
boundary excitations:
\begin{align}
\cC^n & = \Hom(\sC^{n+1},\sC^{n+1}),
\nonumber\\
\mathrm{where}\ \ 
\sC^{n+1} &\in \Hom(\tto^{n+2},\sM^{n+2})
.
\end{align}
\end{Proposition}
\noindent
The above result can be rephrased.  Let us denote the fusion
$\npo$-category of the bulk excitations as $\cM^{n+1}$ (which is given by
$\Hom(\sM^{n+2},\sM^{n+2})$). Then $\cM^{n+1}$ is uniquely determined by a
braided fusion $n$-category $\eM^n=\Om \cM^{n+1}$, via delooping:
$\cM^{n+1}=\Si \eM^n$, since the bulk topological order is anomaly-free.  
\begin{Proposition}
\label{Z1CMp}
The braided fusion $n$-category $\eM^n$ is uniquely determined by $\cC^n$ 
\begin{align}
  \eM^n=Z_1(\cC^n),
\end{align}
where boundary-bulk relation $Z_1$ is called $Z_1$ center (the Drinfeld center
when $n=1$). Thus $\cC^n$ uniquely determines $\cM^{n+1}$ via
\begin{align}
 \cM^{n+1} =\Si  Z_1(\cC^n).
\end{align}
\end{Proposition}
\noindent
Mathematically, the above result is phrased as
\begin{Proposition}
From any  fusion $n$-category $\cC^n$, we can always construct
a unique braided fusion $n$-category $Z_1(\cC^n)$,
which is the maximal one equipped with a \textbf{central} monoidal functor $\eM^n
\xrightarrow{F_{\cC^n}} \cC^n$, \ie for any $\v x$ in  $\eM^n$ and $\v y$ in $\cC^n$
\begin{align}
 F_{\cC^n}(\v x) \otimes \v y \simeq \v y\otimes F_{\cC^n}(\v x) ,
\end{align}
such that $Z_1(\cC^n)$ is the category of codimension-2 excitations in the bulk
of $\cC^n$
\begin{align}
  \eM^n=Z_1(\cC^n).
\end{align}
\end{Proposition}
\noindent
Such central functor $F_{\cC^n}:Z_1(\cC^n)\to \cC^n$ is also referred to as the
forgetful functor, since by construction, objects in $Z_1(\cC^n)$ can be viewed as objects in
$\cC^n$ equipped with additional half braiding structures.

\begin{figure}[t]
  \centering
  \includegraphics[scale=0.8]{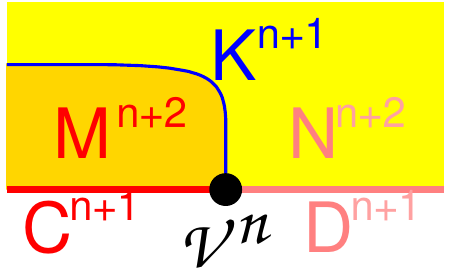}
%14
  \caption{The ``bulk''  of a domain wall on the boundary\cite{KZ150201690}.}
  \label{fig:bimodfun}
\end{figure}

We now explain an explicit construction of $Z_1$. To do this, consider a
slightly more complicated configuration as in Fig.~\ref{fig:bimodfun}, where $\sC^{n+1} \in
\Hom(\tto^{n+2},\sM^{n+2})$, $\sD^{n+1} \in
\Hom(\tto^{n+2},\sN^{n+2})$, $\sK^{n+1} \in
\Hom(\sM^{n+2},\sN^{n+2})$, and $\cV^n=\Hom(\sD^{n+1},\sK^{n+1}\ot{\sM^{n+2}} \sC^{n+1})$. We view $\cV^n$ as a collection of domain
walls between the boundary $\sC^{n+1}$ and $\sD^{n+1}$ and it uniquely
determines the ``bulk'' $\sK^{n+1}$, which is a domain wall between the bulk of $\v
C^{n+1}$, and the bulk of $\sD^{n+1}$, namely between $\sM^{n+2}$ and $\sN^{n+2}$. 

Observe that all three fusion $n$-categories, $\cK^n=\Hom(\sK^{n+1},\sK^{n+1})$, $\cC^n=\Hom(\sC^{n+1},\sC^{n+1})$, and $\cD^n=\Hom(\sD^{n+1},\sD^{n+1})$, act on $\cV^n$. Moreover, the three actions commute with each other.
Here we want to separate the action of $\cK^n$ from those of $\cC^n$ and
$\cD^n$. 

Let us introduce $\Fun(\cV^n,\cV^n)$ as a collection of  endofunctors of
the linear $n$-category $\cV^n$, or more precisely, a category of linear functors $f: \cV\to
\cV$. In other words, for objects $\v v,\v w\in \cV$, the functor $f$
satisfies
\begin{align}
 f(\v v\oplus \v w) \simeq f(\v v) \oplus f(\v w), 
\end{align}
Note that these functors are higher functors between higher categories, and
consist of many structures at different levels of morphisms. In this paper, we are not
giving rigorous descriptions, but only listing 
the structures at the object level for illustration. The structures on higher
morphisms are similar. 

$\Fun(\cV^n,\cV^n)$ is naturally a linear monoidal category since, for $f,g \in
\Fun(\cV^n,\cV^n)$, we can define
\begin{align}
 (f\otimes g)(\v v) = f(g(\v v)),\ \ \ \
 (f\oplus g)(\v v) = f(\v v) \oplus g(\v v) .
\end{align}
Now we can see that an action of $\cK^n$ on $\cV^n$ is the same as a monoidal
functor $\cK^n\to \Fun(\cV^n,\cV^n)$, in other words an object $\v k \in \cK^n$
corresponds to a functor $f_{\v k}\in \Fun(\cV^n,\cV^n)$:
\begin{align}
  f_{\v k}(\v v) = \v k \ot{\cK^n} \v v, \ \ \ \v v \in \cV^n,
\end{align}
where $\v k \ot{\cK^n} \v v$ describes the fusion of an object $\v k \in \cK^n$
to an object $\v v \in \cV^n$ along the domain wall $\sK^{n+1}$.

Similarly, we have the actions of $\cC^n$ and $\cD^n$ on $\cV^n$, which commute
with each other and make $\cV^n$ into a $\cC^n$-$\cD^n$-bimodule. Thus the action
of $\cK^n$, that commutes with both actions of $\cC^n$ and $\cD^n$, identifies
$\cK^n$ with the bimodule endofunctors of $\cV^n$, \ie, all the linear functors
that commute with both actions. More precisely, denote the left action of
$\cC^n$ by $\sC\ot{\cC^n} \v  v$ and right action of $\cD^n$ by $\v v\ot{\cD^n}\v  d$
for $\v v\in \cV^n,\ \sC\in \cC^n,\ \v d\in \cD^n$, a bimodule functor is a
functor $f:\cV^n\to \cV^n$ together with natural isomorphisms
\begin{align}
  f(\sC\ot{\cC^n} \v v)\simeq  \sC\ot{\cC^n}f(\v v),\quad 
f(\v v\ot{\cD^n}\v  d)\simeq f(\v v)\ot{\cD^n} \v d,
\end{align}
and other appropriate higher structures. We note that the above are addional
\emph{structures} rather than conditions. Denote the category of all bimodule
functors  by $\Fun_{\cC^n|\cD^n}(\cV^n,\cV^n)$.  The monoidal functor
$\cK^n\to \Fun(\cV^n,\cV^n)$ can be promoted to $\cK^n\to
\Fun_{\cD^n|\cC^n}(\cV^n,\cV^n)$. Following the holographic principle, we expect
such functor to be an equivalence. Thus we have the following boundary-bulk
relation, extended to domain walls on the boundary and in the bulk:
\begin{Proposition}
\label{FunVV}
  Let $ \sC^{n+1}\in
\Hom(\tto^{n+2},\sM^{n+2})$, $\sD^{n+1} \in
\Hom(\tto^{n+2},\sN^{n+2})$, 
 and their
  excitations $\cC^n=\Hom(\sC^{n+1},\sC^{n+1})$, $\cD^n=\Hom(\sD^{n+1},\sD^{n+1})$. A $\cC^n$-$\cD^n$-bimodule $\cV^n$, viewed as a collection of
  domain walls between $\sC^{n+1}$ and $\sD^{n+1}$, uniquely determines a
  domain wall $\sK^{n+1}$ in the bulk, \ie, $\sK^{n+1}\in \Hom(\sM^{n+2},\sN^{n+2})$. In other words, there is a unique $\sK^{n+1}\in \Hom(\sM^{n+2},\sN^{n+2})$ such that $\cV^n=\Hom(
  \sD^{n+1},\sK^{n+1}\ot{\sM^{n+2}} \sC^{n+1})$. The excitations on $\sK^{n+1}$ is given by
    \begin{align}
      \cK^n=\Hom(\sK^{n+1},\sK^{n+1})=\Fun_{\cC^n|\cD^n}(\cV^n,\cV^n).
    \end{align}
\end{Proposition}
\noindent
Objects in
$\cK^n$ correspond to functors in $\Fun_{\cC^n|\cD^n}(\cV^n,\cV^n)$.

As a special case, take $\cV^n=\cD^n=\cC^n$, \ie, we view $\cC^n$ as a
collection of domain walls between $\sC^{n+1}$ and itself. The ``bulk'' of
$\cC^n$ is the trivial domain wall in the bulk of $\sC^{n+1}$ and the
excitations on the trivial domain wall are just the codimension-2 excitations
in the bulk of $\sC^{n+1}$. We obtain the explicit construction
\begin{align}
  \eM^n=Z_1(\cC^n):=\Fun_{\cC^n|\cC^n}(\cC^n,\cC^n).
\end{align}
For a bimodule functor $f\in Z_1(\cC^n)$, and any $\v y \in \cC^n$
\begin{align}
  f(\one_{\cC^n})\otimes \v y\simeq f(\v y)\simeq \v y\otimes f(\one_{\cC^n}).
\end{align}
We see that a bimodule funcor $f$ is the same as an object $f(\one_{\cC^n})$ in
$\cC^n$
together with the half braiding $f(\one_{\cC^n})\otimes \v y\simeq \v y\otimes
f(\one_{\cC^n})$. The forgetful functor is thus
\begin{align}
  F_{\cC^n}: \Fun_{\cC^n|\cC^n}(\cC^n,\cC^n)&= Z_1(\cC^n)\to \cC^n,
  \nonumber\\
  f &\mapsto F_{\cC^n}(f)=f(\one_{\cC^n}).
\end{align}

In this paper, we mainly use the third version of boundary-bulk relation
$Z_1(\cC^n)=\eM^n$: the codimension-1 boundary excitations described by a
fusion $n$-category $\cC^n$ uniquely determines the  codimension-2 bulk
excitations described by a braided fusion $n$-category $\eM^n$. In contrast,
\eqn{BulkCM} is a relation between a boundary topological order (\ie an
anomalous $n$d topological order -- an object in $\Hom(\tto^{n+2},\sM^{n+2})$)
and a bulk topological order (\ie an anomaly-free $\npo$d topological order --
an object in $\afcTO^{n+2}$).

\subsection{Example of topological orders and the corresponding higher
categories}

\subsubsection{Invertible topological orders}
\label{invTop}

The simplest anomaly-free topological orders are invertible topological
orders.\cite{KW1458,F1478,K1459} We use $\icTO^{n+1}$ to denote the category of
all $n$d invertible topological orders.  We believe that there are no
anomalous invertible topological orders:
\begin{Proposition}
Consider a potentially anomalous topological order in $n$-dimensional space:
$\sC^{n+1} \in \Hom(\tto^{n+2},\sM^{n+2})$ for $\sM^{n+2} \in
\afcTO^{n+2}$, if its excitations are the same as those for the trivial
topological order, \ie $\Om\sC^{n+1}
=\Om\tto^{n+1}=n\cVec$, then $\sM^{n+2}=\tto^{n+2}$ (\ie $\sC^{n+1}$ is anomaly-free) and $\sC^{n+1}$ is an invertible topological order.
\end{Proposition}
\noindent
By definition, the invertible topological orders form Abelian groups under the
stacking $\otimes$.  In different dimensions, those groups are given
by\cite{KW1458,F1478,K1459}
\begin{align}
 \bmm
(n+1)\text{D}: & 0+1 & 1+1 & 2+1 & 3+1 & 4+1 \\
\icTO^{n+1}: & 0 & 0 & \Z & 0 & \Z_2\\
\emm
\end{align}

The generator of
$\icTO^3$ is the $E_8$ bosonic quantum Hall state
described by the wave function
\begin{align}
 \Psi(z_i^I) &
= \Big(\prod_{i,j} \prod_{I<J} (z_i^I-z_j^J)^{K_{IJ}} \Big)
\Big( \prod_{i<j} \prod_{I} (z_i^I-z_j^I)^{K_{II}} \Big)
\nonumber\\
& \ \ \ \ 
\ee^{-\frac14 \sum_{i,I} |z_i^I|^2}
\end{align}
where the $K$-matrix is given by
\begin{align}
 K=\begin{pmatrix}
2&1&0&0&0&0&0&0\\
1&2&1&0&0&0&1&0\\
0&1&2&1&0&0&0&0\\
0&0&1&2&1&0&0&0\\
0&0&0&1&2&1&0&0\\
0&0&0&0&1&2&0&0\\
0&1&0&0&0&0&2&1\\
0&0&0&0&0&0&1&2\\
\end{pmatrix}
\end{align}
which satisfies det$(K)=1$.  The generator of $\icTO^5$ is given following
4d bosonic system (described by path integral for cochain
fields\cite{W161201418})
\begin{align}
\label{invT5D}
 Z &=\sum_{a^{\Z_2},b^{\Z_2}} 
\ee^{\ii \pi \int_{M^{4+1}} (\w_2+\dd a^{\Z_2}) (\w_3 +\dd b^{\Z_2})},
\end{align}
where 
$a^{\Z_2}$ is a $\Z_2$-valued 1-cochain, $b^{\Z_2}$ is a $\Z_2$-valued
2-cochain, and $\w_n$ is the $n^\text{th}$ Stiefel-Whitney class of the tangent
bundle of the closed spacetime manifold $M^{4+1}$.  The path integral only
depends on the cohomology classes of $\w_2$ and $\w_3$, since the path integral
is invariant under the following gauge transformation
\begin{align}
 \w_2 &\to \w_2 + \dd \ga, & \w_3 &\to \w_3 + \dd \la ,
\nonumber\\
 a^{\Z_2} &\to a^{\Z_2} + \ga, & b^{\Z_2} &\to b^{\Z_2} + \la. 
\end{align}
The path integral can be calculated exactly
\begin{align}
 Z &=\sum_{a^{\Z_2},b^{\Z_2}} 
\ee^{\ii \pi \int_{M^{4+1}} (\w_2+\dd a^{\Z_2}) (\w_3 +\dd b^{\Z_2})},
\nonumber\\
&= 2^{N_l+N_t} \ee^{\ii \pi \int_{M^{4+1}} \w_2\w_3}
\end{align}
where $N_l$ is the number of the links and $N_t$ the number of the triangles in
the triangulated spacetime $M^{4+1}$.  The non-trivial topological invariant $
\ee^{\ii \pi \int_{M^{4+1}} \w_2\w_3}$ implies that \eqn{invT5D} realize a
non-trivial 4d invertible topological order.

The invertible topological order have no non-trivial \emph{nl-type} of
excitations, \ie no non-trivial \emph{topological} excitations.  All the
excisions are \emph{local}.  The different \emph{types} of local excitations
are described by the trivial fusion $n$-category $n\cVec$ for an $n$d
invertible topological order.  

For example, in 2-dimensional space, the objects in the category of invertible topological
orders $\icTO^3$ form an Abelian group $\Z$.  The morphisms on each object form
a trivial fusion $2$-category $2\cVec$.  Since the $E_8$ quantum Hall state has
no gapped boundary, it is not an \emph{exact} topological order, but is a
\emph{closed} (\ie anomaly-free) topological order.  Therefore, $\icTO^3$ has no
1-morphisms between different objects.  All domain walls between different
objects are gapless.  The 1-morphism that connect the same object is also
trivial.  This is because such a 1-morphism corresponds to an $1+1$D excitation
and there is no non-trivial $1+1$D anomaly-free topological order.

In our attempt to use higher categories to characterize topological orders, the
invertible topological orders are the most difficult ones.  This is because
higher categories mainly describes the excitations, but the excitations on top
of invertible topological orders are identical to those on top of of trivial
product state.  Fortunately, in the category of topological orders, we also
have information on the stacking operation $\otimes$ and the gapped domain
walls between topological orders. This allows us to distinguish invertible
topological orders. The invertible topological orders in 2d are particularly
difficult, since we do not even have any gapped domain walls (\ie no
1-morphisms).  Only the stacking operation $\otimes$ allows us to distinguish
2d invertible topological orders.

\subsubsection{$G$-topological orders}

Another class of topological orders for bosonic systems are called
$G$-topological orders (see Section~\ref{Lgauge}), which are described by gauge
theories with a finite group $G$.  We use $\sGT^{n+1}_G \in \afcTO^{n+1}$ to
denote $G$-topological order in $n$-dimensional space.  We use $\Om\sGT^{n+1}_G$ to denote the fusion $n$-category that describes the
excitations in $\sGT^{n+1}_G$ and use $\Om^2\sGT^{n+1}_G$ to denote the
braided fusion $\nmo$-category that describes the excitations with
codimension-2 and higher in $\sGT^{n+1}_G$.  It is known that $\sGT^{n+1}_G$ is anomaly-free
and has gapped boundary.  An example of $\sGT^3_{\Z_2}$ is given by Example
\ref{excZ2top}.

Let us describe 3d $\Z_2$-topological order $\sGT^4_{\Z_2}$ in more details.
Such a state has two {nl-types} of point-like excitations $\one,e$,  two
{nl-types} of string-like excitation $\one_s,m_s$, and one trivial nl-type
membrane-like excitations.  The $e$-particle has a fusion $e\otimes e=\one$ and
the $m_s$-loop has a fusion $m_s\otimes m_s=\one_s$.  The $\Z_2$-topological
order also has two types of point-like excitations $\one,e$,  four {types} of
string-like excitation $\one_s,m_s, e_s, e_s\otimes m_s$.  The string $e_s$ is
formed by $e$-particles condensing into the $\Z_2$ symmetry breaking state.  The
$e_s$-loop has a fusion $e_s\otimes e_s = 2e_s$.  Those
point-like and string-like excitations form the braided fusion $2$-category
$\Om^2\sGT^4_{\Z_2}$.

The 3d $\Z_2$-topological order $\sGT^4_{\Z_2}$ has infinite types of
membrane-like excitations corresponding to infinite different 2d topological
orders formed by trivial point-like excitations $\one$'s.  $\sGT^4_{\Z_2}$ also
has infinite types of membrane-like excitations corresponding to infinite
different 2d SET orders with $\Z_2$ symmetry, formed by $e$-particles with
mod-2 conservation.  There are third types of membrane-like excitations
corresponding to 2d topological orders formed $m_s$-loops.  The $m_s$-loops has
a mod-2 conservation that corresponds to a $\Z_2$ higher symmetry. Thus, this
kind of 2d topological orders can be viewed as having a spontaneous breaking of
$\Z_2$ 1-symmetry.  Those point-like, string-like, and membrane-like
excitations form the fusion $3$-category $\Om\sGT^4_{\Z_2}$.  The point-like and
string-like excitations form the braided fusion $2$-category $\Om^2\sGT^4_{\Z_2}$.

The above  3d $\Z_2$-topological order $\sGT^4_{\Z_2}$ is anomaly-free, which
means that it can be realized by a bosonic lattice model, as shown in Section
\ref{Lgauge}.  Another way to realize $\sGT^4_{\Z_2}$ is via the path integral of
$\Z_2$-valued 1-cochain fields, $a^{\Z_2}$:\cite{W161201418}
\begin{align}
 Z=\sum_{\dd a^{\Z_2} = 0} 1
\end{align}
where $\sum_{\dd a^{\Z_2}=0}$ is a summation over $\Z_2$-valued 1-cocycles.
One can also realize
$\sGT^4_{\Z_2}$ via the path integral of $\Z_2$-valued 2-cochain
fields, $b^{\Z_2}$:\cite{W161201418}
\begin{align}
 Z=\sum_{\dd b^{\Z_2} = 0} 1
\end{align}
where $\sum_{\dd b^{\Z_2}=0}$ is a summation over $\Z_2$-valued 2-cocycles.

Since $\sGT^4_{\Z_2}$ is anomaly-free, its excitations described by
$\Om\sGT^4_{\Z_2}$ satisfy
\begin{align}
 Z_1(\Om\sGT^4_{\Z_2}) = \Om 4\cVec \equiv 3\eVec.
\end{align}
But the above boundary-bulk relation between fusion higher categories and
braided fusion higher categories only tell us that $\sGT^4_{\Z_2}$ is either
anomaly-free or has invertible anomaly.  The stronger boundary-bulk relation is
given by
\begin{align}
  \Bulk(\sGT^4_{\Z_2}) = \tto^5.
\end{align}
This boundary-bulk relation tells us that $\sGT^4_{\Z_2}$ is anomaly-free. 

We would like to mention that there is also a 3d twisted $\Z_2$-topological order
where the point-like $\Z_2$-charges are fermions.  We denote such a twisted
$\Z_2$-topological order as $\sGT^4_{\Z_2^f}$.  The twisted $\Z_2$-topological
order $\sGT^4_{\Z_2^f}$ is also anomaly-free and can be realized by the path
integral of $\Z_2$-valued 2-cochain fields,
$b^{\Z_2}$:\cite{W161201418,KT170108264,LW180901112}
\begin{align}
 Z=\sum_{\dd b^{\Z_2} = 0} \ee^{\ii \pi \int_{M^{3+1}} b^{\Z_2}b^{\Z_2} }
 =\sum_{\dd b^{\Z_2} = 0} \ee^{\ii \pi \int_{M^{3+1}} b^{\Z_2}\w_2} 
\end{align}
where $\sum_{\dd b^{\Z_2}=0}$ is a summation over $\Z_2$-valued 2-cocycles,
$M^{3+1}$ is a (3+1)-dimensional closed spacetime (with a triangulation), and
$\w_2$ is the second Stiefel-Whitney class of the tangent bundle of $M^{3+1}$.
Here we used a fact that $ b^{\Z_2}b^{\Z_2} +  b^{\Z_2}\w_2$ is a $\Z_2$-valued
coboundary.  The topological term $\ee^{\ii \pi \int_{M^{3+1}} b^{\Z_2}b^{\Z_2}
} = \ee^{\ii \pi \int_{M^{3+1}} b^{\Z_2}\w_2}$ makes the point-like
$\Z_2$-charges to be fermions.

\subsubsection{A 2d anomalous topological order}

Now, let us consider an anomalous topological order in 2d, denoted as $\sC^3_{Z_2}$, which has two \emph{nl-types} of point-like excitations, labeled by
$\one,e$, where $\one$ is a \emph{trivial} point-like excitation and $e$ has a
$\Z_2$ fusion $e\otimes e = \one$.  The anomalous topological order has two
\emph{types} of point-like excitations, which are also given by $\one,e$.  The
anomalous topological order has only one \emph{nl-type} of string-like
excitations, which is a local string-like excitation.  But it has two
\emph{types} of string-like excitations, labeled by $\one_s,e_s$.  The
$e_s$-type of string-like excitation is formed by the $e$-particles, condensing
into a 1d phase of spontaneous $\Z_2$ symmetry breaking state.  The $e_s$ loop
has a fusion $e_s\otimes e_s =2e_s$.  $\one_s,e_s$ are local
string-like excitations, \ie belong to the trivial \emph{nl-type} of
string-like excitations.

The excitations in the anomalous topological order $\sC^3_{\Z_2}$ are described
by a fusion $2$-category $\cC^2_{\Z_2}=\Om\sC^3_{\Z_2}=2\cRep(\Z_2)$.
$\cC^2_{\Z_2}$ has two simple objects $\one_s,e_s$.  On $\one_s$, there are two
simple 1-morphisms $\one,e$.  On $e_s$, there are also two simple 1-morphisms
$\one_{e_s},d_{e_s}$, with a fusion rule $d_{e_s}\otimes d_{e_s} = \one_{e_s}$.
There is one simple 1-morphisms $\sigma\in \Hom(\one_{s},e_s)$ and one simple
$\bar\sigma\in\Hom(e_s,\one_s)$, with fusion rules 
\begin{align}
  \sigma\ot{\one_s}\one=\sigma\ot{\one_s} e
  &=\one_{e_s}\ot{e_s}\sigma=d_{e_s}\ot{e_s}\sigma=\sigma,\nonumber\\
  \one\ot{\one_s}\bar\sigma=e \ot{\one_s}\bar \sigma
  &=\bar\sigma\ot{e_s}\one_{e_s}=\bar\sigma\ot{e_s} d_{e_s}
  =\bar\sigma
  ,\nonumber\\
  \bar \sigma\ot{e_s} \sigma&=\one\oplus e,\nonumber\\
  \sigma\ot{\one_s} \bar \sigma&=\one_{e_s}\oplus d_{e_s}.
\end{align}

The bulk of the anomalous topological order $\sC^3_{\Z_2}$ is the
$\Z_2$-topological order in 3-dimensional space $\sGT^4_{\Z_2}$:
\begin{align}
\label{BZ2a}
 \Bulk(\sC^3_{\Z_2}) = \sGT^4_{\Z_2}.
\end{align}
Since $\sGT^4_{\Z_2}$ is non-trivial, $\sC^3_{\Z_2}$ is  anomalous.  In fact
$\sC^3_{\Z_2}$ is a 2d gapped boundary of the 3d $\Z_2$ topological order $\sGT^4_{\Z_2}$ obtained via condensation of $Z_2$-flux strings.  
We have a similar relation for excitations
\begin{align}
 \label{BZ2b}
 Z_1(\cC^2_{\Z_2}) = \Om^2\sGT^4_{\Z_2}
\end{align}
where $\cC^2_{\Z_2}=\Om\sC^3_{\Z_2}$ is the fusion
$2$-category describing the excitations in $\sC^3_{\Z_2}$.  The relation
\eqn{BZ2a} carries more information than \eqn{BZ2b}.  We would like to remark that
when we stack the two  anomalous topological orders, both the boundaries and
the bulks are stacked:
\begin{align}
  \Bulk(\sC^3_{\Z_2}\otimes \sC^3_{\Z_2}) = \sGT^4_{\Z_2}\otimes \sGT^4_{\Z_2}.
\end{align}

\subsubsection{Anomalous 3d $\Z_2$ topological order}

The anomaly-free  3d $\Z_2$-topological order $\sGT^4_{\Z_2}$ discussed above can
also be realized via the path integral of $\Z_2$-valued 1-cochain and 2-cochain
fields, $a^{\Z_2}$ and $b^{\Z_2}$:\cite{W161201418}
\begin{align}
 Z=\sum_{a^{\Z_2},b^{\Z_2}} \ee^{\ii \pi \int_{M^{3+1}} b^{\Z_2} \dd a^{\Z_2}},
\end{align}
where $\sum_{a^{\Z_2},b^{\Z_2}}$ is a summation over $\Z_2$-valued 1-cochain
and 2-cochain.  
The above path integral has a gauge invariance for closed
$M^{3+1}$
\begin{align}
 a^{\Z_2} \to a^{\Z_2} + \dd \al,\ \ \ \
 b^{\Z_2} \to b^{\Z_2} + \dd \bt .
\end{align}

In this formulation, the twisted 3d $\Z_2$-topological order
$\sGT^4_{\Z_2^f}$ is realized by the path integral
\begin{align}
 Z=\sum_{a^{\Z_2},b^{\Z_2}} \ee^{\ii \pi \int_{M^{3+1}} b^{\Z_2} \dd a^{\Z_2}
+ b^{\Z_2} \w_2 }.
\end{align}
The above path integral is also gauge invariant for closed
$M^{3+1}$
\begin{align}
 a^{\Z_2} &\to a^{\Z_2} + \ga, \ 
 b^{\Z_2} \to b^{\Z_2} + \dd \bt, \ 
 \w_2 \to \w_2 + \dd \ga.
\end{align}
The path integral only depends on the cohomology classes of $\w_2$, So it
describes an anomaly-free theory.

In this section, we are going to study an anomalous 3d $\Z_2$ topological
order, realized by the following path integral
\begin{align}
 Z=\sum_{a^{\Z_2},b^{\Z_2}} \ee^{\ii \pi \int_{M^{3+1}} b^{\Z_2} \dd a^{\Z_2}
+ a^{\Z_2} \w_3 + b^{\Z_2} \w_2 }.
\end{align}
Under the gauge transformation
\begin{align}
 a^{\Z_2} &\to a^{\Z_2} + \ga, &
 b^{\Z_2} &\to b^{\Z_2} + \la, 
\nonumber\\
 \w_2 &\to \w_2 + \dd \ga, & 
 \w_3 &\to \w_3 + \dd \la ,
\end{align}
the above  path integral is not invariant.  The gauge non-invariance can be
fixed by adding a bulk term $ \ee^{\ii \pi \int_{N^5} \w_2\w_3}$ in one higher
dimension, where $ \prt N^5 = M^{3+1}$. The resulting path integral
\begin{align}
\label{Z2A}
 Z=\sum_{a^{\Z_2},b^{\Z_2}} \ee^{\ii \pi \int_{M^{3+1}} b^{\Z_2} \dd a^{\Z_2}
+ a^{\Z_2} \w_3 + b^{\Z_2} \w_2 } \ee^{\ii \pi \int_{N^5} \w_2\w_3}
\end{align}
is gauge invariant, \ie only depends on the cohomology classes of $\w_2$ and
$\w_3$.  Since, the path integral requires a bulk in one higher dimension to be
gauge invariant (\ie only depends on the cohomology classes of $\w_2$ and
$\w_3$), so it describes an anomalous theory.  We denote such a 3d anomalous
$\Z_2$-topological order as $\sGT^{4,\w_2\w_3}_{\Z_2^f}$.

Such a 3d anomalous $\Z_2$-topological order $\sGT^{4,\w_2\w_3}_{\Z_2^f}$ has a
fermionic point-like $\Z_2$ charge.  If the world-sheet for the $\Z_2$ flux
loop is unorientable, there is a world-line that marks the reversal of the
orientation.  Such an orientation-reversal world-line corresponds to a fermion
world-line.  In other words, the anomalous $\Z_2$ topological order has a
special property that a un-orientable world-sheet of the $\Z_2$-flux must 
bind with a world-line of the fermionic point-like $\Z_2$ charge.  Such a
fermionic worldline corresponds to the orientation reversal loop on the
unorientable worldsheet.

The 3d anomalous $\Z_2$-topological order $\sGT^{4,\w_2\w_3}_{\Z_2^f}$ has a
non-trivial bulk.  
The boundary-bulk relation can be written as
\begin{align}
\label{BB1}
 \Bulk(\sGT^{4,\w_2\w_3}_{\Z_2^f}) = \sI^5_{\w_2\w_3} ,
\end{align}
where $\sI^5_{\w_2\w_3}$ is the 4d invertible topological order
characterized by the topological invariant $\ee^{\ii \pi \int_{N^5}
\w_2\w_3}$.\cite{KW1458,F1478,K1459}
The boundary-bulk relation \eq{BB1} implies the following boundary-bulk
relation for the excitations
\begin{align}
\label{BB2}
 Z_1(\Om\sGT^{4,\w_2\w_3}_{\Z_2^f}) &= 3\eVec , \nonumber\\
3\eVec &=\Om\hspace{0.2mm}\sI^5_{\w_2\w_3},
\end{align}
since the excitations in an invertible topological order are described by a
trivial braided fusion higher category.  Despite the right-hand-side of
$Z_1(\Om\sGT^{4,\w_2\w_3}_{\Z_2^f}) = 3\eVec$ is a trivial braided fusion higher category,
as we have mention before, the boundary-bulk relation for fusion higher
categories $Z_1(\Om\sGT^{4,\w_2\w_3}_{\Z_2^f}) = 3\eVec$ does not imply $\sGT^{4,\w_2\w_3}_{\Z_2^f}$ to be anomaly-free.  In fact $\sGT^{4,\w_2\w_3}_{\Z_2^f}$ has an
invertible anomaly, which is a $\Z_2$ global gravitational anomaly.  So
\eqn{BB1} carries more information, which indecates that $\sGT^{4,\w_2\w_3}_{\Z_2^f}$
is anomalous.

\section{An example of algebraic higher symmetries: $G$-gauge theory}
\label{Lgauge}

For a quantum system with usual symmetry, the Hamiltonian commutes with a set
of operators which form a group under the operator product.  In this section,
we construct an example, in which the Hamiltonian commutes with a set of
operators that do not form a group under the operator product.   The
constructed model is an exactly solvable 3d local bosonic model\cite{K032}
whose ground state has a topological order described by a 3d gauge theory of a
finite group $G$.  The operators that commute with the Hamiltonian are the
Wilson line operators.  When $G$ is non-Abelian, the Wilson line operators,
under the operator product, form an algebra, which is not a group.

Our lattice bosonic model is defined on a 3d spatial lattice whose sites are
labeled by $i$.  Physical degrees of freedom live on the links which are
labeled by $ij$.  On an oriented link $ij$, the degrees of freedom are labeled
by $g_{ij} \in G$.  The labels $g_{ij}$'s on links with opposite orientations
satisfy 
\begin{align}
 g_{ij}=g_{ji}^{-1}
\end{align}
The many-body Hilbert space of our lattice bosonic model has the following
local basis
\begin{align}
|\{g_{ij}\}\>, \ \ \ g_{ij} \in G,\ \ \ \
ij \in \text{ links of cubic lattice}. 
\end{align}
The Hamiltonian of the exactly solvable model is expressed in terms of string
operators and point operators.

\subsection{The string operators}

The string operators $B_q(S^1)$ are defined on a closed loop $S^1$
formed by the links of the cubic lattice
and are labeled by $q$, the irreducible representation of the
gauge group $G$:
\begin{align}
 R_q(h_i g_{ij} h_j^{-1}) = R_q(h_i ) R_q( g_{ij} ) R_q^{-1}(h_j)
\end{align}
where $R_q(g_{ij})$ is the matrix of the irreducible representation.  A
$q$-string operator is given by
\begin{align}
\label{string}
 B_q(S^1) |\{g_{ij}\}\> &= 
\Tr \big[\prod_{ij \in S^1} R_q(g_{ij})\big] |\{g_{ij}\}\>.
\end{align}
So $B_q(S^1)$ is diagonal in the basis $|\{g_{ij}\}\>$:
$B_q(S^1) = \Tr \big[\prod_{ij \in S^1} R_q(g_{ij})\big]$.
We note that
\begin{align}
 B_{q}(S^1) B_{s}(S^1) = 
\Tr \prod_{I \in S^1} R_{q}(g_{ij}) \otimes_\C R_{s}(g_{ij}).
\end{align}
(We use $\otimes_\C$ to denote the usual tensor product of matrices or vector
spaces over the complex numbers $\C$, while $\otimes$ to denote the fusion of
excitations.)  Using 
\begin{align}
R_{q} \otimes_\C R_{s} =\bigoplus_{t} N^{qs}_{t} R_{t}, \ \ \ \
N^{qs}_{t} \in \mathbb{N},
\end{align}
we see that
\begin{align}
\label{BBNB}
  B_{q}(S^1) B_{s}(S^1) = \sum_{t} N^{{q}{s}}_{t} B_{t}(S^1).
\end{align}
The ends of the strings are point-like topological excitations and the above
$N^{qs}_{t}$ are the fusion coefficients of those  topological excitations.  
The quantum dimensions of those topological excitations, i.e. $d_q=\text{dim}(R_q)$, satisfy the following identity: 
\begin{align}
\sum_{s} N^{qs}_{t} d_{s} = d_{q}d_{t} .
\end{align}
We see that these string operators form a 
fusion algebra which is not a group when $G$ is non-Abelian.

Let
\begin{align}
 B(S^1) = \sum_q \frac{d_q}{D^2} B_q(S^1),\ \ D^2=\sum_q d_q^2 .
\end{align}
We have
\begin{align}
 B^2 &= \sum_{{q},{s}} \frac{d_{q}d_{s}}{D^4} B_{q}   B_{s}
  = \sum_{{q},{s},{t}} \frac{d_{q}d_{s}}{D^4}N^{{q}{s}}_{t} B_{t} 
  \nonumber\\
  &= \sum_{{q},{t}} \frac{d_{q}d_{q}}{D^4} d_{t} B_{t} =B.
\end{align}
Thus, $B$ is a projection operator.  In fact, it is a projection operator into
the subspace with $\prod_{{ij} \in \text{loop}} g_{ij}=1$.

\subsection{The point operators} 

A point operator is given by its action on the basis:
\begin{align}
 Q_h(i) |\{\cdots,g_{ij} ,g_{ij'}, \cdots\}\> 
= |\{\cdots,hg_{ij} ,hg_{ij'}, \cdots\}\>,
\end{align}
Clearly they satisfy
\begin{align}
 Q_h(i) Q_{h'}(i)= Q_{hh'}(i) .
\end{align}
So for a non-Abelian group $G$, in general
\begin{align}
 Q_h(i) Q_{h'}(i) \neq Q_{h'}(i) Q_h(i). 
\end{align}
But we have
\begin{align}
 Q_h(i) Q_{h'}(j) = Q_{h'}(j) Q_h(i),\ \ \ \ i\neq j. 
\end{align}
Let us introduce
\begin{align}
 C_a(i) = \sum_{h\in \chi_a}  Q_h(i),
\end{align}
where $\chi_a$ is a conjugacy class  labeled by $a$.
We find that
\begin{align}
 C_a(i) C_b(j)=C_b(j)C_a(i)
\end{align}
regardless if $i=j$ or not.

We note that, on a given site $i$,
\begin{align}
\label{CCMC}
 C_a(i)C_b(i) = \sum_{h \in \chi_a} \sum_{h' \in \chi_b}
Q_{h h'}(i)
=\sum_c M^{ab}_c C_c(i) ,
\end{align}
The above expression allows us to see that
$M^{ab}_c$ are non-negative integers.
Using $C_a C_b = C_bC_a$ and $(C_a C_b) C_c= C_a (C_b C_c)$, we find  that
\begin{align}
 M^{ab}_c &= M^{ba}_c, & \sum_d M^{ab}_d M^{dc}_{e} &= \sum_d M^{ad}_e M^{bc}_{d}
\end{align}
Let $(M_a)_{cb} = M^{ab}_c$, and we can rewrite the second equation in the
above as
\begin{align}
 M_c M_a = M_a M_c .
\end{align}

For example, the permutation group of three elements $S_3 = \{(123),(132),(321),(213),(231),(312)\}$ has
three conjugacy classes: $\chi_1 =\{(123)\}$, $\chi_2 =\{(132),(321),(213)\}$,
and $\chi_3 =\{(231),(312)\}$.  We find that
\begin{align}
C_1C_a&=C_a, & C_2C_2&=3C_1+3C_3,
\nonumber\\ 
 C_3C_3&=2C_1+C_3, & C_2C_3&=2C_2.
\end{align}

Let $\sC$ be a particular common eigenvector of $M_a$ whose components are all
non-negative.  (Such common eigenvector exists since the matrix elements of
$M_a$ are all non-negative.) The eigenvalue of such an eigenvector is $\la_a$
for $M_a$.  We choose the scaling factor of $\sC$ to satisfy
\begin{align}
 \sum_a \la_a c_a =1.
\end{align}
In this case we can define $Q_i=\sum_a c_a C_a(i)$ that satisfy
\begin{align}
 Q_i^2=Q_i,
\end{align}
Thus $Q_i$ is a projection operator.
In fact, $Q_i$ is given by
\begin{align}
 Q_i = |G|^{-1}\sum_{h\in G}  Q_h(i),
\end{align}
where $|G|$ is the number of elements in the group $G$.  We can check
explicitly that
\begin{align}
 Q_i^2
&=|G|^{-2} \sum_{h,h'} Q_h(i) Q_{h'}(i)
=|G|^{-2} \sum_{h,h'} Q_{hh'}(i)
\nonumber\\
&= |G|^{-1} \sum_{h} Q_{h}(i) = Q_i .
\end{align}

\subsection{A commuting-projector Hamiltonian}

We note that $Q_h(i)$'s generate the local gauge transformations.  Since the
closed-string operators are gauge invariant, we have
(for closed-string operators)
\begin{align}
 [B_q(S^1_\text{closed}),C_a(i)]&=0,
\nonumber\\
 [B_q(S^1_\text{closed}),B_{q'}({S^{1\prime}_\text{closed}})]&=0,
\nonumber\\
 [C_a(i),C_b(j)]&=0.
\end{align}
Therefore, we can construct the following commuting projector Hamiltonian
\cite{K032,MR1315}
\begin{align}
\label{Ham}
 H=U \sum_i(1-Q_i)+ J\sum_{\<ijkl\>} (1-B_{\<ijkl\>}),
\end{align}
where $U,J >0$,
\begin{align}
B_{\<ijkl\>} = \sum_i \frac{d_q}{D^2} B_q(\<ijkl\>)
\end{align}
and $\<ijkl\>$ labels the loops around the squares of the cubic
lattice.

The ground state of the above  exactly solvable Hamiltonian has a nontrivial
topological order.  The low energy effective theory is the $G$-gauge
theory.\cite{K032,MR1315}

\begin{figure}[t]
\begin{center}
\includegraphics[scale=1.0]{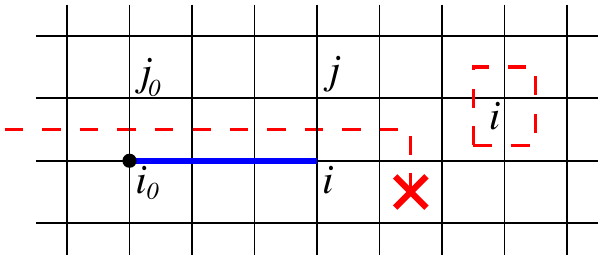} \end{center}
%15
\caption{ 
The red dashlines are membranes and the cross marks the boundary
of the membranes. The blue thick line is the path
$i_0\to i$.
}
\label{memb}
\end{figure}

\subsection{The point-like and string-like excitations}
\label{exc}

What are the excitations for the above Hamiltonian?  There are local point-like
excitations created by local operators.  There are also topological point-like
excitations that cannot be created by local operators.  Two topological
point-like excitations are said to be equivalent if they differ by local
point-like excitations.  The equivalent topological point-like excitations are
said to have the same type.  

We note that the closed string operators $B_q(S^1_\text{closed})$
\eqn{string} commute with the  Hamiltonian \eq{Ham}.  Thus the string operators
act within the ground state subspace.  We see that the ends of the open string
operators create point-like excitations, which are labeled by representations
$R_q$.  The types of topological point-like excitations one-to-one correspond to
the irreducible representations of $G$. In other words, topological point-like
excitations are described by representations of $G$ in a $G$-gauge theory.

Similarly, there are also topological string-like excitations.  They are created
at the boundary of the open membrane operators.  To define the membrane
operators, we point out that a membrane $\t S^2$ is formed by the faces of the
dual lattice, which is also a cubic lattice.  The faces of the dual lattice
correspond to the links in the original lattice and are also labeled by $ij$.
Let us first assume $G$ is Abelian. In this case,
the membrane operators are defined as
\begin{align}
 C_h(\t S^2)|\{g_{ij}\}\>
 =   \prod_{ij \in \t S^2} T_{ij}(h)|\{g_{ij}\}\>,
\end{align}
where the operator $T_{ij}(h)$ acts only on link $ij$ and is defined as
\begin{align}
 T_{ij}(h) |g_{ij}\> = |hg_{ij}\> \ \ \ \text{ or } \ \ \
 T_{ij}(h) |g_{ji}\> = |g_{ji}h^{-1}\> 
.
\end{align}
We see that $ C_h(\t S^2)$ simply shifts $g_{ij}$ on the membrane $\t S^2$ by
$h$.

For non-Abelian $G$, the membrane operators are given by
\begin{align}
\label{memop}
 C_a(\t S^2)|\{g_{ij}\}\>
 = \sum_{h \in \chi_a} \  \prod_{ij \in \t S^2} 
T_{ij}(h_{ij})|\{g_{ij}\}\>,
\end{align}
where $\chi_a$ is the $a^\text{th}$ conjugacy class of $G$.  In the $\prod_{ij
\in \t S^2}$, $i$'s are on one side of the membrane and $j$'s are on the other
side of the membrane (see Fig. \ref{memb}).  Last $h_{ij}$ is a function of
$h$ and $g_{ij}$.  For non-Abelian group $G$, $h_{ij}$ is complicated.  But
when all $g_{ij}=1$, $h_{ij}$ has a simple form $h_{ij}=h$.  For general
$g_{ij}$, we need to choose a base point $i_0$ on one side of the membrane,
and a path $i_0\to i$ on the membrane that connect the base point $i_0$ to any
other point $i$ on the membrane (see Fig. \ref{memb}). Then we can define
$h_{ij}$ as 
\begin{align} 
h_{ij} = (g_{i_0 i'} \cdots g_{i''i})^{-1} h (g_{i_0 i'}\cdots g_{i''i}), 
\end{align} where $(g_{i_0 i'} \cdots g_{i''i})$ is the
product of the link variables along the path $i_0\to i$.

We note that when the closed membrane enclose only one site $i$ (see Fig.
\ref{memb}), the operator $C_a(\t S^2)$ reduces to $C_a(i)$ discussed
before:
\begin{align}
 C_a(\t S^2) = C_a(i).
\end{align}
Thus $ C_a(i)$ can be viewed as a small membrane operator, rather than a point operator.

Let us consider a loop $i_0\to i \to j \to j_0 \to i_0$. The $G$-flux through
such a loop in the ground state $|\Psi_0\>$ is trivial: $(g_{i_0 i'} \cdots
g_{i''i}) g_{ij} (g_{j j'} \cdots g_{j''j_0}) g_{j_0i_0}=1$.  This is because
the ground state $|\Psi_0\>$ satisfies
\begin{align}
 B |\Psi_0\> = |\Psi_0\> .
\end{align}
After we apply the membrane operator \eq{memop}, the $G$-flux through the same
loop becomes
\begin{align}
&\ \ \ \ 
(g_{i_0 i'} \cdots g_{i''i})
h_{ij} g_{ij} (g_{j j'} \cdots g_{j''j_0}) g_{j_0i_0} h^{-1}
\nonumber\\
&=
h (g_{i_0 i'} \cdots g_{i''i})
 g_{ij} (g_{j j'} \cdots g_{j''j_0}) g_{j_0i_0} h^{-1}
=1 ,
\end{align}
which is still trivial.  This allows us to conclude that for states $|\Psi\>$
satisfying $B |\Psi\> = |\Psi\>$ and for closed membrane $\t S^2_\text{closed}$, we have
\begin{align}
\label{BC}
 B C_a(\t S^2_\text{closed})|\Psi\> = C_a(\t S^2_\text{closed})|\Psi\>
\end{align}

We can also show that
\begin{align}
 Q_g(i)  C_a(\t S^2) =  C_a(\t S^2) Q_g(i) .
\end{align}
For example
\begin{align}
 Q_g^{-1}(i_0)  C_a(\t S^2) Q_g(i_0)
& = \sum_{g^{-1}hg \in \chi_a} \prod_{ij \in \t S^2} T_{ij}(h_{ij})
\nonumber\\
&
= C_a(\t S^2) .
\end{align}
Also
\begin{align}
Q_g^{-1}(i) T_{ij}(h_{ij}) Q_g(i)
= T_{ij}(\t h_{ij})
\end{align}
where
\begin{align}
\t h_{ij} =
g^{-1} (g_{i_0 i'} \cdots g_{i''i}g^{-1})^{-1} h (g_{i_0 i'}\cdots g_{i''i}g^{-1}) g =h_{ij}.
\end{align}
Thus in general, we have
\begin{align}
\label{CaCb}
 Q_g^{-1}(i)  C_a(\t S^2) Q_g(i) = C_a(\t S^2) ,
\nonumber \\
 C_b(i)  C_a(\t S^2)  = C_a(\t S^2) C_b(i),
\end{align}
for any $i$, even for open membranes.  The results \eq{BC} and \eq{CaCb} imply
that closed membrane operators $ C_a(\t S^2_\text{closed})$ act within the
ground state subspace of the Hamiltonian \eq{Ham}.  Therefore, the boundary of
the open membrane operators \eq{memop} create string-like excitations, which are
labeled by conjugacy classes $\chi_a$.

\subsection{Exact algebraic higher symmetry}

Since the Hamiltonian  \eq{Ham} commutes with the closed string operators
$B_q(S^1_\text{closed})$: 
\begin{align}
\label{Bsymm}
 [H, B_q(S^1_\text{closed})]=0,
\end{align}
we say that the Hamiltonian has an algebraic 2-symmetry generated by
$B_q(S^1_\text{closed})$ for any closed strings.  Since the composition of the
symmetry transformations satisfies the fusion rule \eq{BBNB}, which is not a group
multiplication rule for non-Abelian $G$.  Thus the $B_q(\text{closed
string})$'s generate an exact algebraic 2-symmetry which is not a higher
2-symmetry.  However, when $G$ is Abelian, $B_q(S^1_\text{closed})$'s generate
a higher 2-symmetry.

There is another way to describe the algebraic 2-symmetry using the open string
operators.\cite{JW191213492} We note that the Hamiltonian is a sum of local
operators $H =\sum_i H_i$, where $H_i$ acts only on the degrees of freedom near
site-$i$.  We find that $H_i$ commutes with open string operators as long as
the ends of the strings is a distance away from the site-$i$:
\begin{align}
\label{BsymmO}
 [H_i, B_q(S^1_\text{open})]=0.
\end{align}

\subsection{Emergent algebraic higher symmetry}

We also note that  the Hamiltonian  \eq{Ham} commutes with $U_h(S^3)$
\begin{align}
\label{Usymm}
 [H,U_h(S^3)]=0,\ \ \ 
U_h = \prod_i Q_h(i),\ \ h\in G.
\end{align}
Thus the  Hamiltonian has a 0-symmetry, \ie a global symmetry described by
symmetry group $G$.  In fact, the Hamiltonian has a much bigger symmetry.  It
has a local symmetry described by group $G^{N_v}$, where $N_v$ is the number of
lattice sites:
\begin{align}
 [H,Q_{h_i}(i)]=0,\ \ \ 
 h_i \in G,
\end{align}

On the other hand, the membrane operators $C_a(\t S^2_\text{closed})$'s do not
commute with the Hamiltonian \eq{Ham}. Thus the Hamiltonian does not have
algebraic 1-symmetries.  However, $C_a(\t S^2_\text{closed})$ acts within
the degenerate ground subspace.  More generally, $C_a(\t S^2_\text{closed})$
and $H$ commute in the subspace where $B_{\<ijkl\>}=1$ (\ie in the finite
energy subspace of $H$ when $J\to +\infty$).  Therefore the Hamiltonian has an
emergent low energy algebraic 1-symmetry generated by $C_a(\t
S^2_\text{closed})$'s when $J\to +\infty$.  Such an emergent  algebraic
1-symmetry is a (group-like) 1-symmetry only when $G$ is Abelian.

\section{Description of algebraic higher symmetry in a symmetric product state}

\label{symprd}

Usually, we use the symmetry transformation, \ie the symmetry group $G$, to
describe a symmetry.  We can also use the symmetry charges, \ie the
representations $\cRep(G)$, to describe a symmetry.  Due to Tannaka duality,
the two descriptions are equivalent.  In last section, we introduced algebraic
higher symmetry via the symmetry transformations.  In this and next sections,
we will develop a similar dual way to describe algebraic higher symmetry, \ie
via the representations of algebraic higher symmetry.  This section will
concentrate on the point of view based lattice model and symmetric product
state.  Next section will present a point view based on higher category.

But what are the representations of algebraic higher symmetry?  Physically, the
representations correspond to the ``charged excitations'' in a symmetric ground
state which has a trivial topological order (\ie be a product state).  So in
the following, we will explain what is ``symmetric state'' (\ie no spontaneous
symmetry breaking) for algebraic higher symmetry?  What kind of algebraic
higher symmetry can have symmetric  ground state with no topological order?
What are the ``charged excitations'' for algebraic higher symmetry?  This
allows us to obtain a representation theory for algebraic higher symmetry, in
terms of local fusion higher category.

\subsection{Spontaneous broken and unbroken algebraic higher symmetry}

In Section~\ref{Lgauge}, we constructed a 3d lattice model that has an exact
algebraic 2-symmetry generated by string operators $B_q(S^1)$.  However, the
ground state of the model \eqn{Ham} spontaneously breaks the algebraic
2-symmetry, which gives us a topological order described by the $G$-gauge theory.

Here, we consider a different model
\begin{align}
\label{HamT}
 H= -V \sum_{ij} \del(g_{ij}) 
+ U \sum_i   (1-Q_i)
+ J\sum_{\<ijkl\>} (1-B_{\<ijkl\>}) 
,  
\end{align}
by including an extra term $-V \del(g_{ij})$ and taking $J\to +\infty$ limit.
Here 
\begin{align}
 \del(g)=\begin{cases}
 1, & \text{ if } g=\id \\
 0, & \text{ otherwise } \\
\end{cases}
.
\end{align}
The model also has the algebraic 2-symmetry $[H,B_q(S_\text{closed}^1)]=0$.  If
we choose the limit $U\ll V$, the ground state is given by $|\{g_{ij}=1\}\>$.
This ground state does not spontaneously break the algebraic 2-symmetry.

For the usual global symmetry, the spontaneous symmetry breaking is defined
via non-zero order parameters.  Here we would like to define the spontaneous symmetry
breaking of algebraic higher symmetry in a different way:
\begin{DefinitionPH}
An algebraic higher symmetry is \textbf{spontaneously broken} if there exists a
close space, such that the symmetry transformations are not proportional the
identity operator in the nearly degenerate ground state subspace on that space.
\end{DefinitionPH}
\noindent

For the Hamiltonian \eqn{HamT}, the ground state is not degenerate on any
closed spaces. Thus the algebraic 2-symmetry is not spontaneously broken.
In contrast, for model \eq{Ham}, the ground states are degenerate on space
$S^1_x\times S^1_y\times S^1_z$.  The different ground states can have
different flux, say $\prod_{(ij) \in S^1_x} g_{ij} =h$.  The symmetry
generator $B_q(S^1_x)$ is not proportional to identity, since $B_q(S^1_x)
=X_q(h)$ on the ground state with flux $h$.  Here 
\begin{align}
X_q(g) = \Tr[R_q(g)],\ \ \ g \in G
\end{align}
is the character of the representation $R_q$.  We see that the ground state of the
model \eq{Ham} spontaneously breaks the algebraic 2-symmetry $B_q(S^1)$.  In
fact, the algebraic 2-symmetry is completely broken, which gives rise to the
topological order described by the $G$-gauge theory.

\subsection{Anomaly-free algebraic higher symmetry}
\label{AFahs}

In this section, we would like to discuss algebraic higher symmetry in the
simplest state -- symmetry unbroken state without topological order.  However,
some algebraic higher symmetries may not allow such a state.  This leads to an
important attribute of algebraic higher symmetry.
There is two ways to describe this attribute:
a microscopic way
\begin{DefinitionPH}
An algebraic higher symmetry in a lattice system is \textbf{anomaly-free} if a
system with the symmetry allows a phase which has a symmetric product state as
its unique gapped ground state.
\end{DefinitionPH}
\noindent
and a macroscopic way
\begin{DefinitionPH}
An algebraic higher symmetry is \textbf{anomaly-free} if a system with the
symmetry allows a phase which has a unique gapped ground state on each closed
space.  Such a phase is also \textbf{symmetric}.
\end{DefinitionPH}

For $0$-symmetry on lattice, we can use on-siteness to define anomaly-free
0-symmetry \cite{W1313}.  Using this definition, we believe that all anomalous
(non-on-site) $0$-symmetry can be realized on a boundary of a system in one
higher dimension with anomaly-free (on-site) $0$-symmetry.\cite{W1313}  For
\emph{finite} symmetries, we believe that there is an one-to-one correspondence
between anomalous $0$-symmetries and the SPT order in one higher dimension
\cite{W1313}. (While for infinite symmetry described by a continuous compact
group,  we do not have the one-to-one correspondence between anomalous
$0$-symmetries and the SPT order in one higher dimension \cite{W1313}.)  As a
result, the finite anomalous $0$-symmetries are classified by the SPT orders in
one higher dimension.  Since we believe that the boundary uniquely determine
the bulk \cite{KW1458,KZ170200673}, the above result also implies that an
anomalous 0-symmetry does not allow a gapped symmetric product state as the
ground state \cite{CLW1141,VS1306}.  Otherwise, the SPT order in one higher
dimension must be trivial as implied by such a symmetric ground state on the
boundary.  

For algebraic higher symmetry, it is hard to define on-siteness.  So we turn
things around, and use the existence of trivial symmetric gapped ground state
to define algebraic higher symmetry (where trivial means product state).  In
this case, algebraic higher symmetry can appear at a boundary of the trivial SPT
phase for algebraic higher symmetry.  The boundary of non-trivial SPT phases for an
algebraic higher symmetry realize an anomalous algebraic higher symmetry.
In this paper, we only consider anomaly-free algebraic higher symmetries.

\subsection{The charge objects and charge creation operators for the exact
algebraic 2-symmetry}

The exact algebraic 2-symmetry in the lattice model \eqn{HamT} is generated by
$B_q(S^1_\text{closed})$.  The algebraic 2-symmetry is anomaly-free since the
model \eqn{HamT} allows symmetric gapped product state $|\{g_{ij}=1\}\>$  as
its unique ground state.

The charge objects of such a 2-symmetry live on 2-dimensional surfaces [just like
the charges of a 0-symmetry (the usual global symmetry) live on 0-dimensional
points]. To construct the 2-dimensional operators that create the charge objects
of the algebraic 2-symmetry, let us review the charge creation operators for
the 0-symmetry in a proper general setting.

A pair of charge and anti-charge of a 0-symmetry is created by an operator
$C(S^0)$ on $S^0$ (\ie on two points $i$ and $j$), for example
\begin{align}
 C(S^0) = \sum_a \psi^\dag_a(i) \psi_a(j),
\end{align}  
where the local operator $\psi_a(i)$ form a unitary representation
$R_{ab}$ for the 0-symmetry group $G$:
\begin{align}
 U_g \psi_a = \sum_b R_{ab}(g) \psi_b U_g,\ \ \ g\in G.
\end{align}
We note that, when the two points in $S^0$ belong to the same connected
component of the space, $C(S^0)$ 
commutes with the algebraic 0-symmetry transformations and creates an neutral
excitation.  On part of $S^0$, the creation operator becomes $\psi_a(i)$ which
creates a non-neutral excitation.

Similarly, a neutral charge object of a $k$-symmetry is created by operators on
closed contractible $k$-dimensional manifolds, such as $S^k$.  Such an operator
on contractible $S^k$ commutes with the algebraic $k$-symmetry transformations
and creates an neutral excitation.
A charge object of $k$-symmetry is created by operators $C(M^k)$ on
open $k$-dimensional manifold $M^k$.  In $n$d,  when the algebraic $k$-symmetry
generator $B_q(S^{n-k}_\text{closed})$ on $n-k$-dimensional sub-manifold
intersects with the submanifold $M^{k}$ at one point, we can detect the
$k$-symmetry charge.  The algebra between symmetry generators
$B_q(S^{n-k}_\text{closed})$ and charge creation operators $C(M^k)$ only depend
on the linking between $S^{n-k}_\text{closed}$ and $\prt M^k$, and do not
depend on the deformations of $S^{n-k}_\text{closed}$ and $\prt M^k$ that do
not change their linking.  Those are key conditions for the charge creation
operators $C(M^k)$ for an algebraic higher symmetry.

For our algebraic 2-symmetry in 3d, the charge creation operator acts on
2-dimensional surfaces with or without boundary.  In fact, such a charge
creation operator is nothing but the membrane operator $C_a(\t S^2)$
discussed in Section~\ref{exc}.  The charge object created by $C_a(\t S^2)$ can
be detected by the $2$-symmetry generator $B_q(S^1_\text{closed})$, when $\t
S^2$ has a boundary, or when $\t S^2$ is closed and non-contractible.

In fact, on the $|\{g_{ij}=1\}\>$ ground state, the creation operator can
have a simpler form
\begin{align}
\t C_h(\t S^2) = \prod_{ij \in \t S^2} T_{ij}(h)
\end{align}
where $\prod_{ij \in \t S^2}$ is over all the links $ij$ of the original
lattice that form the faces in $\t S^2$ of the dual lattice.  Such an operator
just changes $g_{ij}=1$ to $g_{ij}=h$ on links $ij$ of the original lattice
that form the faces in $\t S^2$ of the dual lattice.  $g_{ij}=h$ on $\t S^2$
corresponds to a charged excitation, called a 2-charge object labeled by $h$, of our algebraic 2-symmetry generated by
$B_q(S^1_\text{closed})$.

\begin{figure}[t]
\begin{center}
\includegraphics[scale=0.7]{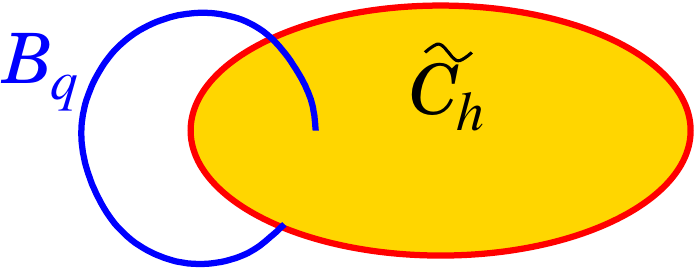} \end{center}
%16
\caption{In 3-dimensional space, a disk-like 2-charge object (a 2-dimensional
excitation) created by $\t C_h(\t S^2)$ can be detected by the algebraic
2-symmetry transformation loop operator $B_q(S^1_\text{closed})$.  }
\label{detsymm}
\end{figure}

If $\t S^2$ is a disk in 3d space, then the 2-charge object created by $\t
C_h(\t D^2)$ can be detected by the algebraic 2-symmetry operator
$B_q(S^1_\text{closed})$ if $S^1_\text{closed}$ is linked with $\prt \t S^2$  --
the boundary of the 2-charge object (see Fig.  \ref{detsymm}).  If fact,
$B_q(S^1_\text{closed}) = \Tr R_q(h)$ in this case when acting on the 2-charge
object.  In comparison, for the ground state $|\{g_{ij}=1\}\>$, the
$2$-symmetry generator is equal to the dimension $d_q$ of the
$q$-representation: $B_q(S^1_\text{closed})=\Tr R_q(1)=d_q$.  We see that the
algebraic 2-symmetry cannot distinguish two 2-charge objects labeled by $h$ and
$h'$ if $h$ and $h'$ belong to the same conjugacy class. So the distinct
algebraic 2-symmetry charges are labeled by the conjugacy classes $\chi_a$ of
$G$. 

We stress that the membrane operator $C_a(\t S^2)$ that creates the
2-dimensional charge object of the algebraic 2-symmetry is an operator that
acts only on the membrane $\t S^2$.  This is a very important general feature.
\begin{Proposition}
\label{localChOb}
On top of a ground state that does not break the symmetry, the $k$-dimensional
charge object of an algebraic $k$-symmetry is created by an operator that act
only on the $k$-dimensional subspace that supports the charge object.
\end{Proposition} 
\noindent

We note that in $J\to \infty$ limit, only 2-charge objects corresponding to
closed surfaces has low energy.  2-charge objects corresponding to surfaces
with boundary cost energy of order $J$ or bigger.  We may consider the low
energy subspace of the model in $J\to \infty$ limit.  In fact, we consider
an even smaller space, the invariant sub-Hilbert space of all the 2-symmetry
transformations generated by $B_q(S^1_\text{closed})$ operators.  The
collection of those created 2-charge objects within the symmetric sub-Hilbert
space, plus their fusion (and braiding) properties, form a higher category.
The 2-charge objects are labeled by $h\in G$ and created by $ \t C_h(\t
S_\text{closed}^2)$.  The fusion of $\t C_h(\t S_\text{closed}^2)$ is given by
\begin{align}
 \t C_h(\t S_\text{closed}^2) \otimes \t C_{h'}(\t S_\text{closed}^2) 
= \t C_{hh'}(\t S_\text{closed}^2) 
\end{align}
The charged membrane-like excitations, labeled by $h\in G$, 
form a \emph{fusion 3-category}
$\cR=3\cVec_G$ (see Def. \ref{fc}), which is also a \emph{local
fusion 3-category} (see Def. \ref{lfus}).  We also refer $\cR=3\cVec_G$ as
the \textbf{representation category} of the algebraic 2-symmetry.  Physically,
$\cR$ is the fusion 3-category that describes the low energy excitations in
model \eqn{HamT}.

But what is a \emph{fusion higher category} and what is a \emph{local fusion
higher category}?  Roughly speaking, a fusion higher category describes the
point-like, string-like, \etc\ excitations above a gapped liquid ground state.
If an excitation can be annihilated by an operator acting on the excitations,
then we say the excitation is local.  Note that the operators may break any
symmetry and may not be local, as long as they act on the support subspace of
the excitation.  The fusion higher category formed by local excitations is a
local fusion higher category.  Since the membrane excitations in $\cR$ can
all be annihilated by operators on the membranes, $\cR$ is a local fusion
higher category.

\begin{table*}[t]
  \caption{Correspondence between concepts in fusion higher category and 
concepts in topological order.\cite{KW1458,KZ150201690}}
  \label{tab:dicF}
  \centering
  \begin{tabular}{|p{2.0in}|p{4.8in}|}
    \hline
    \textbf{~~Concepts in higher category} ~~~ &  ~~\textbf{Concepts in physics}\\
    \hline
    Fusion $n$-category $\cC$ & 
A collection of all the \emph{types} of codimension-1 and higher excitations (plus their fusion and braiding
properties) in an $n$d (potentially anomalous) topological order.\\
    \hline
    Simple objects of $\cC$ & The types of codimension-1 topological excitations. They can fuse.\\
    \hline
    Simple 1-morphisms of $\cC$ & The types of  codimension-2 topological excitations. They can fuse and braid.\\
    \hline
    Simple $\nmt$-morphisms of $\cC$ & The types of string-like topological excitations\\
    \hline
    Simple $\nmo$-morphisms of $\cC$ & The types of point-like topological excitations\\
    \hline
    $n$-morphisms of $\cC$ & The local operators acting on the point-like excitations\\
    \hline
Local fusion $n$-category $\cR$ & The ``charged'' excitations (charge objects)
above a product state of a bosonic system with an
algebraic higher symmetry $\cR$. It is called the representation category of the
algebraic higher symmetry \\
    \hline
  \end{tabular}
\end{table*}

The following discussions use the notions of  topological order higher
categories extensively,\cite{KW1458,KZ150201690,GJ190509566,J200306663} which
are reviewed in Section~\ref{toprev}.  Table \ref{tab:dicF} summarizes some
related concepts in higher category and in topological order.

\section{Local fusion higher category and representations (charge objects) of
anomaly-free algebraic higher symmetry}

\label{lfcat}

In the last section, we described the charged excitations (\ie the charge
objects) in a trivial symmetric ground state with anomaly-free algebraic higher
symmetry.  Here trivial state means a product state.  In the rest of this
paper, we will mainly discuss anomaly-free algebraic higher symmetry, and we
drop ``anomaly-free'' for simplicity.

For a 0-symmetry $G$, we know that its charges are representations of $G$. All
those representations form a symmetric fusion category $\Rep G$.  Due to
Tannaka duality, we can use the local fusion category $\Rep G$ to fully
describe the symmetry group $G$.\cite{LW160205936,LW160205946} To be more
precise, the charges (the representations) of $G$ correspond to point-like
excitations.  Those  point-like charges can condense to form other descendent
excitations. All those excitation are described by a fusion $n$-category, if
the 0-symmetry $G$ lives in $n$-dimensional space.  We denote such a fusion
$n$-category as $n\cRep G$.  In other words, an $n$d  0-symmetry $G$ is fully
characterized by a symmetric fusion $n$-category $n\cRep G$.  We refer
to $n\cRep G$ as the representation category of the 0-symmetry $G$.

In the above, we try to use excitations (trapped by the symmetric traps) to
characterize a symmetry.  Here we would like to stress that the excitations
described by the fusion $n$-category $n\cRep G$ only correspond to the
excitations in the \emph{symmetric sub-Hilbert space} $\cV_\text{symm}$ of the
many-body system.  The fusion $n$-category $n\cRep G$ do not include the
excitations outside the symmetric sub-Hilbert space.  In the thermodynamic
limit, restricting to symmetric sub-Hilbert space does not affect our ability
to understand the properties of a symmetric system.  We would like to use the
similar approach to characterize an algebraic higher symmetry (which is not
characterized by groups or even higher groups): the
representations (\ie the charge objects) of an algebraic higher symmetry
are simply the excitations above a symmetric product state, which are also
described by a category -- a \emph{local fusion higher category}.

\subsection{The excitations in a symmetric state with no topological order}

To have a general understanding of the charge objects, let us consider a local
lattice Hamiltonian $H$ with an algebraic higher symmetry.  We assume the
ground state $|\Psi_\text{grnd}\>$ of $H$ has no topological order nor SPT
order, \ie can be deformed into a product state without a phase transition, via
a symmetry preserving path.  Then how to understand the point-like, string-like
excitations, \etc\ of the above ground state?  Also similar to the 0-symmetry
case, here we only consider the symmetric excitations (\ie those trapped by
symmetric traps) in the symmetric sub-Hilbert space $\cV_\text{symm}$.  We know
that an algebraic higher symmetry  is generated by many operators acting on all
closed submanifolds.  The symmetric sub-Hilbert space is the invariant
sub-Hilbert space of all those symmetry generators.

To understand the excitations, first, let us define excitations more
carefully.  For example, to define string-like excitations, we can add several
trap Hamiltonians $\Del H_\mathrm{str}(S^1_i)$ to $H$ such that $H + \sum_i
\Del H_\mathrm{str}(S^1_i)$ has an energy gap.  $\Del H_\mathrm{str}(S^1_i)$ is
only non-zero along the string $S^1_i$ and \emph{commutes with the generators
of the algebraic higher symmetry}.  We also assume $\Del H_\mathrm{str}(S^1_i)$
to be \emph{stable}: any small symmetric change of $\Del H_\mathrm{str}(S^1_i)$
does not change the ground state degeneracy in the large string $S^1_i$ limit.
The resulting string corresponds to a simple morphism in mathematics.  We also
define two strings labeled by $\Del H_\mathrm{str}(S^1)$ and $\Del \t
H_\mathrm{str}(\t S^1)$ as equivalent, if we can deform $\Del
H_\mathrm{str}(S^1)$ into $\Del \t H_\mathrm{str}(\t S^1)$ without closing the
energy gap while preserving the algebraic higher symmetry.  The equivalent
classes of the strings define the \textbf{types} of the strings (see Def.
\ref{type}).  

In the example in Section~\ref{symprd}, the 2-dimensional charge object of
an algebraic 2-symmetry is created by a membrane operator.  If the membrane is
a closed 2-dimensional subspace, then the membrane operator acts within the
symmetric sub-Hilbert space $\cV_\text{symm}$, and create an excitation in the
fusion higher category.  If the membrane has a boundary, then the membrane
operator does not act within the symmetric sub-Hilbert space, and create an
excitation not in the fusion higher category.  When the membrane has a
boundary, such a boundary is the morphism that connect the membrane excitation
to the trivial excitation.  In the above example, such a boundary (\ie the
morphism) is not allowed, since it breaks the
algebraic 2-symmetry (\ie the membrane with the boundary does not act within
the symmetric sub-Hilbert space).

\subsection{Local fusion higher category } \label{lfc}

Now we are ready to define a \emph{local fusion higher category}, which describes
the collection of excitations (\ie the collection of types) in the system
mentioned above, \ie a system with algebraic higher symmetry whose ground state
is a symmetric bosonic product state without degeneracy. Also, we only consider
excitations within the symmetric sub-Hilber space $\cV_\text{symm}$.  For a
symmetric trivial phase without topological order, it has only local
excitations.  From a categorical point of view, a local excitation can always
be connected to the trivial excitation through a morphism as described above,
if we are willing to break the symmetry.  However, if we preserve the symmetry,
the symmetry breaking morphism is not allowed and some excitations cannot
connect to trivial excitation via symmetry preserving morphisms (\ie symmetry
preserving domain walls).  This leads to the following definition: \emph{A
fusion $n$-category $\cR$ is local if we can add morphisms in a consistent way,
such that all the resulting simple morphisms are isomorphic to the trivial
one.} Physically, this process of ``adding morphisms'' corresponds to explicit
breaking of algebraic higher symmetry.  This because, $\cR$ only has morphisms
that correspond to symmetric operators.  Adding morphisms means including
morphisms that correspond to symmetry breaking operators.  If after breaking
all the symmetry, $\cR$ describes a trivial phase without symmetry, then $\cR$
is a local fusion $n$-category.  The above can be stated more precisely
\begin{Definition}
\label{lfus}
A fusion $n$-category $\cR$ (see Def. \ref{fc}) equipped with a
\textbf{top-faithful} surjective monoidal functor $\bt$ from $\cR$ to the
trivial fusion $n$-category: $\cR \map{\bt} n\cVec$ is called a \textbf{local}
fusion $n$-category.  Here, \textbf{top-faithful} means that the functor $\bt$
is injective when acting on the top morphisms (\ie the $n$-morphism in this
case).
\end{Definition}
\begin{Remark}
  The top-faithful condition means that operators in $\cR$ form a subset of
operators in $n\cVec$, which agrees with the physical interpretation that from
$\cR$ to $n\cVec$ we add symmetry breaking operators.  The functor $\bt$ may
not be faithful when acting on other morphisms.  In other words, every objects
and morphisms in $\cR$ can be viewed as (\ie can be mapped into) objects and
morphisms in $n\cVec $, but the map may not be injective.
\end{Remark}

When we are interested in fermion systems, we need to replace $n\cVec$ for
$n\scVec$. More generally, the building blocks of our physical system may have
even larger intrinsic symmetry (which is unbreakable or we are not willing to
break) besides the fermion number parity. Let $\cV$ denote the
fusion $n$-category formed by the building blocks ($\cV=n\cVec$ for bosons,
  $\cV=n\scVec$ for fermions, and possibly any other $\cV$ for more exotic cases
such as an effective theory built upon anyons). We define the notion of
$\cV$-local fusion $n$-categories.
\begin{Definition}\label{vlocal}
  A $\cV$-local fusion $n$-category is a fusion $n$-category $\cR$ equipped with
  a top-faithful surjective monoidal functor $\beta:\cR\to \cV$.
\end{Definition}

\subsection{Local fusion 1-category $\cRep G$ and $\cVec_G$ }
\label{empLFC}

As an example, let us consider a 1d system with degrees of freedom labeled by
$g_i\in G$ on site $i$, where $G$ is a group.
The Hamiltonian of the system is given by
\begin{align}
 H = - J \sum_i \sum_{h\in G} T_i(h) - V \sum_i  \del(g_{i-1}^{-1}g_i),
\end{align}
where
$T_i(h)$ is an operator
\begin{align}
T_i(h) & |\cdots, g_{i-1}, g_i,g_{i+1},\cdots\>
\nonumber\\
 = & |\cdots, g_{i-1}, hg_i , g_{i+1},\cdots\>, \ \ \ \ h \in G.
\end{align}
The system has a symmetry $G$
\begin{align}
 |\cdots, g_{i-1}, g_i,g_{i+1},\cdots\>
\to
 |\cdots, gg_{i-1}, gg_i,gg_{i+1},\cdots\>.
\end{align}
When $J\gg  |V|$, the ground state is a product state
$|\Psi_\text{grnd}\>=\otimes_i |0\>_i$ where $|0\>_i \equiv
|G|^{-1/2}\sum_g|g\>_i$,  that does not spontaneously break the symmetry.

Note that $\{|g\>_i,g \in G\}$ spans the regular representation of $G$. It can be
further decomposed into irreducible representations. Let $|a\>_i,|b\>_i,\cdots$
be a basis in an irreducible representation.  Under the symmetry transformation $h\in G$,
$|a\>_i$ transforms to $T_i(h)|a\>_i =\sum_b R_{ab}(h) |b\>_i$ where $R_{ab}(h)$
is the matrix representing $h$.
A point-like excitation at site $i$ is created by changing the state $|0\>_i$
on site-$i$ to $|a\>_i= \sum_g \<g|a\>|g\>_i$. Since
\begin{align}
T_i(h) |a\>_i &= \sum_g \<g|a\> |hg\>_i = \sum_g \<h^{-1}g|a\>|g\>_i 
\nonumber\\
& = \sum_b R_{ab}(h) |b\>_i, 
\end{align}
we see $\<h^{-1}g|a\> = \sum_b R_{ab}(h) \<g|b\>$.

Such a ground state plus its excitations are described by a fusion 1-category
$\cRep G$ whose objects correspond to the point-like excitations (\ie the
representations $R$ of $G$).  The 1-morphisms of $\cRep G$ correspond to the
symmetric local operators that act on each site.  We see that the 1-morphisms
directly act on the point-like excitations (the objects).  If we view an
excitation (an object) as a world line in spacetime, an 1-morphism that
changes the excitation can be viewed as a ``domain wall'' on the world line.  For a
symmetric system, all those 1-morphisms should be symmetric operators.
Respecting to those symmetric 1-morphisms, the excitations corresponding to
the irreducible representations are simple objects.  Different irreducible
representations cannot be connected by symmetric operators, \ie different
simple objects cannot be connected by 1-morphisms.

If we add the additional 1-morphisms that correspond to local operators that
break all the symmetry, then the excitations corresponding to the irreducible
representations $R$ are still allowed, but they are no longer simple object,
and split into direct sum of trivial excitations:
\begin{align}
\label{RGC}
 R \to \underbrace{\C\oplus \cdots \oplus \C}_{\text{dim}R\text{ copies}}.
\end{align}
As a result, the fusion 1-category is reduced to the trivial 1-category -- the
category of vector spaces $\cVec$.  Thus the fusion 1-category $\cRep G$ is a
local fusion 1-category.  Indeed, all the point-like excitations can be
annihilated by local operators that may break the symmetry.  

Now  consider a 1d system with symmetry $G$, whose ground state spontaneously
breaks all the symmetry.  In this case, the ground states are $|G|$-fold
degenerate and are labeled by the group elements: $|\Psi_g\>$, $g\in G$.  The
point-like excitations are domain walls, which live on the links and are
labeled by the elements $h$ of the group: $|h\>_{i_0,i_0+1} =|\Psi_{g,i\leq
i_0} \Psi_{hg,i\geq i_0+1}\>$.  Such symmetry breaking state plus its
excitations are described by a fusion 1-category $\cVec_G$, whose objects
correspond to the point-like excitations (the domain walls) discussed above.
We may still choose the 1-morphisms of $\cVec_G$ to be the symmetric local
operators acting on the sites.  However, such a choice is not proper, since
such 1-morphisms cannot be viewed as the ``domain walls'' on the world-lines of
the point-like excitations (the domain walls on the links).  In any case, let
us proceed.  If we add the 1-morphisms that correspond to local operators that
break all the symmetry, then objects (the point-like domain-wall excitations)
are confined (\ie non longer allowed), since the ground state degeneracy
is lifted.  This appears to suggest that the fusion 1-category $\cVec_G$
is not a local fusion 1-category, if we view it as describing domain walls in a
spontaneous symmetry breaking state that breaks a $0$-symmetry of group $G$.
Since our choices of the 1-morphisms is not proper, the above conclusion is
incorrect.

In fact, $\cVec_G$ can also be viewed as a fusion 1-category that describes
excitations on top of a product state with an algebraic 0-symmetry.  The
degrees of freedom on each site $i$ of our 1d model are labeled by group
elements of a finite group $G$.  A basis of the many-body Hilbert space is given by
$|\{g_i\}\>$, $g_i\in G$.  The Hamiltonian is given by
\begin{align}
\label{VJ1d}
 H = - V \sum_i \del(g_i) -  t \sum_{i,h\in G}  T_{i-1,i}(h)
\end{align}
where
$T_{i-1,i}(h)$ is an operator
\begin{align}
T_{i-1,i}(h) & |\cdots, g_{i-1}, g_i,g_{i+1},\cdots\>
\nonumber\\
 = & |\cdots, g_{i-1}h^{-1}, hg_i , g_{i+1},\cdots\>, \ \ \ \ h \in G.
\end{align}
The model has an algebraic 0-symmetry generated by
\begin{align}
 B_q = \Tr \big[ \prod_i R_q(g_i) \big],
\end{align}
where $q$ labels the representations of $G$.
In the $t \to 0$ limit, the ground state is a symmetric product state
$|\{g_i=\id\}\>$.

Above such a ground state, a point-like excitation is generated by changing
$g_i=\id$ to $g_i=h$ on site-$i$.  Thus the excitations are labeled by group
elements $h\in G$, with $h=\id$ corresponding to the ground state.  They fuse as
$h\otimes h' = hh'$.  When the algebraic 0-symmetry operators act on
the excitations $h$, we get $B_q(h) =X_q(h)$, where $X_q$ is the character for
the representation $q$.  Those point-like excitations form a local 1-fusion
category $\cVec_G$.

The operators that break the algebraic 0-symmetry
are given by
\begin{align}
\del H & = T_{i}(h)  |\cdots, g_{i-1}, g_i,g_{i+1},\cdots\>
\nonumber\\
 & =  |\cdots, g_{i-1}, hg_i , g_{i+1},\cdots\>, \ \ \ \ h \in G.
\end{align}
Those operators reduce the  local 1-fusion category $\cVec_G$ to the trivial
1-fusion category $\cVec$, since those operators correspond to new
morphisms $h\to h'$ for any $h,h'\in G$.  Therefore, the 1-fusion category
$\cVec_G$ is local.

We would like to mention that the 3d generalization of the 1d model \eq{VJ1d} was
discussed in Section~\ref{symprd}.  Using a similar reason, we show that
the 3-fusion category $3\cVec_G$ is local.

\subsection{Representation category of algebraic higher symmetry}

Let us summarize the relation between the charge objects of an  algebraic
higher symmetry and a local fusion higher category.
\begin{Proposition}
Consider an $n$d trivial ground state which is a product state with an algebraic
higher symmetry.  The different \emph{types} of the excitations above the
ground state and within the symmetric sub-Hilber space form a local fusion
$n$-category $\cR$ (\ie with a fiber functor $\bt: \cR \to n\cVec$), which is
called the \textbf{representation category} of the algebraic higher symmetry in
$n$-dimensional space.
\end{Proposition}
\noindent
We would like to conjecture that the Tannaka duality can be generalized to algebraic
higher symmetries: 
\begin{Proposition}
There is an one-to-one correspondence between local fusion $n$-categories $\cR$
and algebraic higher symmetries for bosonic systems in $n$-dimensional space.
\end{Proposition}
\noindent
In other words, the algebraic higher symmetries in $n$d bosonic systems are
fully characterized and classified by local fusion $n$-categories.  Since a
local fusion $n$-category $\cR$ fully characterizes an anomaly-free algebraic
higher symmetry, in this paper, an algebraic higher symmetry is denoted
by $\cR$.

We would like to remark that there are anomalous algebraic higher symmetries.
For those symmetries, we cannot have trivial symmetric ground state, 
and it is difficult to define its representation category, since
representation category, by definition, is formed by the charged excitations
above the symmetric product state.

\subsection{Categorical symmetry -- a holographic view of symmetry}
\label{csym}

To gain an even deeper understanding of algebraic higher symmetry, following
\Ref{JW191213492}, we would like to introduce the notion of a categorical
symmetry, which is a holographic point of view of a symmetry.  We know that a
symmetry is simply a restriction on the local operators whose sum gives raise
to the Hamiltonian.  Usually, the restriction is imposed by symmetry
transformations.  But in the holographic point of view, we do not impose
restrictions via symmetric transformations. Instead, we use a topological order
without any symmetry in one higher dimension to encode a symmetry.  In other
words, we use long range entanglement\cite{CGW1038} to encode a symmetry.  Then
the restrictions to local operators is realized via the boundary of the
topological order.

Let us consider an $n$d system with an algebraic higher symmetry $\cR$.  When
we restrict the system to the symmetric sub-Hilbert space $\cV_\text{symm}$ of
the algebraic higher symmetry, the system has a potentially non-invertible
gravitational anomaly,\cite{JW190513279} since $\cV_\text{symm}$ does not have
a tensor product decomposition $\cV_\text{symm} \neq \otimes_i \cV_i$.  This
relates the symmetry to entanglement.  Thus the system (when restricted to the
symmetric sub-Hilbert space $\cV_\text{symm}$) can be viewed as a boundary of
an anomaly-free topological order $\sM$ in one higher dimension.  The
topological order $\sM$ is described by an object in $\afcTO^{n+2}$.

Which topological order $\sM$ in one higher dimension gives rise to the desired
algebraic higher symmetry $\cR$?  We note that $\cR$ is a fusion $n$-category.
We believe that for every fusion $n$-category $\cR$, there is exist a unique
anomaly-free topological order $\sM$ in one higher dimension such that $\sM$
has a boundary whose excitations realize the fusion $n$-category $\cR$ (see
\eqn{bulkCM}). Therefore, we can find $\sM$ from $\cR$ via
\begin{align}
\sM=\bulk(\cR).
\end{align}

As we have discussed in Section~\ref{IIahs}, an $n$d algebraic higher symmetry
$\cR$ selects a set of local operators $\{ O_\cR\}$.  $\{ O_\cR\}$ can be
viewed as a set of lattice local operators that commute with the symmetry
generators, or as a set of local operators that describe all possible short
range interaction between excitations, as well as local operators that create
particle-anti-particle, small loop excitations, \etc, described by $\cR$.  In
Section~\ref{IIcats}, we mentioned that an $n$d categorical symmetry $\sM$ also
selects a local operators $\{ O_\sM\}$, on the boundary of $\npo$d topological
order $\sM$.  If $\sM=\bulk(\cR)$, the two sets $\{ O_\cR\}$ and $\{ O_\sM\}$
have an one-to-one correspondence and the corresponding operators has identical
properties (such as identical algebraic relations for the corresponding
operators).  In order words, the algebraic higher symmetry $\cR$ and the
categorical symmetry $\sM$ are holo-equivalent (see Proposition~\ref{MbulkR}).

Let us examine the algebraic higher symmetry $\cR$ and the  categorical
symmetry $\sM$ in terms of their excitations.  Roughly speaking, the
conservation law from the symmetry is encoded in the fusion rule for the
excitations.  Thus the fusion rule of the excitations with codimension-2 and
higher in $\sM$ encode the categorical symmetry $\sM$.  (A codimension-1
excitation in $\sM$ has codimension-0 on the boundary and cannot be viewed as
an excitation there.)  Those excitations are described by the braided fusion
$n$-category (see Section~\ref{looping}) 
\begin{align}
\eM = \Om \cM = \Om^2\sM,  
\end{align}
where $\cM=\Om\sM$ is the fusion $n$-category describing the bulk excitations
in $\sM$.  As we move a bulk excitation in $\eM$ to the boundary, it may become
some boundary excitations in $\cR$, or it may condense (\ie becomes the trivial
excitation in $\cR$).  So there is a monoidal functor $F_\cR: \eM \to \cR$. The
fusion rule in $\eM$ induces a fusion rule in $\cR$.  Thus the bulk symmetry
encoded in $\eM$ becomes an algebraic symmetry in $\cR$.  However, the bulk
excitations of $\eM=Z_1(\cR)$ are more than that of $\cR$.  In this sense, the
fusion rule of excitations in $\eM$ gives rise to a bigger symmetry than that
from the fusion rule of excitations in $\cR$. This bigger symmetry corresponds
to the categorical symmetry.\cite{JW191213492}

We know that $\sM$ can have many boundaries (denoted by $\sC \in
\cTO_{\sM}^{n+1}$, see Def. \ref{acat}).  The excitations on the boundary is
described by a fusion $n$-category $\cC=\Hom(\sC,\sC)=\Om \sC$, which satisfies
(see Proposition~\ref{Z1CMp})
\begin{align}
 \eM=Z_1(\cC) .
\end{align}
As we move a bulk excitation in $\eM$ to the boundary, it may become some
boundary excitations in $\cC$, or it may condense (\ie becomes the trivial
excitation in $\cC$). So there is a forgetful functor $F_\cC: \eM \to \cC$.
Because some excitations in $\eM$ are condensed on the boundary, we say the
boundary spontaneously breaks part of the categorical symmetry $\eM$.
Different boundaries $\sC$'s may spontaneously break different parts of the
categorical symmetry $\sM$, since the forgetful functor $F_\cC$ may map
different excitations in $\sM$ into the trivial excitations in $\cC$ (\ie
condense different excitations of $\sM$ on the boundary).

We see that all the boundaries have the same categorical symmetry $\sM$, if we
view the boundary as a lattice boundary Hamiltonian. If we view the boundary as
a state, then the categorical symmetry $\sM$ is spontaneously broken down to a
smaller symmetry.  The part of the categorical symmetry $\sM$, described by the
excitations that condense on the boundary, is spontaneously broken.  The
smaller survived symmetry is an algebraic higher symmetry.  We know that the
bulk fusion rule only induces the fusion rule for some boundary excitations
(\ie those in the image of the forgetful functor $F_\cC$).  Thus the image of
$F_\cC$ is related to this algebraic higher symmetry -- the unbroken part of
the categorical symmetry.

One might expect the image of $F_\cC$ to be the local fusion $n$-category that
characterizes the algebraic higher symmetry in $\cC$.  But this impression
is incorrect.  The image of $F_\cC$ may not even be a fusion $n$-category, \ie
there may not be an anomaly-free bulk topological order $\sM$ whose
boundary excitations realize the image of $F_\cC$.  

But what is the algebraic higher symmetry in $\cC$ (the unbroken part of the
categorical symmetry $\sM$)?  First, such an algebraic higher symmetry must be
described by a local fusion $n$-category $\cR$.  Since $\cR$ is the unbroken
part of the categorical symmetry $\sM$, the corresponding categorical symmetry
for $\cR$ should be given by the same $\sM$.  Mathematically, this means that
$\bulk(\cR) = \sM$.  Since $\cR$ is the algebraic higher symmetry in $\cC$,
$\cC$ must contain all the charge objects of $\cR$ as part of excitations in
it.  In other words, $\cR$ can be embedded into $\cC$, \ie there exists an
top-fully faithful functor $\iota: \cR \inj{\iota} \cC$.  Here 
\begin{Definition}
\label{topfully}
\textbf{Top-fully faithful} means the functor is
bijective when acting on top morphisms, and is injective when acting on lower
morphisms and on objects.  
\end{Definition}
\noindent
We know that the $\cR$-symmetry can be explicitly broken, via the functors
$\bt,\ \bt_\cC$, which changes $\cR$ to $n\cVec$ (see Def. \ref{lfus}) and
changes $\cC$ to $\underline{\cC}$.  $\underline{\cC}$ describes the
excitations in the anomaly-free topological order $\sC_0 \in \afcTO^{n+1}
\equiv \npo\cVec$ that are induced from $\sC$ after we explicitly break the
$\cR$-symmetry in $\sC$.  We note that the excitations described by $\cC$
contain both the topological excitations and the symmetry-charge excitations
described by $\cR$ (the charge objects of the algebraic higher symmetry).  One
may roughly understand $\underline{\cC}$ as ``$\cC/\cR$'' \ie ``$\cC$ mod
$\cR$''.  More precisely, $\underline{\cC}$ is the pushout defined in the
following diagram, 
\begin{align}
\label{RinC}
\xymatrix@R=2.0em@C=5.0em{
n\cVec  \ar@{<-}[d]|{\bt} \ar@{^(->}[r]^{\iota_{0}} & \underline{\cC} \ar@{<-}[d]|{\bt_\cC} \\
\cR   \ar@{^(->}[r]^\iota & \cC=\Hom(\sC,\sC)  
}
\end{align}
Moreover, the bulk of $\cR\to n\cVec$ and $ \cC\to \underline{\cC}$ should
coincide, which requires that $\gamma:Z_1(\cR)\simeq  Z_1(\cC)$ satisfies the
condition as later illustrated in \eqref{bulkww}.

To summarize, the different boundaries of $\sM$ all have the same categorical
symmetry as a system.  But the boundary may spontaneously breaks part of the
categorical symmetry when viewed as a state.  Because the charge objects of a
categorical symmetry have non-trivial mutual statistics, the boundary that does
not break the categorical symmetry $\sM$ must be
gapless.\cite{L190309028,JW191213492} For a \emph{gapped} boundary, the
categorical symmetry must be partially (and only partially) broken
spontaneously.  For the boundary $\sC$ discussed above, the categorical
symmetry is spontaneously broken down to the algebraic higher symmetry $\cR$.
We see that
\begin{Proposition}
\label{catsymmprop}
\textbf{Categorical symmetries} in $n$-dimensional space are fully
characterized and classified by an $\npo$d anomaly-free topological orders
$\sM$.  
\end{Proposition}
\noindent
A categorical symmetry $\sM$ may include several different anomaly-free
algebraic higher symmetries $\cR$, where $\cR$ satisfies $\sM=\bulk(\cR)$ (see
Proposition~\ref{Z1CMp}) 

A boundary of $\sM$ is described by a boundary Hamiltonian.  Such a Hamiltonian
always has the full  categorical symmetry $\sM$.  The ground state of the
boundary Hamiltonian, if gapped, is described by a boundary topological order
$\sC$ that satisfies $\Bulk(\sC)=\sM$.  For such a boundary ground state (\ie
the boundary topological order $\sC$), the categorical symmetry is partially
spontaneous broken, down to an algebraic higher symmetry $\cR$ that satisfies
\eqn{RinC}.  We say the categorical symmetry $\sM$ contains the algebraic
higher symmetry $\cR$.

We would like to remark that for an $n$d categorical symmetry, its
corresponding topological order $\sM$ in one higher dimension may have several
different gapped boundaries with different unbroken algebraic higher
symmetries.  Thus, an categorical symmetry can contain several different
algebraic higher symmetries.  The gapped ground \emph{state} of the boundary
Hamiltonian must spontaneously break part of the categorical symmetry, and can
only spontaneously break part of the categorical symmetry.  For example, as
pointed out in \Ref{JW191213492}, an $n$d system with a 0-symmetry described by
a finite group $G$ (or a fusion $n$-category $n\cRep G$) actually has a larger
categorical symmetry. The categorical symmetry is characterized by a $G$-gauge
theory $\sGT_G^{n+1}=\bulk(n\cRep G)$ in one higher dimension.  The categorical
symmetry include both the 0-symmetry $G$ (with $\cR=n\cRep G$) and an algebraic
$\nmo$-symmetry $G^{(n-1)}$ (with $\cR=n\cVec_G$).

\section{Gapped liquid phases with algebraic higher
symmetry} 

In Section~\ref{lfcat}, we discussed gapped liquid state with algebraic higher
symmetry, which is a trivial symmetric product state.  In this section, we are
going to discuss gapped liquid phases with unbroken algebraic higher symmetry,
which may have non-trivial topological orders.  Those states are called SET
states if the topological order is non-trivial (\ie with long range
entanglement), or SPT states if the topological order is trivial (\ie with
short range entanglement).

Let us first summarize some previous results in literature, which represent
some systematic understanding of \emph{gapped liquid
phases}\cite{ZW1490,SM1403} for boson and fermion systems with and without
symmetry (but only for $0$-symmetry).  In 1+1D, all gapped phases are
classified by $(G_H,G_\Psi,\om_2)$\cite{CGW1107,SPC1139}, where $G_H$ is the
on-site symmetry group of the Hamiltonian, $G_\Psi$ the symmetry group of the
ground state $G_\Psi \subset G_H$, and $\om_2\in H^2(G_\Psi,\R/\Z)$ is a group
2-cocycle for the unbroken symmetry group $G_\Psi$.  

In 2+1D, all gapped phases are classified (up to $E_8$ invertible topological
orders and for a finite unitary on-site symmetry $G_H$) by
$(G_H,\Rep(G_\Psi) \subset\eC \subset\eM)$ for bosonic systems and by
$(G_H,\sRep(G_\Psi^f) \subset\eC \subset\eM)$ for fermionic
systems\cite{BBC1440,LW160205946,LW160205936}.  Here $\Rep(G_\Psi)$ is
the symmetric fusion category formed by representations of $G_\Psi$, and
$\sRep(G_\Psi^f)$ is the symmetric fusion category formed by $\Z_2^f$-graded
representations of $G_\Psi^f$, where $\Z_2^f$
is a center of $G_\Psi^f$.  Also $\eC$ is the braided fusion category of
point-like
excitations and $\eM$ is a minimal
modular extension\cite{LW160205946,LW160205936}.

In 3+1D, some gapped phases are \emph{liquid phases} while others are
\emph{non-liquid phases}.  The 3+1D gapped liquid phases without symmetry for
bosonic systems (\ie 3+1D bosonic topological orders) are classified by
Dijkgraak-Witten theories if the point-like excitations are all bosons, by
twisted 2-gauge theory with gauge 2-group $\cB(G,Z_2)$ if some point-like
excitations are fermions and there are no Majorana zero modes, and by a special
class of fusion 2-categories if some point-like excitations are fermions and
there are Majorana zero modes at some triple-string
intersections\cite{LW170404221,LW180108530,ZW180809394}.  Comparing with the
classification of 3+1D SPT orders for bosonic\cite{CGL1314,K1459} and fermioinc
systems\cite{GW1441,KTT1429,GK150505856,FH160406527,KT170108264,WG170310937},
this result suggests that all 3+1D gapped liquid phases (such as SET and SPT
phases) for bosonic and fermionic systems with a finite unitary symmetry
(including trivial symmetry, \ie no symmetry) are classified by partially
gauging the symmetry of the bosonic/fermionic SPT orders\cite{LW180108530}.

We see that the previous approaches are quite different for different
dimensions and are only for  0-symmetries.  In this section, we describe a
classification that works for all dimensions. We also generalize the 0-symmetry
in the above results to algebraic higher symmetry.

\subsection{Partially characterize a symmetric gapped liquid phase using a pair
of fusion higher categories}

\label{pSETSPT}

The classification of gapped liquid phases with algebraic higher symmetry is
quite completecated.  In this section, we state a simple partial result, to
identify the difficulties and the issues in the classification.

For a gapped liquid state with unbroken algebraic higher symmetry $\cR$, there
are point-like, string-like, \etc\ excitations, that correspond to the charge
objects of the symmetry.  Those charge objects are described by the
representation category $\cR$.  In general, the  gapped liquid state also has
extra point-like, string-like, \etc\ excitations, that are beyond those
described by the local fusion higher category $\cR$.  So the total excitations
are described by a bigger fusion higher category $\cC$ that includes $\cR$.
This leads to the following result:
\begin{Proposition}
\label{RC}
An $n$d anomaly-free gapped liquid phase with an unbroken algebraic higher
symmetry described by a fusion $n$-category $\cR$, is fully described by a
potentially anomalous topological order $\sC$ (see Section~\ref{cata}).  The
excitations of $\sC$ are described by a fusion $n$-category $\cC=\Om\sC$ which
admits a top-fully faithful functor $\cR \inj{\iota} \cC$.  Thus we can use the
data $ \cR \inj{\iota} \cC $ to partially classify the gapped liquid phase with
algebraic higher symmetry $\cR$.
\end{Proposition}
\noindent
One way to see the above result is to note that stacking a trivial symmetric
state $\cR$ and the symmetric topological order $\cC$ together give rise to a
fusion $n$-category $\cR\otimes \cC$, if there is no coupling between the
trivial symmetric state $\cR$ and the symmetric topological order $\cC$.  The
$\cR\otimes \cC$ state has a larger algebraic higher symmetry $\cR\otimes \cR$,
one from the trivial symmetric state $\cR$ and the other from the symmetric
topological order $\cC$.  However, we can add the so called ``symmetric
interactions'' between $\cR$ and $\cC$ to reduce the  $\cR\otimes \cR$ symmetry
to the original symmetry $\cR$.  The stacking with such symmetric interactions,
which preserves the diagonal $\cR$ symmetry but break the other
symmetries, is denoted by $\otimes_\cR$ and $\cR\otimes_\cR \cR=\cR$.  Including the ``symmetric interactions''
is similar to adding the symmetry breaking morphisms in our definition of local
fusion higher category (see Def. \ref{lfus}).  Such a process can also be
realized by a condensation of excitations.  Since $\cR$ is local, the
condensation does not confine any excitations in $\cR$, and all the excitations
in $\cR$ become excitations in $\cR\otimes_\cR\cC$.  Physically we require that $\cR\otimes_\cR\cC
=\cC$. Therefore, all the excitations in $\cR$ become excitations in $\cC$.  Thus
there is a functor $\cR\inj{\iota} \cC$, which is faithful (\ie injective) at
each level of morphisms and objects.  Since both $\cR$ and $\cC$ have the same
algebraic higher symmetry $\cR$, the allowed local symmetric operators are the
same.  Thus the faithful functor $\cR\inj{\iota} \cC$ is fully faithful (\ie
bijective) at the top morphisms (which correspond to the local symmetric
operators).

Does every pair of fusion $n$-categories $(\cR, \cC)$ satisfying $\cR
\inj{\iota} \cC$ describes an anomaly-free topological order with an algebraic
higher symmetry?  The answer is no, as implied by some counterexamples when
$\cR$ describes a 0-symmetry.\cite{LW160205936,LW160205946} If the pair $(\cR,
\cC)$ does describe a symmetric topological order, does it uniquely  describe
the symmetric topological order?  The answer is also no.  For example, a pair
of fusion $n$-categories $(\cR, \cR)$ can describe a symmetric trivial product
state.  The same pair $(\cR, \cR)$ can also describe a SPT state of the same
symmetry.  This is because, as we mentioned before, the fusion $n$-category
only describe the excitations which do not contain all the information of a
topological order, and cannot distinguish different invertible topological
orders.  In our case here, the pair $(\cR, \cR)$ cannot distinguish symmetric
trivial product state from non-trivial SPT state with the same anomaly-free
algebraic higher symmetry.

However, the $\cR \inj{\iota} \cC $ description does not miss much.  In the
following, we try to understand which pairs $\cR \inj{\iota} \cC$ can
describe anomaly-free topological orders with an algebraic higher symmetry.  We
also try to seek additional information beyond $\cR \inj{\iota} \cC$ to
fully characterize a symmetric topological order.  One way to achieve both
goals is to use the notion of \emph{categorical symmetries} described in
\Ref{JW191213492} and in Section~\ref{csym}, which is a holographical way to view
a symmetry.  This new way to view a symmetry is most suitable for algebraic
higher symmetries.  It gives an even more general perspective about algebraic
higher symmetries.  So in the next section, we first study gapped liquid
phases in a bosonic system with a categorical symmetry.

\subsection{Classification of gapped liquid phases in bosonic systems with a
categorical symmetry} 
\label{glpCS}

\begin{figure}[t]
\begin{center}
\includegraphics[scale=0.7]{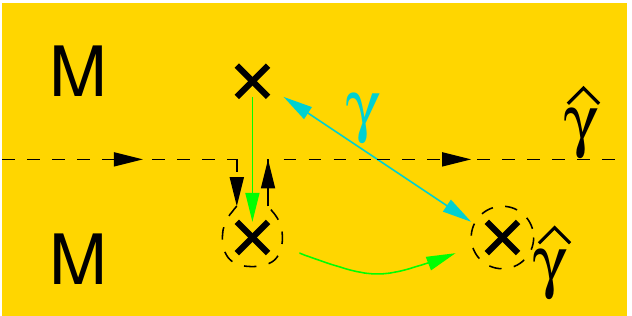} \end{center}
%17
\caption{Consider a topological order $\sM$ separated by a invertible domain
wall $\hat\ga$. Moving an excitation (with codimension-2 or higher) across the
invertible domain wall $\hat\ga$ (with codimension-1) induces an braided
automorphism $\ga$ of the braided fusion higher category $\eM=\Om^2\sM$: $\eM
\seq{\ga} \eM$.  } \label{WA}
\end{figure}

Here, we will address the difficulties and the issues identified in the last
section via a holographic approach.  We will first consider a modified problem:
classification of gapped liquid phases in bosonic systems with a categorical
symmetry.  Note that the  gapped liquid phases do not have the full categorical
symmetry, since gapped phases always particially break the categorical symmetry
spontaneously.  After this discussion, we identify a key difficulty in the
classification. 

Let us consider an $n$d bosonic lattice Hamiltonian with an algebraic higher
symmetry $\cR$.  Such a lattice Hamiltonian can also be regarded as having a
categorical symmetry characterized by an anomaly-free topological order
$\sM=\bulk(\cR)$ in one higher dimension, since algebraic higher symmetry $\cR$
and categorical symmetry $\sM=\bulk(\cR)$ are holo-equivalent.  One may ask,
what are the gapped liquid phases that have this categorical symmetry $\sM$?
The answer is that there is no such phases, since gapped phases in $n$d lattice
Hamiltonians with a non-trivial categorical symmetry $\sM$ must partially break
and only partially break the categorical symmetry
spontaneously.\cite{L190309028,JW191213492} This is because a gapped phase in
an $n$d bosonic lattice Hamiltonian with a categorical symmetry $\sM$
corresponds to a gapped boundary of a $\npo$d anomaly-free topological order
$\sM$.  The gapped boundary comes from the condensation of some of the
excitations in $\sM$, and thus part of the categorical symmetry is
spontaneously broken.  In fact, the condensing excitations form a Lagrangian
condensable algebra, which corresponds to spontaneously breaking part of the
categorical symmetry.  Also, the excitations that can condense (\ie those in
the Lagrangian condensable algebra) must have trivial mutual statistics between
them.  Therefore, we cannot condense all the excitations in $\sM$
simultaneously.  This is why we cannot completely break a categorical symmetry
spontaneously.  (Certainly, we can always partially or completely break a
categorical symmetry explicitly.) This picture leads to the following result
(see Fig.  \ref{CwM}a): 
\begin{Proposition} \label{CM1} 
For $n$d bosonic lattice Hamiltonians with a categorical symmetry $\sM$, their
gapped liquid phases are classified by the gapped boundaries of $\npo$d
anomaly-free topological orders $\sM$.  In other words, the gapped liquid
phases are classified by (potentially anomalous) topological orders $\sC$'s
(objects in $\cTO_{\sM}^{n+1}$'s, see Section~\ref{cata}) satisfying the
condition:
\begin{align} 
\sM=\Bulk(\sC) .  
\end{align} 
\end{Proposition} 
\noindent 
In light of Props. \ref{BulkCMp} and \ref{bulkCMp}, the above result implies
that (see Fig.  \ref{CwM}a) 
\begin{Proposition}
\label{CM2} For $n$d bosonic lattice Hamiltonians with a categorical symmetry
$\sM$, the excitations in their gapped liquid phases are described by fusion
$n$-categories $\cC$ such that 
\begin{align} 
\bulk(\cC) = \sM.  
\end{align} 
For every fusion $n$-category $\cC$ satisfying $\bulk(\cC) = \sM$, there are
one or more gapped liquid phases, $\sC$'s, to realize it: $\cC=\Om\sC$.
\end{Proposition} 
\noindent 
Let us remark the second part of the above result.  Given a fusion $n$-category
$\cC$ describing the excitations in a gapped phase of $n$d bosonic lattice
Hamiltonian with categorical symmetry $\sM$, do we have another gapped phase of
another $n$d bosonic lattice Hamiltonian with categorical symmetry $\sM$, that
also have excitations described by $\cC$.  In general, the answer is yes.  This
is because two gapped phases, differing by stacking of invertible topological
orders or SPT orders, have the same set of excitations.

To summarize 
\begin{Proposition} \label{CM3} 
The gapped liquid phases in $n$d bosonic lattice Hamiltonians with a
categorical symmetry $\sM$ are \textbf{partially} classified by the fusion
$n$-categories $\cC$ that satisfy $\bulk(\cC)=\sM$.
\end{Proposition} \noindent 
Here \emph{partially} means that the classification is one-to-many: the same
fusion $n$-category $\cC$ may corresponds to several different gapped liquid
phases, $\sC$'s, of systems with the categorical symmetry $\sM$.  As we have
mentioned before, here, the gapped liquid phases must break only part of the
categorical symmetry $\sM$ spontaneously.  

To get a full one-to-one classification, we need to find extra information
beyond the excitations, \ie the fusion $n$-category $\cC$, to characterize the
gapped liquid states.  One way to get extra information is to study the
boundaries of the gapped liquid states. This will be done later.  

Here, we consider another type of extra information.  As we have mentioned
above, stacking invertible topological orders or SPT orders to a gapped phase
does not change the excitations.  Let us consider a boundary of $\sM$ with
excitations $\cC$ (see Fig. \ref{CwM}).  We can use the trivial excitations in
$\sM$ to form an $n$d invertible topological order, which is a domain wall in
$\sM$.  We can also use the topological excitations in $\sM$ to form an $n$d
SPT order, which is also a domain wall in $\sM$. The protecting symmetry of the
SPT order comes from the fusion rule (the conservation law) of the topological
excitations that form the SPT order.  Both kinds of domain walls are invertible
domain walls.  There are also invertible domain walls that correspond to
braided automorphisms of the braided fusion $n$-category $\eM=\Om^2\sM$
describing the excitations in $\sM$.  In fact, each invertible domain wall
$\hat\ga$ corresponds to a braided automorphism $\gamma$ of $\eM$ (see Fig.
\ref{WA}).  Thus, stacking an invertible domain wall $\hat\ga$
to the boundary $\cC$ give us a boundary $(\cC,\hat\ga)$ that is related to the
boundary $\cC$ via an automorphism $\gamma$.  So the two boundaries are described
by fusion $n$-categories that are equivalent to $\cC$. However, the boundaries $(\cC,\hat\ga)$,
with different invertible domain walls $\hat\ga$, may correspond to different
boundary phases, \ie inequivalent $\sC$'s.  We conjecture that all different
boundary phases, $\sC$'s, with sets of boundary excitations equivalent to
$\cC$ can be
obtained this way: $\sC=(\cC,\hat\ga)$. % However, the correspondence between $\sC$'s and $(\cC,\hat\ga)$'s is not one-to-one.  It is possible that inequivalent $(\cC,\hat\ga)$'s correspond to equivalent $\sC$'s. 
This leads to the following
result (see Fig.  \ref{CwM}b)
\begin{Proposition}
\label{Cga}
For $n$d bosonic lattice Hamiltonians with a categorical symmetry $\sM$, their
gapped liquid phases are classified %(in a many-to-one fashion)
by a pair
$(\cC,\hat\ga)$, where $\cC$ is a fusion $n$-category $\cC$ that satisfies
$\bulk(\cC)\seq{\hat\ga} \sM$, and $\hat\ga$ is an invertible domain wall 
between $\bulk(\cC)$ and $\sM$.
\end{Proposition}
\noindent

The possible invertible domain wall $\hat\ga$ is of course not unique.
However, when we are considering gapped liquid phases with an algebraic higher
symmetry $\cR$ (instead of  gapped liquid phases that may spontaneously break
the categorical symmetry $\sM=\bulk(\cR)$), the different invertible domain
walls may have different physical meanings with respect to $\cR$. An invertible
domain wall may either preserve the algebraic higher symmetry $\cR$, or
(partially or completely) break $\cR$.  To classify  gapped liquid phases with
an algebraic higher symmetry $\cR$, we need to select $\hat\ga$'s that preserve
the algebraic symmetry $\cR$.  

How to select $\hat\ga$'s is a key difficulty in the classification.  In the
next section, we give several (hopefully equivalent) criteria when an
invertible domain wall $\hat\ga$ preserves an algebraic symmetry $\cR$.

\subsection{Classification of SET orders and SPT orders with an algebraic
higher symmetry}

\label{ahsTop}

\subsubsection{A simple result}

\label{ahsTopS}

Let us first give a simple partial result by ignoring the invertible domain
wall $\hat\ga$ (\ie by overlooking the key difficulty).  Given an algebraic
higher symmetry $\cR$, there is an $\npo$d anomaly-free topological order $\sM
=\bulk(\cR)$ (\ie the holo-equivalent categorical symmetry) that has one boundary
with excitations described by $\cR=\Om\sR$.  The boundary topological order
$\sR$ corresponds to a trivial product state with the algebraic higher symmetry
$\cR$. (More precisely, the trivial product state with the algebraic higher
symmetry $\cR$, plus its excitations, when restricted to the symmetric
sub-Hilbert space $\cV_\text{symm}$ correspond to the boundary topological
order $\sR$.) 

\begin{figure}[t]
\begin{center}
\includegraphics[scale=0.7]{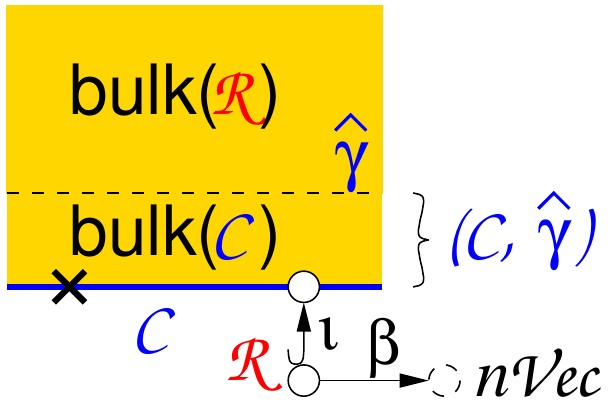} \end{center}
%18
\caption{ A gapped liquid phase with an algebraic higher symmetry $\cR$ can be
characterized by a pair $(\cC,\hat\ga)$, or more precisely, by
$\cC\ot{\bulk(\cC)}\hat\ga\ot{\bulk(\cR)}$, where $\hat\ga$ is an invertible domain
wall between $\bulk(\cR)$ and $\bulk(\cC)$: $\bulk(\cR)\seq{\hat\ga}
\bulk(\cC)$.  The fusion $n$-category $\cC$ on the boundary describes the
excitations on top of the gapped liquid state, which include $\cR$, \ie $\cR
\inj{\iota} \cC$.  Since $\cR$ is local, we also has a fiber functor $\bt: \cR
\to n\cVec$.  Not every invertible domain wall can be included. Finding the
condition to select the proper invertible domain walls $\hat\ga$ is the key to
classify gapped liquid phases with symmetry $\cR$.  The cross is a generic
excitation in $\cC$.  The white-circles are $\cR$-charge objects in $\cC$.  }
\label{CgaR}
\end{figure}

Now consider another boundary $\sC$ of $\sM$, with the excitations described by
a fusion $n$-category $\cC=\Om\sC)$.  However, the boundary $\sC$ may
spontaneously break the algebraic higher symmetry $\cR$.  Here we would like to
classify gapped liquid phases, $\sC$'s, that do not spontaneously break the
algebraic higher symmetry $\cR$.  To do so, we just need to  select $\sC$'s
such that its excitations $\cC$ contain $\cR$.  We believe that all the
anomaly-free topological orders with the algebraic higher symmetry $\cR$ can be
viewed in this way.  If we replace $\sC$ by $\cC$ to get a partial
classification, we obtain (see Fig. \ref{CgaR})
\begin{Proposition}
Anomaly-free gapped liquid phases with an algebraic higher
symmetry $\cR$ in $n$-dimensional space are partially classified by fusion
$n$-categories $\cC$, that satisfy $\bulk(\cC) \simeq \bulk(\cR)$ and admit a
top-fully faithful functor $\cR\inj{\iota} \cC$.
\end{Proposition}
\noindent
Here $\bulk(\cC) \simeq \bulk(\cR)$ means that the two bulk topological orders,
$\bulk(\cC)$ and $\bulk(\cR)$, are connected by an invertible domain wall (see
Fig. \ref{WA}). We used a more relaxed condition, just requiring $\bulk(\cC)$
and $\bulk(\cR)$ to be equivalent rather than equal.  This is because we are
not considering different boundaries of a fixed bulk. Here we are considering
different boundaries and their bulks, and hoping the bulks to be the same.  But
when we compare two bulks to see if their are the ``same'', the best we can do
is to see whether they are equivalent, \ie whether they are connected by an
invertible domain wall.

We would like to point out that the above $\cC$ classifies the excitations in
the anomaly-free topological order $\sC$ with an algebraic higher symmetry
$\cR$ (\ie $\sC$ can be realized by a bosonic lattice model in the same
dimensions with the symmetry $\cR$).  We know that topological orders differ by
invertible gapped liquids have the same excitations. Thus the above $\cC$'s
cannot distinguish different invertible gapped liquids, \ie  different
invertible topological orders and SPT orders.  The above $\cC$'s only classify
anomaly-free topological orders with the algebraic higher symmetry $\cR$, up to
invertible topological orders and SPT orders for symmetry $\cR$.

To obtain a more complete classification, \ie to include the SPT orders with
symmetry $\cR$, we should include the invertible domain walls $\hat\ga:
\bulk(\cR)\seq{}\bulk(\cC)$ as our data as we discussed in Section~\ref{glpCS} (see
Fig. \ref{CgaR}): 
\begin{Proposal}
\label{CgaZR}
Bosonic anomaly-free gapped liquid phases in $n$-dimensional space with an
anomaly-free algebraic higher symmetry $\cR$ are classified 
by data $(\cC,\ \iota: \cR\inj{}\cC,\ \hat\ga: \bulk(\cR)\simeq
\bulk(\cC))$, where $\cC$ is a fusion $n$-category that includes $\cR$ (\ie
$\iota: \cR \inj{} \cC$ is a top-fully faithful functor, see Proposition~\ref{RC}),
and $\hat\ga: \bulk(\cR)\simeq \bulk(\cC)$ is a invertible domain wall
between $\bulk(\cR)$ and $\bulk(\cC)$.  
\end{Proposal}
\noindent
However, the above proposal is incorrect.  We cannot use an arbitrary
invertible domain wall $\hat\ga: \bulk(\cR)\simeq \bulk(\cC)$.  The reason is
that $\cC$ contains the symmetry $\cR$ via the embedding $\iota: \cR\inj{}\cC$.
$\bulk(\cC)$ also contains the symmetry $\cR$ via the forgetful functor $F_\cC:
\Om^2\bulk(\cC)= Z_1(\cC)\to \cC$.  $\bulk(\cR)$ also contains the symmetry
$\cR$ via the forgetful functor $F_\cR: \Om^2\bulk(\cR)= Z_1(\cR)\to \cR$.  If
we allow an arbitrary invertible domain wall $\hat\ga$ which induces an
arbitrary braided equivalence $\ga: Z_1(\cR)\simeq Z_1(\cC)$, then the $\cR$
symmetry contained in $Z_1(\cC)$ and $Z_1(\cC)$ may not be compatible (\ie may
not be matched by $\ga$).  Thus the key is to find proper conditions to select
proper $\hat\ga$'s.  In the following, we describe several, hopefully
equivalent, ways to do so.

\begin{figure}[t]
  \centering
  \includegraphics[scale=0.7]{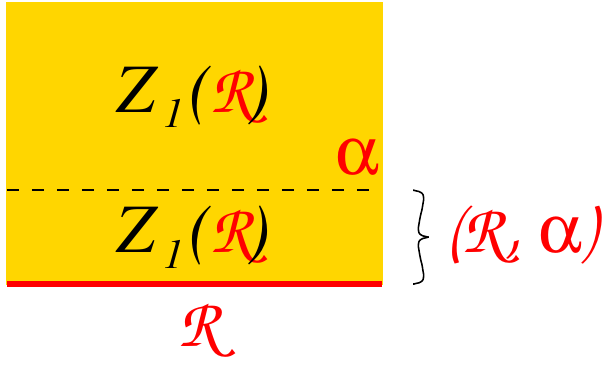}
%19
\caption{Similar to Fig. \ref{CgaR}, with $\cC=\cR$.  An SPT state
is characterized by a pair $(\cR,\al)$, or more precisely, by the stacking of
$\cR$ and $\al$ through the bulk $Z_1(\cR)$:
$\cR\ot{Z_1(\cR)}\al\ot{Z_1(\cR)}$. Here $\al$ is a braided automorphism of
$Z_1(\cR)$.  Not every automorphism can be included.  Finding the condition to
select the proper automorphisms $\al$ is the key to classify $\cR$-SPT orders.
} \label{RalR}
\end{figure}

When $\cC=\cR$, the above reduces to a classification of SPT phases with
algebraic higher symmetry $\cR$ via a pair $(\cR,\al)$, where $\al$ is an
automorphism of $Z_1(\cR)$ (see Fig. \ref{RalR}).  To describe this
classification of SPT phases in more detail, we like to remark that the map
from the invertible domain walls $\hat\al$ in $\bulk(\cR)$ to the braided
automorphisms $\al$ of $\Om^2\bulk(\cR)=Z_1(\cR)$ may not one-to-one.  We
conjecture that
\begin{Conjecture}
the kernel of the map $\hat\al \to \al$ is the set of invertible domain walls
in $\bulk(\cR)$ that correspond to invertible topological orders formed by
trivial excitations in $\Om^2 \bulk(\cR)= Z_1(\cR)$.
\end{Conjecture}
\noindent
This conjecture is based on the belief that any properties of excitations
$Z_1(\cR)$ cannot see invertible topological orders.

Because the classification of SPT phases do not include invertible topological
orders, we can replace the invertible domain walls $\hat\al$ by the braided
automorphisms $\al$. This allows us to obtain the following proposal (see Fig.
\ref{RalR})
\begin{Proposal}
\label{RgaZR}
Bosonic SPT phases in $n$-dimensional space with an anomaly-free algebraic
higher symmetry $\cR$ are classified by the braided automorphisms $\al$ of
$Z_1(\cR)$.
\end{Proposal}
\noindent
Again, the above proposal is not correct since some braided automorphisms $\al$
may break the symmetry $\cR$ (\ie change the symmetry $\cR$ contained in
$Z_1(\cR)$).  To make it correct, we need to find proper conditions to select
proper  braided automorphisms $\al$, that do not break the symmetry $\cR$.

The above discussions reveal the key difficulty in classifying gapped liquid
phases with an algebraic higher symmetry. In the next a few subsections, we
propose several approaches to address this issue, which leads to several,
hopefully equivalent, classification results.

\subsubsection{A classification assuming $\cR$ to be symmetric}

\begin{figure}[t]
\begin{center}
\includegraphics[scale=0.7]{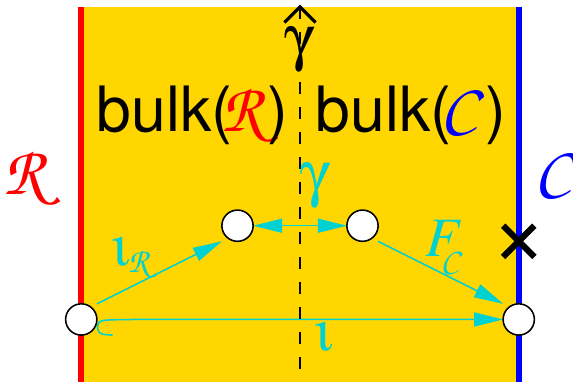} \end{center}
%20
\caption{ Similar to Fig. \ref{CgaR}, but now we assume $\cR$ to be symmetric.
In this case, the $\cR$-charges in $\cR$ can be lifted into the bulk via the
embedding $\iota_\cR$.  The equivalence $\ga$ should be compatible with the
lifting $\iota_\cR$ of $\cR$, the embedding $\iota: \cR \inj{} \cC$, and the
bulk-to-boundary functor $F_\cR$.  } \label{RCMiF}
\end{figure}
\begin{figure}[t]
\begin{center}
\includegraphics[scale=0.7]{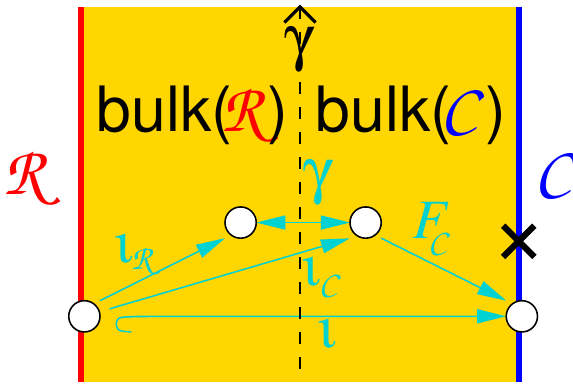} \end{center}
%21
\caption{ 
The graphic representation of \eqn{RCii} which is equivalent to
Fig. \ref{RCMiF}.  We assume $\cR$ to be symmetric.  In
this case, the $\cR$-charges in $\cR$ and in $\cC$ can be lifted into the bulk
via the embedding $\iota_\cR$ and $\iota_\cC$. 
}
\label{RCMi}
\end{figure}

\begin{figure}[t]
\begin{center}
\includegraphics[scale=0.65]{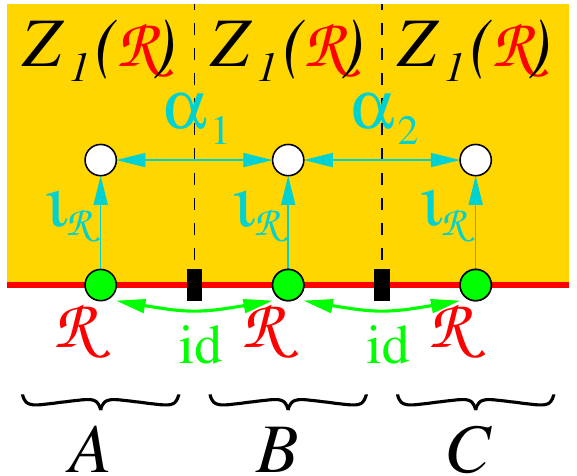} \end{center}
%22
\caption{ 
Three SPT phases $A,\ B,\ C$ with symmetry $\cR$.  $A$ and $B$ differ by an
invertible domain wall characterized by the equivalence $\al_1$ (satisfying
\eqn{RCM2R}).  $B$ and $C$ differ by an invertible domain wall $\al_2$.  Then
$A$ and $C$ differ by an invertible domain wall characterized by the
composite equivalences $\al_1\circ \al_2$.  We see that the SPT phases are
classified by the invertible domain walls, \ie by the  equivalences
$\al:Z_1(\cR)\seq{\al}Z_1(\cR)$ satisfying \eqn{RCM2R}.
}
\label{RCM2SPT}
\end{figure}

When $\cR$ is symmetric, it can be lifted to the bulk $Z_1(\cR)$ via a
canonical braided embedding $\iota_\cR:\cR\hookrightarrow Z_1(\cR)$. In this
case, we have a simple criteria for $\gamma$ to make the two $\cR$ symmetries
in $\cC$ and $Z_1(\cR)$ compatible (\ie to preserve the $\cR$ symmetry, see
Fig.  \ref{RCMiF}):
\begin{Proposition}
\label{RCM2}
Anomaly-free gapped liquid phases in $n$-dimensional space with an anomaly-free
algebraic higher symmetry $\cR$ (which is assumed to be symmetric) are
classified by data $(\cC,\ \iota:
\cR\inj{}\cC,\ \hat\ga: \bulk(\cR)\simeq \bulk(\cC))$, where $\cC$ is a fusion
$n$-category that includes $\cR$ (\ie $\iota: \cR \inj{} \cC$ is a top-fully
faithful functor), and $\hat\ga $ is an invertible domain wall
rendering the following diagram commutative (up to a natural isomorphism):
\begin{align}
\label{gaembed}
\xymatrix@R=2.0em@C=2.0em{
 \cR \ar@{_(->}[d]|{\iota_\cR} \ar@{^(->}[rr]^{\iota}  &&\cC \\
     Z_1(\cR) \ar@{<->}[rr]^\ga_\simeq &&  Z_1(\cC)\ar[u]|{F_\cC} 
}
\end{align}
where $\ga$ is the braided equivalence $\ga: Z_1(\cR)\simeq Z_1(\cC))$ induced
by the  invertible domain wall $\hat\ga: \bulk(\cR)\simeq \bulk(\cC))$ and
$F_\cC:Z_1(\cC)\to \cC$ is the forgetful functor.
\end{Proposition}
\noindent
\Ref{KZ200308898} proposed this result  in a slightly
different manner. An embedding $\iota_\cC:\cR\hookrightarrow Z_1(\cC)$ is
considered as the data for gapped liquid phases instead of $\iota:\cR\hookrightarrow \cC$, and it
is required that $F_\cC\circ \iota_\cC :\cR\to \cC$ is an embedding, thus
reproduce the data $\iota$. Then, \eqn{gaembed} is replaced by (see Fig.  \ref{RCMi})
\begin{align}
\label{RCii}
\xymatrix@R=2.0em@C=2.0em{
& \cR \ar@{_(->}[dl]|{\iota_\cR} \ar@{^(->}[dr]|{\iota_\cC} &\\
     Z_1(\cR) \ar@{<->}[rr]^\ga_\simeq &&  Z_1(\cC)
}
\end{align}
When $\cC=\cR$ the above result reduces to a classification of SPT orders with
symmetry $\cR$ (see Fig.  \ref{RCM2SPT}):
\begin{Proposition}
\label{RCM2spt}
SPT phases in $n$-dimensional space with an anomaly-free algebraic higher
symmetry $\cR$ (which is assumed to be symmetric) are classified by data
$(\cR,\al)$, where $\al: Z_1(\cR)\seq{}  Z_1(\cR)$ is a braided  equivalence
rendering the following diagram commutative (up to a natural isomorphism):
\begin{align}
\label{RCM2R}
\xymatrix@R=2.0em@C=2.0em{
 \cR \ar@{_(->}[d]|{\iota_\cR} \ar@{=}[rr]^{\id_\cR}  &&\cR \\
   Z_1(\cR) \ar@{<->}[rr]^\al_\simeq &&  Z_1(\cR)\ar[u]|{F_\cR} 
}
\end{align}
\end{Proposition}
\noindent
\begin{Remark} Note that $F_\cR \circ \alpha \circ \iota_\cR = \id_\cR
  =F_\cR\circ\iota_\cR$ is a central functor, where
  the central structure comes from the symmetric structure of $\cR$. By the
  universal property of center \cite{KZ170200673},
  \eqref{RCM2R} is equivalent to
  \begin{align}\label{sptembed}
\xymatrix@R=2.0em@C=2.0em{
 &\cR \ar@{_(->}[dl]|{\iota_\cR} \ar@{^(->}[rd]|{\iota_\cR} & \\
   Z_1(\cR) \ar@{<->}[rr]^\al_\simeq &&  Z_1(\cR) 
}
  \end{align}
\end{Remark}

\subsubsection{First version of general classification}
\label{cl1}

Now we discuss a classification for more general algebraic higher symmetry
where $\cR$ may not be symmetric.  To do so, we need a very different approach.
Let us first consider the classification of bosonic SPT orders 
with an algebraic higher symmetry $\cR$ in $n$-dimensional space.  Those SPT
orders all have excitations described by the same local fusion $n$-category
$\cR$.  To distinguish different SPT orders, we need to include extra
information beyond $\cR$, and to use pairs $(\cR,\al)$ to describe the SPT
orders, where $\al$ is an automorphism of $Z_1(\cR)$.  To identify the proper
$\al$'s, we notice that the physical way to distinguish different SPT orders is
to include the boundary of a SPT state.  Here we consider the canonical
boundary that spontaneously breaks all the symmetry $\cR$.

\begin{figure}[t]
\begin{center}
\includegraphics[scale=0.7]{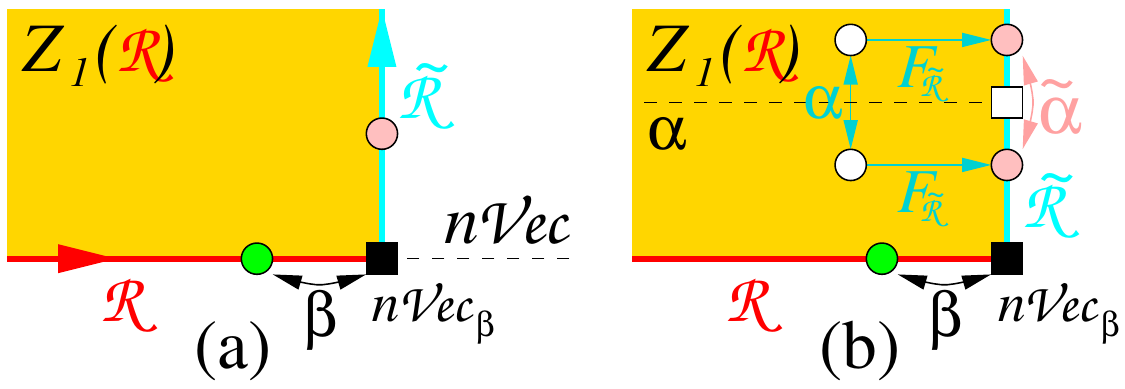} \end{center}
%23
\caption{ (a) A SPT state with symmetry $\cR$ (the red-line) can be viewed as a
gapped boundary of a topological order in one higher dimension with excitations
$Z_1(\cR)$. The boundary $n\cVec_\bt$ of SPT state $\cR$ that breaks
all the $\cR$-symmetry also has a bulk, which can be viewed as a gapped
boundary of $Z_1(\cR)$ for the dual symmetry $\dual\cR$.  (b) The
automorphism $\al$ of $Z_1(\cR)$ correspond to an invertible domain wall in the
bulk (the dash-line), which also has an invertible boundary (the white square).
The boundary-bulk relation between $Z_1(\cR)$ and $\dual\cR$ is described by
the bulk-to-boundary functor $F_{\dual\cR}$.  Such a boundary-bulk relation is
preserved by the automorphisms $\al,\ \dual\al$, which classify the $\cR$-SPT
orders.  } \label{RdRMspt} 
\end{figure}

In the following, we develop a theory for the canonical boundary that break all
the symmetry $\cR$, using a holographic point of view of the symmetry $\cR$,
\ie using a topological order with excitations $Z_1(\cR)$ in one higher
dimension to describe the symmetry $\cR$.  In other words, we need to use the
holographic point of view to describe the boundary that breaks all the symmetry
$\cR$.  Such a symmetry breaking boundary also has a bulk in one higher
dimension.  Such a bulk has a different set of excitations described by another
local fusion $n$-category, denoted as $\dual\cR$. In fact $\dual\cR$ can be
viewed as another gapped boundary of the bulk $Z_1(\cR)$ (see Fig.
\ref{RdRMspt}a), therefore, $Z_1(\cR)\simeq Z_1(\dual\cR)$.  $\dual\cR$ is
nothing but the local fusion $n$-category that describes the dual symmetry of
$\cR$ (see Section~\ref{dsymm}).  For example, when $\cR=n\cRep G$, the symmetry
is the 0-symmetry described by the group $G$.  The dual symmetry is an
algebraic higher symmetry described by $\dual\cR=n\cVec_G$.  

We know that the bulk $Z_1(\cR)$ and the boundaries, when viewed as systems,
have a categorical symmetry $Z_1(\cR)$ that includes both the symmetry $\cR$
and the dual symmetry $\dual\cR$.  The boundary $\cR$ in Fig.  \ref{RdRMspt}
has the symmetry $\cR$ but spontaneously breaks the dual symmetry $\dual\cR$,
while the boundary $\dual\cR$ has the dual symmetry $\dual\cR$ but
spontaneously breaks the symmetry $\cR$. The intersection $n\cVec_\bt$  of the
two boundaries breaks both the symmetry $\cR$ and the dual symmetry $\dual\cR$
(see Fig.  \ref{RdRMspt}a).  

The ``bulk'' of the canonical boundary $n\cVec_\bt$ of $\cR$ ($n\cVec_\bt$ is
  the same $n$-category as $n\cVec$ with $\cR$-module structure induced by
$\beta:\cR\to n\cVec$), which is also a
``boundary'' of the bulk of $\cR$, gives us the criteria when the automorphism
$\al:Z_1(\cR)\simeq Z_1(\cR)$ preserves the symmetry $\cR$ and thus represents
an $\cR$ SPT order (see Fig.~\ref{RdRMspt}a).  To identify the proper
automorphisms, we note that $\al$ can be viewed as an invertible domain wall in
the bulk $Z_1(\cR)$ (see Fig.~\ref{RdRMspt}b).  Such an invertible domain wall
has a boundary on the boundary $\dual\cR$ (the white square in
Fig.~\ref{RdRMspt}b).  Since the difference between SPT orders are invertible,
the boundary of the invertible domain wall should also be invertible.  This
motivates us to conjecture that the boundary of the invertible domain wall
$\al$ corresponds to an automorphism $\dual\al$ of $\dual\cR$.  The
automorphisms $\al$ for the bulk $Z_1(\cR)$ and $\dual\al$ for the boundary
$\dual\cR$ should preserve the whole structure of $\cR$ and its boundary
$n\cVec_\bt$ (the red-line and the black-box in  Fig. \ref{RdRMspt}b).  This can
be achieved by requiring $\al,\ \dual\al$ to preserve the bulk-boundary
relation described by the bulk to boundary functor $F_{\dual\cR}: Z_1(\cR) \to
\dual\cR$.  This leads to the following result:
\begin{Proposition}
\label{RdRMSPT}
Bosonic SPT orders with an anomaly-free algebraic higher symmetry $\cR$ in
$n$-dimensional space are classified by braided equivalence $\al: Z_1(\cR)
\seq{} Z_1(\cR)$ and monoidal equivalence $\dual\al: \dual\cR \seq{} \dual\cR$, such that the
following diagram is commutative (up to a natural isomorphism):
\begin{align}
\xymatrix@R=2.0em@C=4.0em{
Z_1(\cR) \ar@{<->}[r]^\al_\simeq &  Z_1(\cR) \\
\dual\cR \ar@{<-}[u]|{F_{\dual\cR}} \ar@{<->}[r]^{\dual\al}_\simeq & \dual\cR \ar@{<-}[u]|{F_{\dual\cR}} 
}
\end{align}
\end{Proposition}
\begin{Remark}\label{dualal}
  Note that $\dual\al$ in the above contains some redundant information. This
  can be seen from the fact that even when $\al=\id_{Z_1(\cR)}$, there can still be
  nontrivial $\dual\al$. We believe that such extra $\dual\al$'s are higher
  structures (such as lower dimensional SPT or invertible phases) and should be
  quotiented out when considering the classification of SPT/SET orders. See also
  Remark~\ref{mu}.
\end{Remark}
\noindent

The above is just for SPT orders.  In the following, we use the similar
approach to develop a more general categorical theory to classify both SPT and
SET orders. We also allow more general algebraic higher symmetry, by allowing
$\cR$ to be $\cV$-local (recall Definition~\ref{vlocal}), to include at least
both boson and fermion systems.  When $\cV=n\cVec$, $\cR$ describes the
algebraic higher symmetry in bosonic systems.  When $\cV=n\scVec$, the fusion
$n$-category of super vector spaces, $\cR$ describes the algebraic higher
symmetry in fermionic systems.

\begin{Remark}
Physically, we think $\cV$ as the building blocks of our system. $\cR$ is built
upon $\cV$ with some additional symmetry that can be totally broken. $\bt:\cR\to
\cV$ exactly describes the symmetry breaking that leaves only the intrinsic
symmetry $\cV$ of the building blocks which is not physically breakable (for
example, fermion parity).
\end{Remark}
\noindent

A gapped liquid state with symmetry $\cR$ has excitations $\cC$ that is
equipped with a top-fully faithful monoidal functor $\iota:\cR\to \cC$. There
is an anomaly-free topological order $\underline{\cC}$ underlying $\cC$ by
breaking all the $\cR$ symmetry. Mathematically, we may define
$\underline{\cC}$ to be the pushout (i.e., the colimit) of
$\cV\xleftarrow{\beta} \cR\xrightarrow{\iota} \cC$ in the category of fusion
$n$-categories, 
\begin{align}
\xymatrix{
  \cR \ar@{^(->}[r]^\iota\ar[d]^\beta  & \cC\ar[d]^{\bt_\cC }\\ 
  \cV\ar@{^(->}[r]^{\iota_0}          &  \underline{\cC}
}
\end{align}
As a colimit, $\underline{\cC}$, $\bt_\cC$ and $\iota_0$ are uniquely
determined by $\cV\xleftarrow{\beta} \cR\xrightarrow{\iota} \cC$ up to
isomorphisms.  In particular, for SPT orders, we take $\cC=\cR$,
$\iota=\id_\cR$, and then $\underline{\cC}=\cV$, $\bt_\cC=\beta$,
$\iota_0=\id_\cV$:
\begin{align}
\xymatrix{
  \cR \ar@{=}[r]^{\id_\cR}\ar[d]^\beta  & \cR\ar[d]^{\bt}\\ 
  \cV\ar@{=}[r]^{\id_\cV}         &  \cV
}
\end{align} 
Alternatively, $\beta$ can be consider as condensing some
excitations (which form an algebra $A_\beta$, and condensing means taking the
modules over this algebra) in $\cR$. Condensing the same excitations in $\cC$
(identified via $\iota$), gives $\underline{\cC}$. 

$\underline{\cC}$ constitutes a symmetry breaking domain wall between
$\underline{\cC}$ and $\cC$.  Mathematically, $\underline{\cC}$ is
$\underline{\cC}$-$\cC$-bimodule; the left action is by fusion in
$\underline{\cC}$ and right action is by first mapping $\cC$ into
$\underline{\cC}$ via $\bt_\cC$ and then fusion in $\underline{\cC}$. To
emphasize, we denote the bimodule by $\underline{\cC}_{\bt_\cC}$. The bulk of
$\underline{\cC}$, $\cC$, as well as the domain wall
$\underline{\cC}_{\bt_\cC}$, can be defined via bimodule functors (see
Section~\ref{BBrel}). More
precisely,
\begin{itemize}
  \item The bulk of
    $\underline{\cC}$ is $Z_1(\underline{\cC}):=\Fun_{\underline{\cC}|\underline{\cC}}(\underline{\cC},\underline{\cC})$;
  \item The bulk of $\cC$ is $Z_1(\cC):=\Fun_{\cC|\cC}(\cC,\cC)$;
  \item The bulk of the domain wall $\underline{\cC}_{\bt_\cC}$ is
\begin{align}
\label{tcC}
    \dual\cR_\cC:=\Fun_{\underline{\cC}|\cC}(\underline{\cC}_{\bt_\cC},\underline{\cC}_{\bt_\cC}). 
\end{align}
 $\dual\cR_\cC$ is also a symmetry breaking
    domain wall in the bulk, between $Z_1(\underline{\cC})$ and $Z_1(\cC)$.
  \item There are also bulk to wall functors  
    \begin{align*}
      F'_{\dual\cR_\cC}: Z_1(\underline{\cC})=\Fun_{\underline{\cC}|\underline{\cC}}(\underline{\cC},\underline{\cC}) & \to
      \Fun_{\underline{\cC}|\cC}(\underline{\cC}_{\bt_\cC},\underline{\cC}_{\bt_\cC})=\dual\cR_{\cC}\\
      f &\mapsto f(\one_{\underline\cC})\ot{}-,\\
      F_{\dual\cR_\cC}: Z_1(\cC)=\Fun_{\cC|\cC}(\cC,\cC) & \to
      \Fun_{\underline{\cC}|\cC}(\underline{\cC}_{\bt_\cC},\underline{\cC}_{\bt_\cC})=\dual\cR_{\cC}\\
      f &\mapsto - \ot{} \bt_\cC(f(\one_\cC)),\\
    \end{align*}
\end{itemize}

\begin{figure}[t]
  \centering
  \includegraphics[scale=0.75]{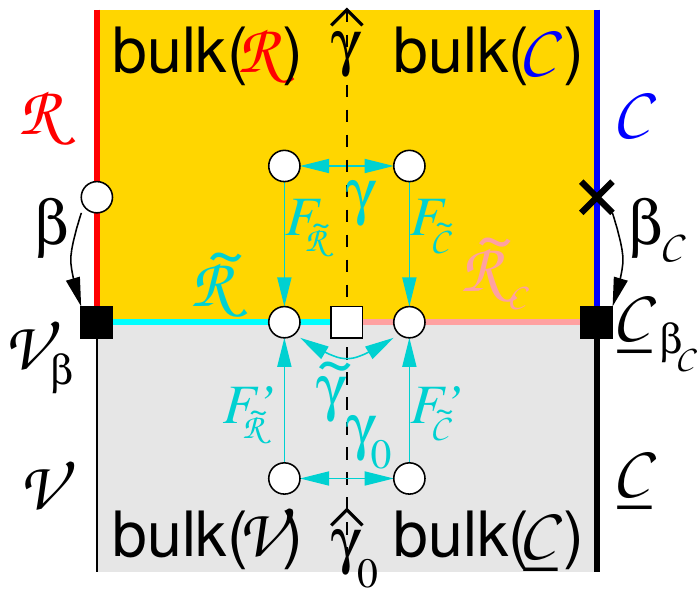}
%24
  \caption{ $\cR$, $\cV$ and their domain wall $\cV_\bt$ describe a trivial
product state with symmetry $\cR$ (as well as the symmetry breaking $\cR
\map{\bt} \cV$).  $\cC$, $\underline{\cC}$ and their domain wall
$\underline{\cC}_{\bt_\cC}$ describe a gapped liquid state with symmetry $\cR$.
The bulk of two structures (the categorical symmetry and its breaking) must
coincide.  This is just the anomaly matching condition, since the categorical
symmetry can be viewed as anomaly.  As a result, the $\cR$-symmetric gapped
liquid phase and the $\cR$-symmetric trivial phase can be two phases of the
same system.  The data $(\cR \map{\bt} \cV, \cR \inj{\iota} \cC,
\hat\ga,\dual\ga,\hat\ga_0)$ classify the bosonic gapped liquid phases with
$\cR$-symmetry when $\cV=n\cVec$.  They classify the fermionic gapped liquid
phases with $\cR$-symmetry when $\cV=n\scVec$.  } \label{RdRC}
\end{figure}

In conclusion, if we view $\underline{\cC}, \cC$ with a domain wall
$\underline{\cC}_{\bt_\cC}$ between them as a whole, the bulk of this structure
is given by
$Z_1(\underline{\cC})\xrightarrow{F'_{\dual\cR_\cC}} \dual\cR_\cC \xleftarrow{F_{\dual\cR_\cC}}
Z_1(\cC)$ (see Fig.~\ref{RdRC}), \ie, two non-degenerate braided fusion
$n$-categories $Z_1(\underline{\cC})$ and $Z_1(\cC)$, a fusion $n$-category
$\dual\cR_\cC$ and two bulk to wall functors $F'_{\dual\cR_\cC}$ and $F_{\dual\cR_\cC}$.
Physically, $Z_1(\underline{\cC})\xrightarrow{F'_{\dual\cR_\cC}} \dual\cR_\cC
\xleftarrow{F_{\dual\cR_\cC}} Z_1(\cC)$ is the categorical symmetry and how it breaks
down corresponding to $\beta: \cR\to\cV$.

Clearly, the bulk of the product state with the $\cR$-symmetry (and its
breaking $\cR \map{\bt}\cV$) is given by $Z_1(\cV)\xrightarrow{F'_{\dual\cR}}
\dual\cR \xleftarrow{F_{\dual\cR}} Z_1(\cR)$. As before, we require that the bulk of
$\cC$ coincide with the $\cR$-symmetric product state, which amounts to say
that there should be three equivalences $\gamma:Z_1(\cR)\simeq Z_1(\cC),$
$\dual\gamma: \dual\cR \simeq \dual\cR_\cC,$ and $\gamma_0: Z_1(\cV)\simeq
Z_1(\underline{\cC})$. The criteria for these equivalences to preserve $\cR$ is
given below:
\begin{Proposition}
\label{vset}
An anomaly-free gapped liquid phase in $n$-dimensional space with a generalized
anomaly-free algebraic higher symmetry described by a $\cV$-local $\cR$ is
characterized by the data $\iota:\cR\hookrightarrow \cC$, and $\hat\gamma,\dual\gamma,\hat\gamma_0$, where $\cC$ is a
fusion $n$-category,
$\hat\gamma$ is an invertible domain wall between $\bulk(\cR)$ and $\bulk(\cC)$,
$\hat\gamma_0$ is an invertible domain wall between $\bulk(\cV)$ and
$\bulk(\underline{\cC})$ and $\dual\gamma:\dual \cR\simeq \dual\cR_\cC$ is a monoidal equivalence.  The two invertible domain walls induce braided
equivalences $\gamma:Z_1(\cR)\simeq Z_1(\cC)$ and $\gamma_0: Z_1(\cV)\simeq
Z_1(\underline{\cC})$.  The three equivalences $\gamma,\dual\gamma,\gamma_0$
must render the following diagram commutative.  
\begin{align}
\label{bulkww}
    \xymatrix@R=1.7em@C=4.0em{
      Z_1(\cR) \ar[r]^\gamma_\simeq \ar[d]^{F_{\dual\cR}} & Z_1(\cC)\ar[d]^{F_{\dual\cR_\cC}}\\
      \dual\cR \ar[r]^{\dual\gamma}_\simeq &\dual\cR_\cC\\
      Z_1(\cV) \ar[r]^{\gamma_0}_\simeq \ar[u]_{F'_{\dual\cR}} & Z_1(\underline{\cC}) \ar[u]_{F'_{\dual\cR_\cC}}
    }
\end{align}
\end{Proposition}
\noindent
In the above, we have used the invertible domain walls $\hat\ga$ and
$\hat\ga_0$ to capture invertible topological orders.  We use the equivalences
$\ga$ and $\ga_0$ of braided fusion higher categories, induced by the
invertible domain walls, to formulate the condition to select the proper domain
walls.

In particular, taking $\cC=\cR$ and then
$\underline{\cC}=\cV,\dual\cR_\cC=\dual\cR$, we obtain a classification of
$\cR$-SPT orders:
\begin{Proposition}
\label{vspt}
An SPT phase in $n$-dimensional space with a generalized anomaly-free algebraic
higher symmetry described by a $\cV$-local $\cR$ is characterized by the three
automorphisms $\al:Z_1(\cR)\simeq Z_1(\cR),\ \dual\al: \dual\cR\simeq
\dual\cR,\ \al_0:Z_1(\cV)\simeq Z_1(\cV)$ rendering the following diagram
commutative
\begin{align}
\label{bulkwwSPT}
    \xymatrix@R=1.7em@C=4.0em{
      Z_1(\cR) \ar[r]^\al_\simeq \ar[d]^{F_{\dual\cR}} & Z_1(\cR)\ar[d]^{F_{\dual\cR}}\\
      \dual\cR \ar[r]^{\dual\al}_\simeq &\dual\cR\\
      Z_1(\cV) \ar[r]^{\al_0}_\simeq \ar[u]_{F'_{\dual\cR}} & Z_1(\cV) \ar[u]_{F'_{\t{\cR}}}
    }
\end{align}
The above triples of automorphisms $(\al,\dual\al,\al_0)$, that label different
$\cR$-SPT orders, can be composed, which correspond to the stacking of the SPT
orders.
\end{Proposition}
\noindent
If we choose $\cV=n\cVec$, the above classifies SET/SPT orders for bosonic
systems with an algebraic higher symmetry.  If we choose $\cV=n\scVec$, the
above classifies SET/SPT orders for fermionic systems with a generalized
algebraic higher symmetry. Again, as pointed out in Remarks~\ref{dualal} and
\ref{mu}, it is
very likely that differenct choices $\dual \ga,\ga_0$ or $\dual \al,\al_0$
correspond to the same SET/SPT order, and thus only $\ga$ or $\al$ needs to be kept.

In this formulation, there is no need to assume that $\cR$ is symmetric or even
braided.  But assuming $\cR,$ $\cV$ and $\beta:\cR\to \cV$ are braided, we want
to show that the Proposition~\ref{vset} and Proposition~\ref{RCM2} are
equivalent. We sketch a tentative proof here.  There is a canonical braided
embedding $\iota_\cR:\cR\hookrightarrow Z_1(\cR)$. Then consider the pushout of
$ \cV \xleftarrow{\beta} \cR\xrightarrow{\iota_\cR} Z_1(\cR)$. In the category
of fusion $n$-categories, the pushout is just $\dual  \cR$.
\begin{align}
  \xymatrix{\cR \ar@{^{(}->}[r]^{\iota_\cR}\ar[d]^\beta& Z_1(\cR) \ar[d]^{F_{\dual\cR} }\\
      \cV\ar@{^{(}->}[r] &\dual \cR}
\end{align}
Indeed, $\beta$ can be considered as condensing some excitations $A_\beta$ in $\cR$.
Condensing the same excitations in $Z_1(\cR)$ (identified via $\iota_\cR$) gives
$\dual\cR$. Moreover, $Z_1(\cV)$ should be a full subcategory of $\dual\cR$
corresponding to the deconfined excitations and $F'_{\dual\cR}$ is the embedding.
Therefore, the embedding $\iota_\cR$ determines all the other structures
$Z_1(\cV),\dual\cR, F_{\dual\cR}, F'_{\dual\cR}$. Also $\gamma:Z_1(\cR)\simeq Z_1(\cC)$ with
such embedding $\iota_\cR$ determines $\dual\gamma$ and $\gamma_0$.
Then it should be straightforword to verify that \eqref{bulkww} is equivalent to
\eqref{gaembed}.

\begin{Example}
  For $\cR=\cC=\cRep G$, $\cV=\cVec$, $\bt:\cRep G\to \cVec$ the forgetful
functor, we have $\dual\cR=\cVec_G$. Since $F_{\dual\cR}$ corresponds to condensing
the algebra $\Fun(G)\in \cRep G$, preserving the embedding $\cRep
G\hookrightarrow Z_1(\cRep G)$ is the same as preserving $\Fun(G)$ and thus $\t
F_\cR$. The $G=\Z_2\times \Z_2$ case is explicitly calculated in the Section
\ref{Z2Z2}.
\end{Example}

\subsubsection{Second version of classification based on condensable algebra}

\label{cl2}

In this section, we are going to describe another version of classification.
Let us first consider an $\cR$-SPT state characterized by a pair $(\cR,\al)$,
where $\al$ is a braided automorphism of $Z_1(\cR)$ (see Fig.  \ref{RalR}). We
would like to explore other equivalent ways to select proper $\al$'s. We first
restrict to bosonic systems for simplicity, assuming that $\cR$ is a local
fusion $n$-category. A key feature of SPT order is that a SPT state has no
topological order, \ie it becomes a product state if we break the symmetry.
How to impose such a condition, when we use the holographic point view of the
symmetry $\cR$?

Here we would like to point out that if we stack $\cR$ and its dual $\dual\cR$
through their common bulk $\eM=Z_1(\cR)$, denoted as
$\cR\ot{\eM}\dual\cR^\rev$, we get a trivial product state $n\cVec$ (see Fig.
\ref{RdR}).  We may use this property to define a more general notion of dual
symmetry.
\begin{Definition}
\label{RdRdef}
Let $\eM$ be the braided fusion $n$-category describing excitations in a
$\npo$d anomaly-free bulk topological order, and  $\cR$ and $\cB$ (together
with bulk to boundary functors $\eM\to \cR,\eM\to \cB$) be two $n$d boundaries
of $\eM$. $\cR$ and $\cB$ are said dual to each other if
\begin{align}
\label{RMB}
    \cR\ot{\eM}\cB^\rev =n\cVec.
\end{align}
\end{Definition}
\noindent
In fact, there is an one-to-one correspondence between dual
symmetries of $\cR$ and the monoidal functors $\cR\to n\cVec$. If a boundary
$\cB$ of $Z_1(\cR)$ satisfies $\cR\ot{Z_1(\cR)}\cB^\rev =n\cVec$, we may define 
\begin{align}
  \beta_{\cB}: \cR &\to n\cVec=\cR\ot{Z_1(\cR)}\cB^\rev ,\nonumber\\
  x&\mapsto x \ot{Z_1(\cR)} \one_\cB.
\end{align}
Recall that for the given $\beta:\cR\to n\cVec$, the dual symmetry is defined
by $\dual\cR=\Fun_{\cR|n\cVec}(n\cVec_\beta,n\cVec_\beta)$ (see Fig.
\ref{RdRMspt}a).  We have the following correspondence:
\begin{Proposition}
\label{RB}
The maps $\cB\mapsto \beta_{\cB}$ and $\beta\mapsto
\Fun_{\cR|n\cVec}(n\cVec_\beta,n\cVec_\beta)$ are inverse to each other. 
\end{Proposition}
\noindent
In particular, $\cR$ also defines a monoidal functor $\cB\to n\cVec$. In other
words, 
\begin{Proposition}
\label{RBlocal}
if $Z_1(\cR) = Z_1(\cB)$ and $\cR \ot{Z_1(\cR)} \cB^\rev = n\cVec$, then both
$\cR$ and $\cB$ are local fusion $n$-categories.  
\end{Proposition}

Physically,  when $\cR$ and $\dual\cR$ are dual to each other, every excitation
in $\eM=Z_1(\cR)$, either condenses to $\cR$-boundary or condenses to
$\dual\cR$-boundary, or both.  In this case, every excitation in $\eM$ is
condensed and the resulting state is trivial.  We know that $\cR$ can be
obtained from $\eM$ via a Lagrangian condensable algebra $A_\cR$ in $\eM$.
Similarly, $\dual\cR$ can be obtained from $\eM$ via another Lagrangian
condensable algebra $A_{\dual\cR}$.  Roughly speaking a condensable algebra is
formed by excitations with trivial mutual statistics with each other, and those
excitations can all condense simultaneously to form a gapped boundary (see Fig.
\ref{RdR}).  Thus in the $\cR$ boundary, the  excitations in $A_\cR$ condense.
The non-condensing excitations become the boundary excitations that is
described by $\cR$.  So roughly speaking $\cR=\eM/A_\cR$.  (In precise
mathematical language, $\cR$ identifies with the category of modules over
$A_\cR$ in $\eM$.) Similarly, $\dual\cR=\eM/A_{\dual\cR}$.  When $\cR$ and
$\dual\cR$ are dual to each other, the overlap of $A_{\cR}$ and $A_{\dual\cR}$
is minimal and is given by the trivial excitations. Also  $A_{\cR}$ and
$A_{\dual\cR}$ together generate the whole $\eM$.  (More precisely, any
excitation in $\eM$ is contained in $A_\cR\otimes A_{\dual\cR}$.) Thus roughly
speaking, $A_\cR\otimes A_{\dual\cR} \sim \eM $.  We see that, $A_\cR$ is
formed by excitations in $\dual\cR$ and $A_{\dual\cR}$ is formed by excitations
in $\cR$. 

\begin{figure}[t]
\begin{center}
\includegraphics[scale=0.7]{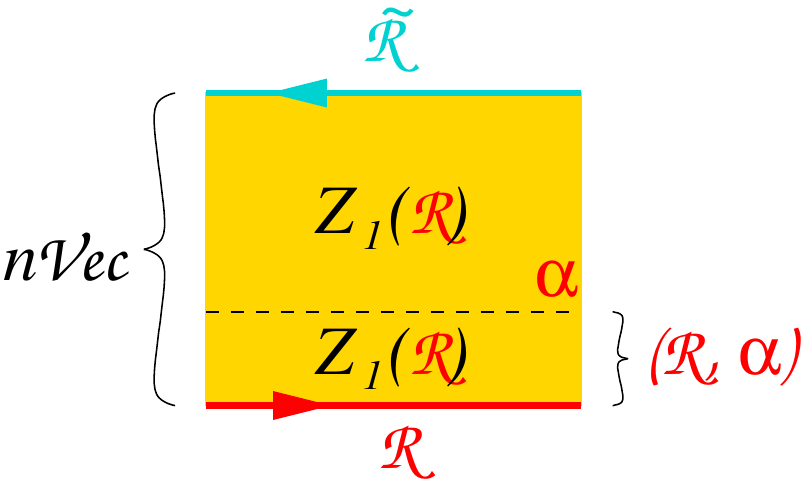} \end{center}
%25
\caption{ $(\cR,\al)\equiv \cR\ot{Z_1(\cR)}\al\ot{Z_1(\cR)}$ describes an SPT
state only if it can be canceled by $\dual\cR$ (\ie producing a product state
$n\cVec$).  } \label{RaldR}
\end{figure}

Proposition~\ref{RB} tells us that dual symmetry is an equivalent way to
describe symmetry breaking. If $\cR$ can be canceled by its dual:
$\cR\ot{\eM}\dual\cR^\rev =n\cVec$, then $\cR$ (as well as $\dual\cR$) is a local fusion
$n$-category, \ie $\cR$ can be reduced to the trivial product state if we break
the symmetry: $\cR \map{\bt} n\cVec$.  Since $\cR$ can be viewed as
$(\cR,\al=\id)$ in Fig.  \ref{RalR}, we see that $(\cR,\id)$  can be cannceled
by $\dual\cR$, which implies that $(\cR,\id)$ is a product state if we break the
symmetry.  This implies that if we do not break the symmetry, then $(\cR,\id)$
is a SPT state.  Therefore, to see if $(\cR,\al)$ is a SPT state or not, we can
just check if it can be cancelled by $\dual\cR$ or not.  This allows us to obtain
(see Fig. \ref{RaldR})
\begin{Proposition}
$(\cR,\al)\equiv \cR\ot{Z_1(\cR)}\al\ot{Z_1(\cR)}$ describes a bosonic
$\cR$-SPT state if the automorphism $\al$ of $Z_1(\cR)$ satisfies
\begin{align}
\label{RaldRVec}
 \cR \ot{Z_1(\cR)}\al \ot{Z_1(\cR)} \dual\cR^\rev  \simeq n\cVec.
\end{align}
\end{Proposition}
\noindent

\begin{figure}[t]
\begin{center}
\includegraphics[scale=0.7]{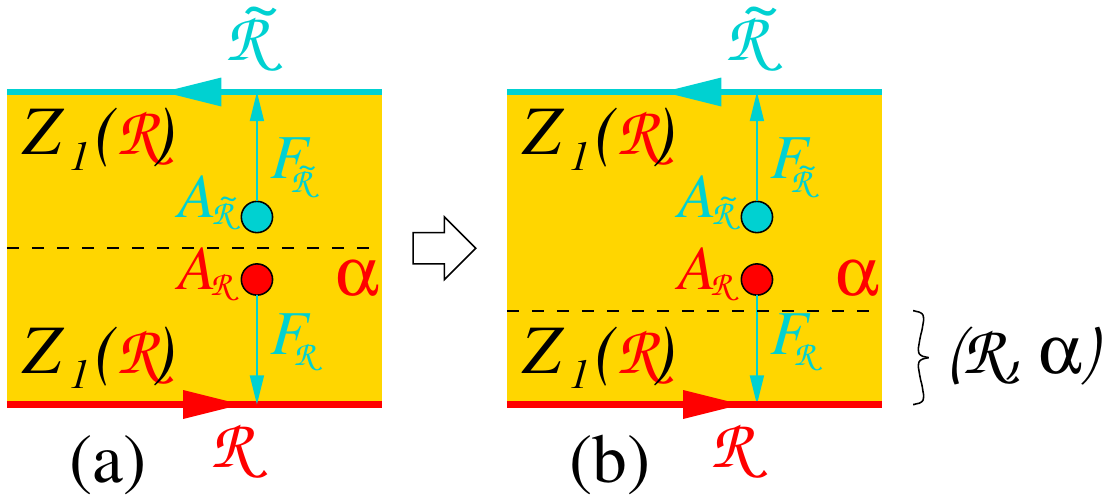} \end{center}
%26
\caption{ If $\al$ keeps the $A_\cR$ unchanged, then $(\cR,\al)\equiv
\cR\ot{Z_1(\cR)}\al\ot{Z_1(\cR)}$ is also determined by the condensable algebra
$A_\cR$ and is equivalent to $\cR\ot{Z_1(\cR)}$.  } \label{RaldR1}
\end{figure}

Using the condensable algebra, we find that one class of the solutions of
\eqn{RaldRVec} are given by $\al$'s that keep the $A_\cR$ unchanged (see Fig.
\ref{RaldR1}):
\begin{align}
\label{ARAR}
\al(A_\cR) \seq{\mu} A_\cR,
\end{align}
where $\mu$ is an algebra isomorphism.  In this case, Fig.  \ref{RaldR1}a can
be deformed into Fig. \ref{RaldR1}b.  By comparing  Fig.  \ref{RaldR1}b with
Fig.  \ref{RaldR}, we see that both $\cR\ot{Z_1(\cR)}\al\ot{Z_1(\cR)}$ and
$\cR\ot{Z_1(\cR)}$ are determined by the same condensable algebra $A_\cR$, and
hence are equivalent: $\cR\ot{Z_1(\cR)}\al\ot{Z_1(\cR)} \simeq
\cR\ot{Z_1(\cR)}$.  This allows us to show that 
\begin{align}
\cR \ot{Z_1(\cR)}\al \ot{Z_1(\cR)} \dual\cR^\rev  
\simeq \cR \ot{Z_1(\cR)} \dual\cR^\rev  \simeq n\cVec
.
\end{align}
$\cR\ot{Z_1(\cR)}\al\ot{Z_1(\cR)} \simeq \cR\ot{Z_1(\cR)}$ also tell us that the SPT state
described by $\cR\ot{Z_1(\cR)}\al\ot{Z_1(\cR)}$ is equivalent to the SPT state described by
$\cR\ot{Z_1(\cR)}$, which is the trivial SPT state.
In fact, $\al$ that keeps $A_\cR \sim \dual\cR$ unchanged may change $\cR$.  Thus
we believe that $\cR\ot{Z_1(\cR)}\al\ot{Z_1(\cR)}$ and $\cR\ot{Z_1(\cR)}$ only
differ by a relabeling of the symmetry (\ie differ by an automorphism of
$\cR$), and thus correspond to the same SPT state.

\begin{figure}[t]
\begin{center}
\includegraphics[scale=0.7]{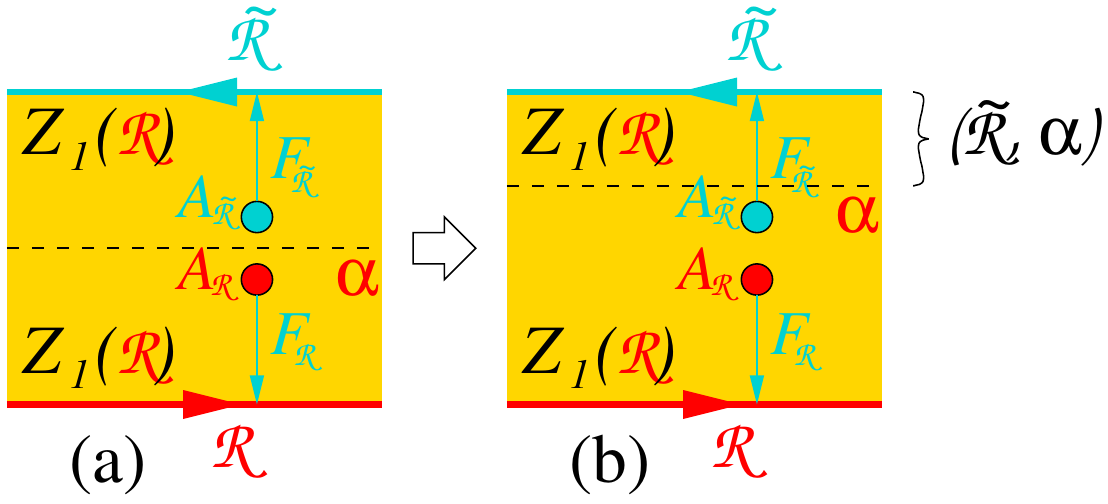} \end{center}
%27
\caption{ 
If $\al$ keeps the $A_{\dual\cR}$ unchanged, then $(\dual\cR,\al)\equiv
\ot{Z_1(\cR)}\al\ot{Z_1(\cR)}\dual\cR^\rev  \simeq \ot{Z_1(\cR)}\dual\cR^\rev $.  } \label{RaldR2}
\end{figure}

\begin{figure}[t]
\begin{center}
\includegraphics[scale=0.7]{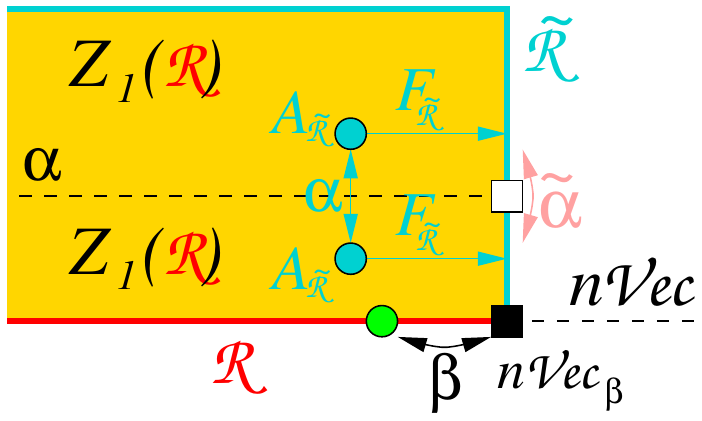} \end{center}
%28
\caption{ The $\cR$-SPT orders are classified by the automorphisms $\al$ of
$Z_1(\cR)$ that keep $A_{\dual\cR}$ unchanged.  An $\al$ corresponds to an
invertible domain wall in the bulk (the dash-line), which also has an
invertible boundary (the white square).  The boundary-bulk relation between
$Z_1(\cR)$ and $\dual\cR$ is described by the bulk-to-boundary functor
$F_{\dual\cR}$, which maps the condesible algebra $A_{\dual\cR}$ to the trivial
excitioins on the $\dual\cR$ boundary.  } \label{RaldRB} 
\end{figure}

\Eqn{RaldRVec} has another class of solutions, given by $\al$'s that keep
$A_{\dual\cR}$ unchanged:
\begin{align}
\label{dARAR}
\al(A_{\dual\cR}) \seq{\mu} A_{\dual\cR}.
\end{align}
By comparing  Fig.  \ref{RaldR2}b with Fig.  \ref{RaldR}, we see that
$\ot{Z_1(\cR)}\al\ot{Z_1(\cR)}\dual\cR^\rev  \simeq \ot{Z_1(\cR)}\dual\cR^\rev $.  This allows us to show that, indeed, 
\begin{align}
\cR \ot{Z_1(\cR)}\al \ot{Z_1(\cR)} \dual\cR^\rev  
\simeq \cR \ot{Z_1(\cR)} \dual\cR^\rev  \simeq n\cVec
.
\end{align}
As we have mentioned, the condensable algebra $A_{\dual\cR}$ is formed by
excitations in $\cR$.  So keeping $A_{\dual\cR}$ part of $\eM$ unchanged
corresponds to keeping $\cR$ part of $\eM$ unchanged.  Therefore, the
automorphisms $\al$, that satisfy \eqn{dARAR}, do not change the $\cR$
symmetry.  But $\al$'s generate non-trivial automorphisms of $\dual\cR$,
$\dual\al: \dual\cR \seq{} \dual\cR$ (see Fig.  \ref{RaldRB}).

Fig.  \ref{RaldRB} also
describes the canonical symmetry-breaking boundary, 
$n\cVec_\bt\ot{\dual\cR} \dual\al \ot{\dual\cR}$, 
of the SPT state $ \cR\ot{Z_1(\cR)} \al\ot{Z_1(\cR)} $.  We see
that different pairs of automorphisms $(\al,\dual\al)$ give rise to different
boundaries (due to different $\dual\al$'s).  This leads to the following result:
\begin{Proposition}
\label{RCMcaspt}
Bosonic SPT phases in $n$-dimensional space with an anomaly-free algebraic
higher symmetry $\cR$ are classified (up to invertible topological orders) by
braided equivalences $\al: Z_1(\cR)\seq{}  Z_1(\cR)$ together with algebra
isomorphisms $\mu:\al(A_{\dual\cR})\simeq A_{\dual\cR}$.
\end{Proposition}
\noindent
We would like to remark that, for a given $\al$, different choices of $\mu$
differ by automorphisms of the condensable algebra $A_{\dual\cR}$.  Those
different $\mu$'s may lead to the same SPT order. This is because if we gauge
the $\cR$-symmetry in a SPT state, the resulting topogecal order does not
depend on $\mu$ (see Remark \ref{mu}).  Thus, the bosonic SPT phases with a
$\cR$-symmetry may actually be classified by $\al$'s, rather than the pairs
$(\al,\mu)$'s.

\begin{figure}[t]
\begin{center}
\includegraphics[scale=0.75]{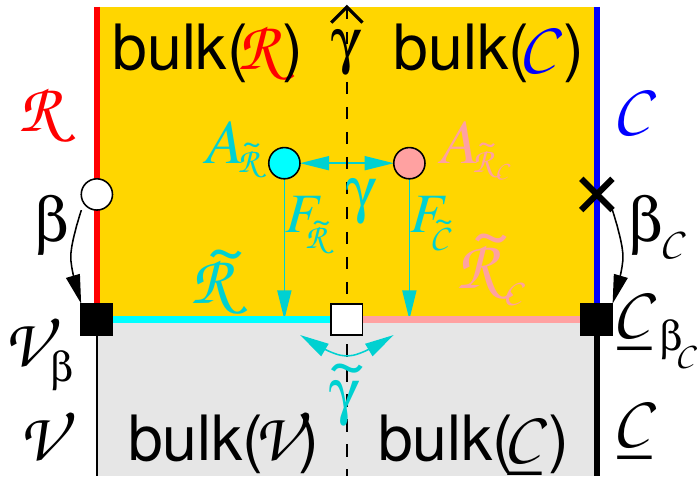} \end{center}
%29
\caption{ Similar to Fig. \ref{RdRC}, but here the invertible domain wall
$\hat\ga$ induces an automorphism $\ga$, which is required to map the two
condensable algebras, $A_\cR$ and $A_{\dual\cR_\cC}$, into each other:
$\mu:\gamma(A_{\dual\cR})\simeq A_{\dual\cR_\cC}$. The data $(\cR
\map{\bt}\cV,\cR \inj{\iota} \cC, \hat\ga,\mu)$ classify the bosonic gapped
liquid phases with $\cR$-symmetry when $\cV=n\cVec$.  They classify the
fermionic gapped liquid phases with $\cR$-symmetry when $\cV=n\scVec$.  }
\label{RCMca}
\end{figure}

We can generalize the above result to include SET orders with $\cR$-symmetry
for bosonic or fermionic systems, as well as invertible topological orders by
using invertible domain walls (see Fig.  \ref{RCMca}):
\begin{Proposition}
\label{RCMcaset}
Anomaly-free gapped liquid phases in $n$-dimensional space with a generalized
anomaly-free algebraic higher symmetry $\cR$ are classified by the data $(\cR
\inj{\iota} \cC,\ \hat\ga,\ \mu)$, where $\cR$ is a $\cV$-local fusion
$n$-category, $\cC$ is a fusion $n$-category that includes $\cR$, $\hat\ga$ is
a invertible domain wall between $\bulk(\cR)$ and $\bulk(\cC)$ and $\mu:
\ga(A_{\dual\cR})\simeq A_{\dual\cR_\cC}$ is an algebra isomorphism.  Here
$\ga: Z_1(\cR)\seq{}  Z_1(\cC)$ is a braided equivalence, $\dual\cR_\cC$ is
defined in Proposition~\ref{vset} and $A_{\dual\cR}$, $A_{\dual\cR_\cC}$ are the
condensable algebras in $Z_1(\cR)$, $Z_1(\cC)$ that produces the $\dual\cR$,
$\dual\cR_\cC$ domain walls between $Z_1(\cR)$, $Z_1(\cC) $ and $Z_1(\cV),
Z_1(\underline{\cC})$, respectively. 
\end{Proposition}
\noindent
Although we have included $\mu$ in the above, it is possible that different
choices of $\mu$ correspond to the same gappled liquid phases, as we discussed
later in Remark \ref{mu}.  Thus the different gapped liquid phases may actually
be classified by the data $(\cR \inj{\iota} \cC,\ \hat\ga)$ such that
$\ga(A_{\dual\cR})\simeq A_{\dual\cR_\cC}$.

When $\cV=n\cVec$, the above classifies SET/SPT
orders in bosonic systems.  When $\cV=n\scVec$, the above classifies SET/SPT
orders in fermionic systems.

\begin{Remark}
Here we would like to sketch the reasoning that Proposition~\ref{vset} and
Proposition~\ref{RCMcaset} are equivalent.  From the condensation point of
view, condensing the algebra $A_\beta$ in $\cR$ induces the symmetry breaking
$\beta: \cR\to \cV$. Similarly, the symmetry breaking $F_{\dual\cR}
:Z_1(\cR)\to \dual \cR$ in the bulk is induced by condensing the algebra
$A_{\dual\cR}$ in $Z_1(\cR)$. $A_{\dual\cR}$ is the lift of $A_\beta$ in the
bulk, \ie, $A_\beta= F_{\cR}(A_{\dual\cR})$, where $F_{\cR}: Z_1(\cR)\to \cR$
is the forgetful functor. Intuitively, we can think that $A_{\dual\cR}$
replaces the role of embedding $\cR\hookrightarrow Z_1(\cR)$; instead of
embedding $\cR$ into the bulk which is only possible when $\cR$ is braided, we
lift the algebra $A_\beta$ in $\cR$ to the algebra $A_{\dual\cR}$ in
$Z_1(\cR)$. $A_\beta$ consists of all objects in $\cR$ when $\cV=n\cVec$.
Mathematically, $\dual\cR$ should be the category of modules over
$A_{\dual\cR}$ in $Z_1(\cR)$ while $Z_1(\cV)$ should be the full subcategory of
local modules, with $F'_{\dual\cR}$ being the embedding (see Fig.~\ref{RdRC}).
(Physically, $Z_1(\cV)$  corresponds the deconfined excitations after
condensation while $\dual\cR$ includes both confined and deconfined
excitations.) Therefore, $\gamma$ together with $\mu:\gamma(A_{\dual\cR})\simeq
A_{\dual\cR_\cC}$ determines $\dual\gamma$ as an equivalence functor between
the categories of modules over $A_{\dual\cR}$ and $A_{\dual\cR_\cC}$;
$\gamma_0$ is the restriction of $\dual\gamma$ to $Z_1(\cV)$.  \Eqn{bulkww} is
equivalent to saying that $\gamma$ preserves the lifted algebra
$\gamma(A_{\dual\cR})\seq{\mu}A_{\dual\cR_\cC}$.

\end{Remark}

\subsubsection{$\cR$-gauge theory obtained by ``gauging'' the algebraic higher
symmetry $\cR$}
\label{Rgauge}

\begin{figure}[t]
\begin{center}
\includegraphics[scale=0.63]{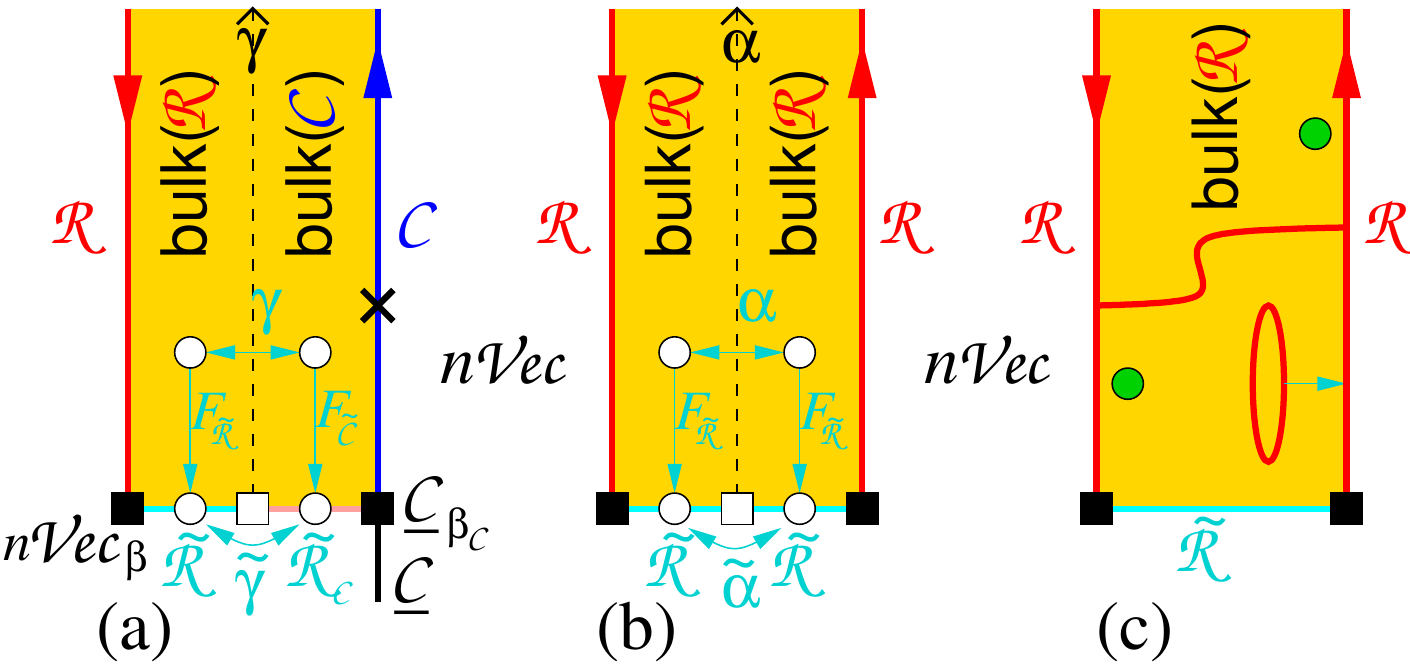} \end{center}
%30
\caption{ (a) Same as Fig. \ref{RdRC}, assuming $\cV=n\cVec$.  The stacking of
the two boundaries $\cR$ and $\cC$ through the bulk $\bulk(\cR)\seq{\hat\ga}
\bulk(\cC)$ with an invertible domain wall $\hat\ga$ (denoted by
$\cR\ot{\bulk(\cR)} \hat\ga\ot{\bulk(\cC)} \cC^\rev $) gives rise to an
anomaly-free topological order with no
symmetry, which is obtained by gauging the algebraic higher symmetry $\cR$ in
the gapped liquid state (SET or SPT state) characterized by the data $(\cC,\
\iota: \cR\inj{}\cC,\ \hat\ga: \bulk(\cR)\simeq \bulk(\cC))$.  (b) When
$\cC=\cR$ and $\hat\ga=\hat\al$, $\cR\ot{\bulk(\cR)}\hat\al\ot{\bulk(\cR)}
\cR^\rev $ describes the twisted $\cR$-gauge theory $\sGT_{\cR,\al}^{n+1}$
obtained by gauging the symmetry $\cR$ in the SPT state stacked with tirivial
or non-trivial invertible topological orders. (c) When $\cC=\cR$ and $\hat\ga=$
trivial, $\cR\ot{\bulk(\cR)} \cR^\rev $ describes the $\cR$-gauge theory
$\sGT_\cR^{n+1}$ obtained from a symmetric product state by gauging its
symmetry $\cR$.  } \label{gauge}
\end{figure}

Using the data $(\cC,\ \iota: \cR\inj{}\cC,\ \hat\ga: \bulk(\cR)\simeq
\bulk(\cC))$, we can explicitly construct the corresponding gapped liquid state
with anomaly-free symmetry $\cR$ that the data describes.  This is done in Fig.
\ref{CgaR}.  Since the gapped liquid state has the symmetry $\cR$, we can gauge
the symmetry $\cR$ to obtain a new topologically ordered stated with no
symmetry.  This is achieved in Fig.  \ref{gauge}a, by stacking $\cR$ and $\cC$
through $\bulk(\cR) \seq{\hat\ga} \bulk(\cC)$, with an invertible domain wall
$\hat\ga$ in the middle.  We denote such a stacking by $\cR \ot{\bulk(\cR)}
\hat\ga \ot{\bulk(\cC)} \cC^\rev $.  The resulting topologically ordered state
is anomaly-free since it is surrounded by the trivial product state (with its
codimension-2 excitations described by $n\eVec$).  As a bonus, such a gauging
picture leads to third version of classification, which will be described in
the next subsection.

Both 0-symmetries and higher symmetries have an holonomy interpretation, which
allows us to gauge them via a geometric approach.  In contrast, the above
proposal to ``gauge'' algebraic higher symmetries (which include  0-symmetries
and higher symmetries) is a purely algebraic approach.  No geometric
interpretation is used.

To understand such a proposal, let us consider a very simple case, by assuming
$\hat\ga=$ trivial, $n=2$, and $\cC=\cR=2\cRep G$.  So the boundary $\cR$ is in
2d while the bulk $\bulk(\cR)$ is in 3d.  The resulting state
$\cR\ot{\bulk(\cR)} \cR^\rev $ given by Fig.  \ref{gauge}c is actually a 2d
gauge theory with group $G$ (\ie the 2d topological order $\sGT_G^3$).  To see
this, we note that the bulk $\bulk(\cR)$ is the 3d gauge theory with group $G$
(\ie the 3d topological order $\sGT_G^4 =\bulk(2\cRep G)$).  The 2d boundary
$\cR=2\cRep G$ is obtained from the 3d $G$-gauge theory $\sGT_G^4$ by
condensing the $G$-flux loops. Thus a $G$-flux loop in the bulk corresponds to
a trivial excitation in the 2d $G$-gauge theory order $\sGT_G^3$.  A $G$-flux
string connecting two boundaries corresponds to a point-like $G$-flux
excitation in the 2d $G$-gauge theory $\sGT_G^3$.  The point-like $G$-charges
in the  3d $G$-gauge theory $\sGT_G^4$ becomes the point-like $G$-charges in
the  2d $G$-gauge theory $\sGT_G^3$.  This suggests that, in general,
$\cR\ot{\bulk(\cR)} \cR^\rev $ in Fig.  \ref{gauge}c is a $\cR$-gauge theory in
$n$-dimensional space.  When $\cR=n\cRep G$, $\cR\ot{\bulk(\cR)} \cR^\rev $ is
an $n$d $G$-gauge theory.  When $\cR$ describes a higher symmetry,
$\cR\ot{\bulk(\cR)} \cR^\rev $ is a higher gauge theory.  But when $\cR$
describes an algebraic higher symmetry, $\cR\ot{\bulk(\cR)} \cR^\rev $ is
something new, which is called a gauge theory from ``gauging'' the algebraic
higher symmetry $\cR$ in a product state.

When $\cC=\cR$ and $\hat\ga=\hat\al\neq $ trivial, the resulting state $\cR
\ot{\bulk(\cR)} \hat\ga \ot{\bulk(\cR)} \cR^\rev $ in  Fig. \ref{gauge}b is a
twisted $\cR$-gauge theory, obtained from ``gauging'' the algebraic higher
symmetry $\cR$ in a SPT state stacked with trivial or non-trivial invertible
topological order.  So gauged SPT states stacked with trivial or non-trivial
invertible topological order are classified by $(\cR,\ \hat\al)$.  If we just
consider gauged SPT states, by ignoring the stacked with trivial or non-trivial
invertible topological order, we can replace the invertible domain wall
$\hat\al$ by its induced automorphism $\al$ of $Z_1(\cR)$.  We see that  gauged
SPT states are described by $(\cR,\ \al:Z_1(\cR)\simeq Z_1(\cR))$ (see Fig.
\ref{RalRgauge}, where $\cR$ is replaced by $\Si\cR$).

When $\cC\neq \cR$, the resulting state $\cR \ot{\bulk(\cR)} \hat\ga
\ot{\bulk(\cC)} \cC^\rev $ in  Fig. \ref{gauge}a is a topological order
obtained from ``gauging'' the algebraic higher symmetry $\cR$ in the gapped
liquid state (SET or SPT state) characterized by data $(\cC,\ \iota:
\cR\inj{}\cC,\ \hat\ga: \bulk(\cR)\simeq \bulk(\cC))$.

\begin{Remark}
\label{mu}
We note that the algebra isomorphism $\mu:\gamma(A_{\dual\cR})\simeq
A_{\dual\cR_\cC}$ and the equivalence functor
$\dual\gamma:\dual\cR\simeq\dual\cR_\cC$ are similar data that are additional
to $\gamma:Z_1(\cR)\simeq Z_1(\cC)$. The fact that these additional data are
not manifestly visible in the gauged theory, may suggest that they are fixed by
$\gamma$ up to certain natural higher structures (such as lower dimensional SPT
or invertible phases). As an analogy, $\mu$ or $\dual\gamma$ are similar to the
$n$-coboundaries generated by $\nmo$-cochains that should be mod out when
considering the $n$th cohomology. But the exact physical meaning of $\mu$ and
$\dual\gamma$ is unclear to us for now. Moreover, $\mu$ and $\dual\gamma$ may
need to satisfy some additional conditions that involve even higher structures,
so on and so forth until the top morphisms. The study of these higher
structures is beyond our current scope, and will be left for future work.
\end{Remark}

\subsubsection{Third version of classification based on gauging the
$\cR$-symmetry}

\label{cl3}

\begin{figure}[t]
\begin{center}
\includegraphics[scale=0.7]{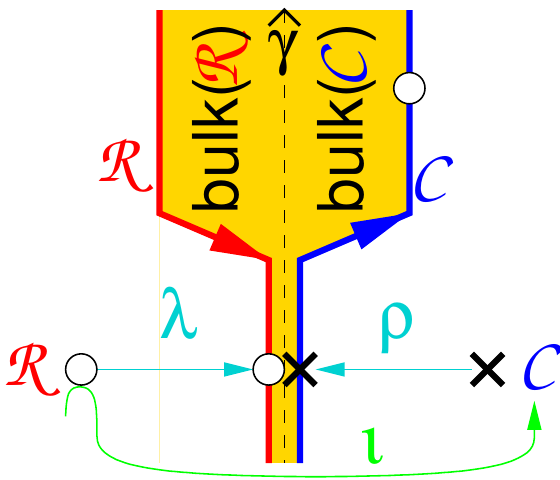} \end{center}
%31
\caption{{The data $\cR \map{\bt}\cV$, $\cR\inj{\iota} \cC$ and $\bulk(\cR)
\seq{\hat\ga} \bulk(\cC)$ classify gapped liquid phases (SET or SPT phases)
with generalized algebraic higher symmetry $\cR$, provided that invertible
domain wall $\hat\ga$ is chosen (see Proposition~\ref{RCMgauge}) to make the
stacking $\cR\ot{\bulk(\cR)} \hat\ga \ot{\bulk(\cC)} \cC^\rev$ to describe the
gauging of the $\cR$-symmetry.  The condensations of the excitations in
$\bulk(\cR)$ and $\bulk(\cC)$ form the $\cR$ and $\cC$ boundaries, which leave
parts of the two categorical symmetries, $\bulk(\cR)$ and $\bulk(\cC)$,
unbroken.  $\hat\ga$ is chosen so that the two unbroken parts of the two
categorical symmetries match with the same $\cR$, giving rise to an unbroken
$\cR$ symmetry.  }  } \label{RCMG}
\end{figure}

We can also use the gauging of the $\cR$-symmetry, and the resulting
topological order $\cR\ot{\bulk(\cR)} \hat\ga \ot{\bulk(\cC)} \cC^\rev $ in
Fig. \ref{RCMG}, to obtain $\hat\ga$ that ``keeps the $\cR$ part in
$\bulk(\cR)$ unchanged'', which leads to another version of classification.  We
note that the excitations in the  topological order $\cR\ot{\bulk(\cR)} \hat\ga
\ot{\bulk(\cC)} \cC^\rev $ is described a fusion $n$-category $\cR\ot{Z_1(\cR)}
\ga \ot{Z_1(\cC)} \cC^\rev $, where $\ga$ is a braided equivalence $Z_1(\cR)
\seq{\ga} Z_1(\cC)$ induced by the invertible domain wall $\hat\ga$.  To make
sense of the above statement, let us consider the natural functors 
from $\cR$ to $\cR\ot{Z_1(\cR)} \hat\ga \ot{Z_1(\cC)} \cC^\rev $:
\begin{align}
  \lambda:\cR &\to \cR\ot{Z_1(\cR)} \hat\ga \ot{Z_1(\cC)} \cC^\rev ,
\nonumber \\
  x &\mapsto x \ot{Z_1(\cR)} \one_{\hat\ga} \ot{Z_1(\cC)} \one_\cC ,
\end{align}
and from $\cC$ to $\cR\ot{Z_1(\cR)} \hat\ga \ot{Z_1(\cC)} \cC^\rev $:
\begin{align}
  \rho:\cC  &\to \cR\ot{Z_1(\cR)} \hat\ga \ot{Z_1(\cC)} \cC^\rev ,
\nonumber \\
  x &\mapsto \one_\cR \ot{Z_1(\cR)} \one_{\hat\ga} \ot{Z_1(\cC)} x.
\end{align}
Here $\lambda$ means mapping from the ``left'' boundary and $\rho$ means
mapping from the ``right'' boundary ($\rho$ may not be monoidal here, but it is
monoidal when restricted to a subcategory, as we will show later).  The above
gives two ways to map $\cR$ into $\cR\ot{Z_1(\cR)}\hat\ga\ot{Z_1(\cC )} \cC^\rev$,
namely $\lambda$ and $\rho \circ \iota$.  They correspond to observing the
$\cR$ symmetry from the left $\cR$ boundary, and from the right $\cC$ boundary
as in Fig. \ref{RCMG}.  Thus we expect that $\lambda$ and $\rho\circ\iota$
coincide. However, recall that in \eqref{bulkww} $\hat\gamma$ is only required to
preserve the ``breakable'' symmetry with respect to $\beta:\cR\to\cV$.
Similarly, we only require the ``breakable'' symmetry to agree on left and
right boundaries of the gauged theory. Let $\ker \beta$ be the preimage of the
trivial excitation in $\cV$. More precisely, if condensing $A_\beta$ gives
$\beta:\cR\to \cV$ ($A_\beta$ consists of all the excitations that becomes
trivial in $\cV$), $\ker\beta$ is the smallest fusion subcategory of $\cR$
containing $A_\beta$.  $\ker\beta$ is then the ``breakable'' symmetry.  The
restriction $\lambda|_{\ker\beta}$ and $(\rho\circ\iota)|_{\ker\beta}$ should
agree.

Besides, there is a natural ``half-braiding" between the excitations from the
left boundary and those from the right boundary. After mapped into the gauged
theory:
\begin{align}
  \lambda(x)\otimes \rho(y)\simeq \rho(y)\otimes\lambda(x).
\end{align}
On the image $\lambda(\ker\beta)=\rho\circ\iota(\ker\beta)$, the above further
defines a braiding:
\begin{align}
  &\lambda(x)\otimes \lambda(y)=\lambda(x)\otimes \rho(\iota(y))
  \nonumber\\
  &\simeq
  \rho(\iota(y))\otimes \lambda(x)=\lambda(y)\otimes \lambda(x).
\end{align}
Such braiding makes $\rho$ a monoidal functor when restricted to
$\iota(\ker\beta)$.  We require the above braiding to be trivial, in the sense
that there exists a braided monoidal functor $\lambda(\ker\beta)\to n\eVec$.

These considerations lead to another version of classification (see Fig.
\ref{RCMG}):
\begin{Proposition}
\label{RCMgauge}
Let $\cR$ with $\beta:\cR\to\cV$ be a $\cV$-local fusion $n$-category.
Anomaly-free gapped liquid phases in $n$-dimensional space with an anomaly-free
algebraic higher symmetry $\cR$ are classified by data $(  \cR \inj{\iota} \cC
,\hat\ga) $, where $\cC$ is a fusion $n$-category that includes $\cR$ (\ie
$\iota: \cR \inj{\iota} \cC$ is a top-fully faithful functor), and $\hat\ga:
\bulk(\cR) \seq{} \bulk(\cC)$ an invertible domain wall between  $\bulk(\cR)$
and $\bulk(\cC)$.  $\hat\ga$ induces a braided equivalence $\ga: Z_1(\cR)
\seq{\ga} Z_1(\cC)$  such that the following diagram is commutative (up to a
natural isomorphism):
\begin{align}
\label{RCembed}
    \xymatrix@R=2.0em@C=1.0em{\ker\beta \ar@{^(->}[rr]^\iota \ar@{->}[rd]^{\lambda} && 
    \cC\ar@{->}[ld]_{\rho}  \\
    & \cR\ot{Z_1(\cR)}\ga \ot{Z_1(\cC)} \cC^\rev &}
\end{align}
and the braiding in the image $\lambda(\ker\beta)$ defined above is trivial. 
\end{Proposition}
\noindent
When $\cC=\cR$, the above gives a classification of SPT
phases with symmetry $\cR$.

\subsubsection{A simple example for $\Z_2\times \Z_2$ symmetry in 1-dimensional
space} \label{Z2Z2}

We would like to apply the above results to compute the $\Z_2\times \Z_2$ SPT
phases in 1-dimensional space.  This leads to new and deeper understanding of
SPT order.

Let $\cR=\cRep(\Z_2\times \Z_2)$, $\beta:\cR\to \cVec$ the forgetful functor,
$\dual\cR=\cVec_{\Z_2\times \Z_2}$, $\eM=Z_1(\cR)$.  We would like to compute
the automorphisms of $\eM$ that preserves the embedding
$\iota_\cR:\cR\hookrightarrow \eM$ or the bulk-to-boundary functor
$F_{\dual\cR}:\eM \to \dual\cR$.

$\eM$ is pointed. It is most efficiently represented by a metric group
$(\Z_2^4,\theta)$, where $\theta$ is a nondegenerate quadratic form which is
physically the topological spin. Denote elements in $\Z_2^4$ by four-component
mod 2 integer vectors $(a,b,c,d)$. We pick $\theta$ to be
\begin{align}
  \theta(a,b,c,d)=(-1)^{ac+bd}.
\end{align}
In other words, (1,0,1,0) and (0,1,0,1) are fermions. If one views $\eM$ as a
double-layer toric code, the generators are identified as the following
\begin{align}
  (1,0,0,0) \sim e\one,\quad & (0,1,0,0) \sim \one e, \nonumber\\
  (0,0,1,0) \sim m\one,\quad & (0,0,0,1) \sim \one m.
\end{align}
$\cR=\cRep(\Z_2\times \Z_2)$ is generated by four simple objects
$\one\one,e\one,\one e, ee$.  Thus the embedding is
\begin{align}
  \iota_\cR(a,b)=(a,b,0,0). 
\end{align}
$\dual\cR=\cVec_{\Z_2\times \Z_2}$ is generated by four simple objects
$\one\one,m\one,\one m, mm$.  Thus, the bulk-to-boundary functor is
\begin{align}
  F_{\dual\cR}(a,b,c,d)=(c,d).
\end{align}

An automorphism of $\eM$ is the same as a group automorphism $\al$ of $\Z_2^4$
that preserves $\theta$, i.e.
\begin{align}\label{gaspin}
  \theta(\al(a,b,c,d))=(-1)^{ac+bd}.
\end{align}

\noindent
{\bf Case 1: $\al$ preserves embedding}
as in \eqref{sptembed}:\\
We require that
\begin{align}
  \al(a,b,0,0)=(a,b,0,0).
\end{align}
Thus 
\begin{align}
  \al(a,b,c,d)=(a,b,0,0)+\al(0,0,c,0)+\al(0,0,0,d).
\end{align}
Let $\al(0,0,c,0)=c(x_1,x_2,x_3,x_4)$ and
$\al(0,0,0,d)=d(y_1,y_2,y_3,y_4)$. Since $\al$ should preserve spin 
\eqref{gaspin}, we have
\begin{align}
  ac+bd&=(a+c x_1+d y_1)(c x_3+d y_3)
\nonumber\\
& +(b+ c x_2+d y_2)(c x_4+d y_4) \ \ \mod \ \ 2.
\end{align}
Rearrange the terms to obtain
\begin{align}
  & ac(1+x_3)+ady_3+bd(1+y_4)+bcx_4
  \nonumber\\
  &+c^2(x_1x_3+x_2x_4)+d^2(y_1y_3+y_2y_4)
\nonumber\\
  &+cd(x_1y_3+x_3y_1+x_2y_4+x_4y_2)=0\ \
  \mod \ \ 2.
\end{align}
One must have $x_3=y_4=1$, $y_3=x_4=0$. Then
\begin{align}
  &c^2x_1+d^2y_2
  +cd(y_1+x_2)=0 \ \ \mod \ \ 2.
\end{align}
Thus $x_1=y_2=0$, $y_1=x_2$. We got two solutions
\begin{align}
  \al_0(a,b,c,d)&=(a,b,c,d),\\
  \al_1(a,b,c,d)&=(a+d,b+c,c,d).
\end{align}

\noindent
{\bf Case 2: $\al$ preserves bulk-to-boundary functor}
as in \eqref{bulkww}:\\
Now we require that
\begin{align}
  F_{\dual\cR}\al(a,b,c,d)=(c,d).
\end{align}
In other words,
\begin{align}
  \al(a,b,c,d)=(*,*,c,d).
\end{align}
Let $\al(a,0,0,0)=a(p_1,p_2,0,0),$ $\al(0,b,0,0)=b(q_1,q_2,0,0),$
$\al(0,0,c,0)=c(r_1,r_2,1,0),$ $\al(0,0,0,d)=d(s_1,s_2,0,1)$. $\al$
preserves spin \eqref{gaspin} and gives
\begin{align}
  ac+bd &=(a p_1+b q_1+c r_1 +d s_1)c
\nonumber\\
&+(a p_2+ b q_2+c r_2 +d s_2)d \ \ \mod\ \ 2.
\end{align}
One must have $p_2=q_1=r_1=s_2=0,$ $p_1=q_2=1$, $s_1=r_2$. We also have two
solutions
\begin{align}
  \al_0(a,b,c,d)&=(a,b,c,d),\\
  \al_1(a,b,c,d)&=(a+d,b+c,c,d).
\end{align}

We see that the two approaches indeed give rise to the same solutions.

Although for pointed modular tensor categories (metric groups), the
automorphism is fully determined by the map on objects, it is not the case for
the automorphisms on fusion categories. Below we briefly explain the
nontrivial structures of $\dual\alpha: \dual\cR\to\dual\cR$. By Lemma 2.1.5 in \Ref{DS170402401},
we know that $F_{\dual\cR}$ and $F_{\dual\cR} \circ \al_1$ differs by a nontrivial automorphism
$\dual\alpha$ of $\dual\cR=\cVec_{\Z_2\times \Z_2}=\{ \one\one,m\one,\one
m, mm\}$, corresponding to the nontrivial
cohomology class in $H^2(\Z_2\times\Z_2,U(1))=\Z_2$.  Such automorphism is identity
on objects $\dual\al(g)=g$ but has nontrivial tensor structures, namely
$\dual\al(g)\otimes \dual\al(h)\xrightarrow{\omega(g,h)}\dual\al(gh)$, where
$\omega(g,h)\in H^2(\Z_2\times\Z_2,U(1))$ is nontrivial.

The nontrivial cohomology class in $H^2(\Z_2\times\Z_2,U(1))$ can be
represented by $\omega((c_1,d_1),(c_2,d_2))=(-1)^{c_1d_2}$. We can also see the
nontrivial tensor structure of $\al_1$. Denote the tensor structure of
$\al_1$ by $u(x,y): \al_1(x)\otimes \al_1(y)\to \al_1(x\otimes y)$.
It needs to preserve braiding, namely
\begin{align}
  \xymatrix{
    \al_1(x)\otimes\al_1(y) \ar[rr]^{c_{\al_1(x),\al_1(y)}} 
    \ar[d]^{u(x,y)}
    && \al_1(y) \otimes \al_1(x) \ar[d]^{u(y,x)}\\
    \al_1(x\otimes y) \ar[rr]^{\al_1(c_{x,y})} &&\al_1(y\otimes x)
  }
\end{align}
Let $x=(0,0,1,0)\sim m\one$ and $y=(0,0,0,1)\sim \one m$. $c_{x,y}=1$ since it
braids $m$ in different layers.  $\al_1(x)=(0,1,1,0)\sim me$ and
$\al_1(y)=(1,0,0,1)\sim em$. Therefore, $c_{\al_1(x),\al_1(y)}=-1$
since it means braiding $m$ with $e$ in the first layer and braiding $e$ with
$m$ in the second layer, thus in total a full braiding between $e$ and $m$.
Clearly the values of these two special braidings are independent of gauge.  We
conclude that, independent of gauge,
$u((0,0,1,0),(0,0,0,1))=-u((0,0,0,1),(0,0,1,0))$, which means $u$ can not be
cohomologically trivial. It is not hard to check that
$u((0,0,1,0),(0,0,0,1))=-u((0,0,0,1),(0,0,1,0))$ agrees with
$\omega((c_1,d_1),(c_2,d_2))=(-1)^{c_1d_2}$.
This way, we show that
\begin{align}
\xymatrix@R=2.5em@C=3.0em{
Z_1(\cR) \ar@{<->}[r]^\al_\simeq &  Z_1(\cR) \\
\dual\cR \ar@{<-}[u]|{F_{\dual\cR}} \ar@{<->}[r]^{\dual\al}_\simeq & \dual\cR \ar@{<-}[u]|{F_{\dual\cR}} 
}
\end{align}
which is an example of Proposition~\ref{vset}.

Next we examine the condensable algebras $A_\cR$ and $A_{\dual\cR}$. By
definition, $A_{\dual\cR}$ is the direct sum of anyons that maps to trivial under
$F_{\dual\cR}$. It is easy to see that 
\begin{align}
  A_{\dual\cR}=\one\one\oplus e\one\oplus \one e\oplus
  ee&=\oplus_{ab}(a,b,0,0),\nonumber\\
  A_{\cR}=\one\one\oplus m\one\oplus \one m\oplus
  mm&=\oplus_{cd}(0,0,c,d).
\end{align}
One can check that the overlap of $A_{\dual\cR}$ and $A_{\cR}$ is $(0,0,0,0)$,
which implies that $\cR\ot{\eM}\dual\cR^\rev=\cVec$. This result can be verified
explicitly using the techniques developed in \Ref{LWW1414,LW191108470}.  Also
$A_{\dual\cR}\otimes A_\cR=\oplus_{abcd}(a,b,c,d)$.

It is obvious that an automorphism preserving $A_{\dual\cR}$ is the same as
preserving the embedding $\cR\inj{\iota_\cR} \eM$, and also the same as
preserving the bulk-to-boundary functor $F_{\dual\cR}:\eM\to \dual\cR$. 

\section{Emergent low energy effective algebraic higher symmetry and
categorical symmetry} 

\label{effCS}

\subsection{Emergent of categorical symmetry from energy scale separation}

In real $n$d condensed matter systems, we usually have 0-symmetry described by
a group $G$ and the associated categorical symmetry $\sM=\bulk(n\cRep G)$
(which is also denoted as $G\vee \dual G^{(n-1)}$).  But it is hard to have
higher symmetry and algebraic higher symmetries, unless we fine tune the
lattice model (if we do not include dynamical electromagnetic
field\cite{W181202517}).  However, emergent algebraic higher symmetries and
associated categorical symmetries can appear at low energies, if our models
have an energy scale separation.\cite{W181202517}  This is a practical way to
realize algebraic higher symmetries and associated categorical symmetries,
which makes the results of this paper useful.

In this subsection, we will discuss how to compute the emergent algebraic
higher symmetries and the categorical symmetries.  It turns out we just need to
compute the emergence of categorical symmetries $\sM$.  The emergent algebraic
higher symmetries $\cR$ can be determined from the emergent categorical
symmetries directly, by solving two equations $\bulk(\cR)\simeq \sM$ and $\cR
\map{\bt}n\cVec$. The solutions are usually not unique.  But the different
solutions are holo-equivalent.

Let us consider a gapped liquid state in $n$-dimensional lattice.  We assume
the excitations in the gapped state has a large separation of energy scale.
The low energy excitations (point-like, string-like, \etc) are closed under
fusion and form a fusion $n$-category $\cC^\text{low}$.  All other topological
excitations have very high energies which are assumed to be infinity.  Now we
add interactions among those low energy excitations to drive phase transitions
by condensing the low energy excitations, and to form gapless states, \etc. We
assume that, in such process, the high energy excitations remain to have high
energies (\ie infinite energy).  We would like to ask what are the possible
phases and gapless states?

Some constraints to the low energy physics come from the underlying symmetry,
while other constraints come from the fusion and statistics of those low energy
topological excitations. It looks hard to understand the effects of all those
different constraints.  But it turns out that the holographic point of view and
the associated categorical symmetry can help us to solve this problem.

We know that some excitations in $\cC^\text{low}$ are topological excitations,
while others are charge objects of the underlying symmetry.  To use the
holographic point of view and to use categorical symmetry, we restrict to
symmetric sub-Hilbert space of the underlying symmetry.  In this case, every
excitations in  $\cC^\text{low}$ can be viewed as topological excitations in a
hypothetical system without symmetry.  However, the fusion $n$-category
$\cC^\text{low}$ that describes those excitations is in general anomalous, \ie
it cannot be realized by a lattice system in the same dimension without
symmetry.  But it can be realized as a boundary of a topological order
$\sM^\text{low}=\bulk(\cC^\text{low})$ in one higher dimension (see
\eqn{bulkCM}).  In fact, $\sM^\text{low}$ is nothing but the emergent
categorical symmetry, which provides all the constraints to the low energy
physics and solves our problem.  

We see that the only input is the low energy excitations $\cC^\text{low}$.
So we do not need to have a lattice model. The above discussion remains valid
for field theories without a given or known lattice regularization.  (In this
paper, we use the term \emph{field theory} to mean theory  without a given or
known lattice regularization.) Thus 
\begin{Proposition}
for a lattice system or a field theory with low energy excitations
$\cC^\text{low}$, the system has a low energy effective (\ie emergent)
categorical symmetry given by $\sM^\text{low} = \bulk(\cC^\text{low})$, that
provides all the constraints to the low energy physics.  
\end{Proposition}
\noindent
Such a low energy effective categorical symmetry $\sM^\text{low}$.  is present
even when low energy excitations condense, undergo phase transitions, \etc, as
long as all other higher energy excitations remain to have very high energies.
The emergent categorical symmetry controls all the low energy behaviors of the
system, including allowed phases, allowed phase transitions, allowed critical
points, \etc.  This is because the allowed phases, allowed phase transitions,
allowed critical points, \etc\ are one-to-one correspond to different
boundaries of $\sM^\text{low}$ -- the categorical symmetry.  In some sense,
$\sM^\text{low}$ is a ``topological invariant'' of low energy physics, and, we
believe, is a complete ``topological invariant''.  All other low energy
topological invariants can be obtained from $\sM^\text{low}$.

Such an emergent categorical symmetry is the most practical and useful
application of the notion of categorical symmetry and the holographic point of
view.  For example, 
\begin{Proposition}
Consider a gapped liquid state in $n$-dimensional space whose low energy energy
excitations are described by a fusion $n$-category $\cC^\text{low}$.  When all
other excitations remain to have higher energies, the gapped liquid phases
formed by low energy energy excitations in $\cC^\text{low}$ must have
excitations described by a fusion $n$-category $\cC$ that satisfy
$\bulk(\cC)\simeq \bulk(\cC^\text{low})$.  
\end{Proposition}
\noindent
In fact, $\bulk(\cC)\simeq \bulk(\cC^\text{low})$ is nothing but the anomaly
matching condition, since the categorical symmetries $\bulk(\cC)$ and
$\bulk(\cC^\text{low})$, as topological orders in one higher dimension, are the
effective non-invertible gravitationa l
anomalies\cite{W1313,KW1458}, after we view the charge objects of the symmetry
as topological excitations.  

\begin{Remark}
We like to point out that the effective gravitational anomaly here is more
general then the usual gravitational anomaly from the non-invariance of the
path integral.  The usual gravitational anomaly is invertible, while our
effective gravitational anomaly, as topological order in one higher dimension,
is in general non-invertible.\cite{KW1458,KZ150201690,JW190513279} Since the
usual gravitational anomaly is invertible, it corresponds to invertible
topological order in one higher dimension, which contains no non-trivial
topological excitations.  Thus the usual gravitational anomaly does not encode
any conservation law, since the conservation law must come from the fusion rule
of excitations for the topological order in one higher dimension. In contrast,
a non-invertible gravitational anomaly does encode a conservation law, since
its corresponding topological order in one higher dimension has non-trivial
excitations and non-trivial fusion rule.  Therefore, a non-invertible
gravitational anomaly can be viewed as a symmetry.  This is why we also refer
to non-invertible gravitational anomaly as categorical symmetry, to stress its
connection to symmetry.
\end{Remark}

\subsection{States with the full categorical symmetry}

Since all the gapped liquid states in systems with an (emergent) categorical
symmetry must spontaneously break part of the categorical symmetry, the states
with the full unbroken categorical symmetry must be
gapless.  A system  with a categorical symmetry $\sM$, may have many
different symmetric gapless states.  Those gapless states may have additional
emergent categorical symmetry.  So here we would like to ask, what is the
minimal gapless state with the categorical symmetry $\sM$?  To define the
notion of `` minimal gapless state'' in $n$-dimensional space, we assume that
the gapless excitations all have the same linear dispersion $\om=v k$.  The low
temperature specific heat of the gapless state has a form
\begin{align}
 c_V = c \ga_n T^n
\end{align}
where
\begin{align}
 \ga_n = (n+1)k_B\Big(\frac{k_B}{v}\Big)^n\int \frac{\dd^n \v k}{(2\pi)^n}  \frac{|\v k|}{\ee^{  |\v k|} -1}.
\end{align}
\noindent
For a system described by a single gapless real scalar field, we find that
$c=1$.  The minimal gapless state has minimal $c$.

From the above discussions, we see that minimal gapless states with the
categorical symmetry $\sM$ are actually minimal gapless boundary of topological
order with excitations described by $\sM$ in one higher dimension.
\Ref{JW190513279,KZ190504924,KZ191201760} discussed how to obtain gapless
boundaries for 2d topological orders, using modular covariant partition
functions or topological Wick rotation.  Those gapless boundaries do not break
the categorical symmetry $\sM$.  Those approach also allow us to obtain the
minimal gapless boundaries with minimal central charge.  However, for a given
categorical symmetry, it is not clear whether its minimal gapless state is
unique or not.\cite{JW191213492}

\section{Examples}

In the section, we discuss some gapped liquid phases. In particular, we
identify their algebraic higher symmetry and categorical symmetry.  We also
discuss low energy effective (\ie emergent) categorical symmetry when some
topological excitations have low energies.

\subsection{The category of 0d topological orders}

The category of 0d topological orders $\cTO^1$ is the category of 0d gapped
phases with no symmetry.  In 0d, a stable gapped phase has non-degenerate ground
state, which corresponds to a simple object in the category of 0d gapped
phases, denoted as $\cTO^1$.  This is the only simple object in $\cTO^1$, and is
the unit object of stacking operation $\otimes$, which is the tensor product
of vector spaces. We denote this unit object as $\one$.  There are accidental
degenerate ground states, which corresponds to a composite object
$\underbrace{\one\oplus\one \oplus \cdots \oplus \one}_{m \text{ copies}} =
m\one$.  In $\cTO^1$, an 1-morphism from $m\one$ to $n\one$ is an $n\times m$
complex matrix $M$: $m\one \xrightarrow{M} n\one$.  Such a fusion 1-category
happen to be $1\cVec$.  We see that $\cTO^1 =1\cVec\equiv \cVec$.

\subsection{2d topological order described by $\Z_2$ gauge theory}

The 2d $\Z_2$ topological order described by the $\Z_2$ gauge theory is
denoted by $\sGT^3_{\Z_2}$. Codimension-2 excitations are described 
by the following braided fusion 1-category $\Om^2\sGT^3_{\Z_2}$, which
has four simple objects (the point-like excitations): $\one,e,m,f$ with the following $\Z_2$ fusion rule
\begin{align}
e\otimes e = m\otimes m = f\otimes f = \one . 
\end{align}
$\one$ is the trivial excitation.  $e,m,f$ are topological excitations which
have mutual $\pi$-statistics between them.  $e,m$ are bosons, and $f$ is a
fermion.  Such a topological order $\sGT^3_{\Z_2}$ can be realized by lattice
models in the same dimension (see \Ref{RS9173,W9164,K032}).  Therefore, the
bulk of $\sGT^3_{\Z_2}$ is a 3d product state, \ie
$\Bulk(\sGT^3_{\Z_2})=\tto^4$ (see \eqn{BulkCM}). 
The 2d topological order $\sGT^3_{\Z_2}$ has no categorical symmetry since
$\Bulk(\sGT^3_{\Z_2})=\tto^4$ or $\bulk(\Om\sGT^3_{\Z_2})=\tto^4$.

Next, we consider the situation when $e$ particles have low energies, and $m,f$
particles have very high energies.  The low energy excitations form a fusion
2-category $2\cRep\Z_2$ (after condensation completion), which simply describes
2d bosons with mod-2 conservation.  In this limit, we have a low energy
effective categorical symmetry characterized by the 3d $\Z_2$ gauge theory
$\sGT^4_{\Z_2} = \bulk(2\cRep\Z_2)$.  

$2\cRep\Z_2$ describes the excitations in a system with $\Z_2$ symmetry in the
$\Z_2$ symmetric phase (within the symmetric sub Hilbert space).  $2\cRep\Z_2$
also describes the excitations in one of the gapped boundaries of 3d $\Z_2$
gauge theory $\sGT^4_{\Z_2}$, obtained by condensing the $\Z_2$-flux lines in
$\sGT^4_{\Z_2}$ at the boundary.  The 3d $\Z_2$ gauge theory $\sGT^4_{\Z_2}$
has another gapped boundary whose excitations are described by the fusion
2-category $2\cVec_{\Z_2}$ (a $\Z_2$ symmetry breaking phase with $e$ boson
condensation), obtained from condensing the $\Z_2$ charges in $\sGT^4_{\Z_2}$
at the boundary.  The second boundary correspond to another gapped phase of the
system with the $\Z_2$ symmetry -- the spontaneous symmetry breaking phase.
The continuous phase transition between the two gapped phases is described by a
critical point which has the full categorical symmetry characterized by
$\sGT^4_{\Z_2}$ (the 3d $\Z_2$ gauge theory).  This critical point is the same
as the  critical point of 2d quantum Ising model (or 3d statistical Ising
model), which has the same categorical symmetry $\sGT^4_{\Z_2}$, as discussed
in \Ref{JW191213492}.

When $e$ bosons have low energies, the resulting $\Z_2$ symmetric system can
have infinity many different symmetric gapped phases, and one of them is the 2d
$\Z_2$-SPT  phase.  The critical point at the continuous transition from the
$\Z_2$-SPT  phase to the $\Z_2$ spontaneous symmetry breaking phase is
described by the same critical point discussed above, this is because the
transition is also described by the same $\Z_2$ charge condensation.

Last, we consider the situation when $f$ particles have low energies, and $e,m$
particles have very high energies.  The low energy excitations form a fusion
2-category $2\scVec$, which simply describes 2d fermions with mod-2
conservation. There is a $\Z_2^f$ symmetry from the mod-2 conservation of the
fermions. In this limit, we have a low energy effective categorical symmetry
characterized by 3d twisted $\Z_2$ gauge theory with fermionic $\Z_2$ charge,
denoted by $\sGT^4_{\Z_2^f}$. 
(The 3d twisted $\Z_2$ gauge theory $\sGT^4_{\Z_2^f}$ is obtained by gauging
$\Z_2^f$, a $\Z_2$ symmetry with fermionic $\Z_2$ charge.) The categorical
symmetry $\sGT^4_{\Z_2^f}$ is different from the categorical symmetry
$\sGT^4_{\Z_2}$ discussed above.  So when $f$  fermions have low energies, our
system has different properties from when $e$ bosons have low energies.

When $f$ fermions have low energies, our system can have 16 gapped phases (up
to $E_8$ 2d bosonic invertible topological order) labeled by $\al \in \Z_{16}$,
which correspond to 2d fermionic invertible topological orders.  The continuous
transition between $\al$ and $\al+1$ phases is described by the following 2d
non-interacting Majorana fermion theory:\cite{RG0067,W0050} 
\begin{align}
 H &= \int \dd^2 \v x \big[
\la^\top(\v x) \ga^i\prt_i \la(\v x) +  m \la^\top(\v x)  \ii\si^2 \la(\v x)
\big]
\nonumber\\
\la &=
\begin{pmatrix}
 \la_{1}\\
 \la_{2}\\
\end{pmatrix},\ \   \la^* = \la,\ \
\ga^1=\si^1,\ \ 
\ga^2=\si^3.
\end{align}
where $\si^i$ is the Pauli matrix.  The transition happens when $m$ change
sign, which change the chiral central charge of the edge state by
$1/2$.\cite{RG0067,W0050} The gapless state at $m=0$ have the full categorical
symmetry $\sGT^4_{\Z_2^f}$.

\subsection{3d topological order described by $\Z_2$ gauge theory}

The 3d $\Z_2$ topological order $\sGT^4_{\Z_2}$ (described by the $\Z_2$ gauge theory) has codimension-2 and codimension-3 excitations described by
 the braided fusion 2-category $\Om^2\sGT^4_{\Z_2}$:  The
simple objects (the string-like excitations) are labeled by 
$\one_s,m_s,e_s, m_s\otimes e_s$, with
the following symmetric fusion
\begin{align}
\one_s\otimes m_s &= m_s , &
\one_s\otimes e_s &= e_s ,
\nonumber\\
m_s\otimes m_s &= \one_s , &
e_s\otimes e_s &= 2e_s ,
. 
\end{align}
$\one_s$ is the trivial string.  $m_s$ is a bosonic topological string-like
excitation, that corresponds to the $\Z_2$-flux string. 

The simple 1-morphisms (the point-like excitations), that connect $\one_s \to
\one_s$, are labeled by $\one_p,e_p$, with the following $\Z_2$ fusion
\begin{align}
e_p\otimes e_p = \one_p . 
\end{align}
$\one_p$ is the trivial particle.  $e_p$ is a bosonic topological excitation
with trivial mutual statistics.  However, $e_p$ and $m_s$ has a non-trivial
mutual $\pi$-statistics between them.  We also have simple 1-morphisms that
connect $m_s \to m_s$, which are labeled by $\one_{m_s},e_{m_s}$ with the
following $\Z_2$ fusion
\begin{align}
e_{m_s}\otimes e_{m_s}= \one_{m_s} . 
\end{align}
They correspond to the point-like excitations on the string $m_s$.  

The $e_s$ string mentioned above is a descendent excitation, formed by
condensing $e_p$ point-like excitations along the string.  Since $e_p$ has a
mod 2 conservation, the $e_p$ condensed state is a spontaneously $\Z_2$
symmetry breaking state. This leads to the fusion rule $e_s\otimes e_s =2 e_s$.

Such a $\sGT^4_{\Z_2}$ topological order has a trivial categorical symmetry
since $\Bulk(\sGT^4_{\Z_2})=\tto^5$ or $\bulk(\Om\sGT^4_{\Z_2})=\tto^5$ (where
$\Om\sGT^4_{\Z_2}$ describes the excitations in $\sGT^4_{\Z_2}$).  However,
when some excitations have low energy and other have high energies, the system
may have a low energy effective categorical symmetry.

When $e_p$ particles have low energies and $m_s$ strings have very high
energies, the low energy excitations are described by a fusion 3-category
$3\cRep\Z_2$ generated by $e_p$ particles.  In this limit, the low energy
effective categorical symmetry is $\bulk(3\cRep\Z_2)=\sGT^5_{\Z_2}$, which is
nothing but the 4d $\Z_2$ gauge theory.  Such a categorical symmetry has
following two gapped phases (plus many others):\\ 
(1) a phase with low energy excitations $3\cRep\Z_2$ (corresponding to the symmetric phase
of 3d quantum Ising model);\\
(2) a phase with low energy excitations $3\cVec_{\Z_2}$ 
(corresponding to the spontaneous symmetry breaking phase
of 3d quantum Ising model).\\
The transition between the two gapped phase
is Higgs transition of the 3d $\Z_2$ gauge theory.
The critical point has the full
categorical symmetry $\sGT^5_{\Z_2}$.  Such a critical point is the same as the
critical point in 3d quantum Ising model or 4d statistical Ising model, which
is described by non-interacting massless real scaler field
\begin{align}
 S = \int \dd t\dd^3\v x\; \big[ \frac12 (\prt_t \phi)^2 +\frac12 v^2 (\prt_{\v x} \phi)^2 \big]
\end{align}

When $m_s$ strings have low energies and $e_p$ particles have very high
energies, the low energy excitations are described by a fusion 3-category
generated by $m_s$ strings, which is denoted as $3\cRep\Z_2^{(1)}$.  Ignoring
the descendant excitations, $3\cRep\Z_2^{(1)}$ has only a single trivial
object, two simple 1-morphisms: trivial string $\one_s$ and $\Z_2$ flux string
$m_s$, and  a single trivial 2-morphism.  In this limit, the low energy
effective categorical symmetry is
$\bulk(3\cRep\Z_2^{(1)})=\sGT^5_{\Z_2^{(1)}}$, where $\sGT^5_{\Z_2^{(1)}}$ is
the 4d $\Z_2$ 2-gauge theory obtained by gauging $\Z_2^{(1)}$ 1-symmetry.  The
4d $\Z_2^{(1)}$ 2-gauge theory has a string-like $\Z_2$ charge and  string-like
$\Z_2$ flux.  The $\Z_2$ string-charge and the $\Z_2$ string-flux has mutual
$\pi$-statistics.  Such a categorical symmetry has two gapped phases:\\
(1) a phase with low energy excitations $3\cRep\Z_2^{(1)}$;\\  
(2) another phase also with low energy excitations $3\cRep\Z_2^{(1)}$.\\  
The transition between the two phases is the confinement transition of the 3d
$\Z_2$ gauge theory.  The critical point of the transition has the full
categorical symmetry $\sGT^5_{\Z_2^{(1)}}$.  Such a critical point is different
from the Higgs transition critical point which has a categorical symmetry
$\sGT^5_{\Z_2}$ (for details, see \Ref{JW191213492}).

\subsection{3d topological order described by twisted $\Z_2$ gauge theory}

The 3d topological order described by the twisted $\Z_2$ gauge theory (\ie 3d
$\Z_2$ gauge theory with fermionic point-like $\Z_2$ charge) is denoted as
$\sGT^4_{\Z_2}$, Its excitations are described by a braided fusion 2-category
$\Om^2\sGT^4_{\Z_2^f}$, which is similar to $\Om^2\sGT^4_{\Z_2}$, except now
the $Z_2$ charge $e_p$ is a fermion.  Many results discussed above remain
unchanged.  In particular, when the $\Z_2$ flux strings $m_s$ have low energies
and $\Z_2$ point charges $e_p$ have high energies, the system have an low
energy effective categorical symmetry $\sGT^5_{\Z_2^{(1)}}$ as discussed above.

But  when $\Z_2$ point charges $e_p$ have low energies and the $\Z_2$ flux
strings $m_s$ have high energies, the system has a very different behavior
since the $\Z_2$ point charges are fermions.  In this limit, the low energy
excitations are described by fusion 3-category $3\cRep\Z_2^f$ (with trivial
object, trivial 1-morphism, and $\Z_2^f$ 2-morphisms which contain fermions).
The categorical symmetry is $\sGT^5_{\Z_2^f}=\bulk(3\cRep\Z_2^f)$ (the 4d
$\Z_2$ gauge theory with fermionic $\Z_2$ point charge).  What are the gapped
liquid phases in a system with $\sGT^5_{\Z_2^f}$ categorical symmetry?  There
is no 3d fermionic invertible topological order. So there is only one gapped
state (up to stacking of bosonic topological orders with no symmetry) that
break the categorical symmetry $\sGT^5_{\Z_2^f}$ down to $\Z_2$ fermionic
0-symmetry.  There should also be gapless states with the full
$\sGT^5_{\Z_2^f}$ categorical symmetry.  

We like to point out that the $\Z_2$ fermionic 0-symmetry is not an algebraic
higher symmetry described by a local fusion higher category $\cR$ (\ie not a
bosonic  algebraic higher symmetry).  The  categorical symmetry
$\sGT^5_{\Z_2^f}$ is not associated with any bosonic  algebraic higher
symmetries, since $\sGT^5_{\Z_2^f}\simeq \bulk(\cR)$ and $\cR\map{\bt} 3\cVec$
has no solution.  The  $\Z_2$ fermionic 0-symmetry described by
$\cR=3\cRep\Z_2^f$ satisfies $\sGT^5_{\Z_2^f}\simeq \bulk(\cR)$, by does not
satisfy $\cR\map{\bt} 3\cVec$.

One particular realization of the gapped phase is via a Majorana fermion field
theory. Here we use a single Weyl fermion field $\psi$ (with two complex
components) to describe a single Majorana fermion field (with four real
components):
\begin{align}
\label{Hmf}
 H &= \int \dd^3 \v x\
\psi^\dag \ii \si^i \prt_i \psi
+ (m \psi^\top \eps  \psi + h.c.)
\end{align}
where $\psi^\dag = (\psi^\top)^*$ and $\eps\equiv \ii\si^2$.  The mass $m$ can
be complex.  The gapless state at $m=0$ should have the full $\sGT^5_{\Z_2^f}$
categorical symmetry.  However, it is not clear if it is the minimal gapless
state with the full $\sGT^5_{\Z_2^f}$ categorical symmetry.

\subsection{$n$d bosonic systems with $S_3$ symmetry}

We consider the class of bosonic $n$d lattice Hamiltonians $\{H_{S_3}\}$ with
$S_3 =\Z_3\rtimes Z_2$ symmetry.  We also consider the class of boundary
Hamiltonians $\{H_{S_3}^\text{bndry}\}$ of $(n+1)$d $S_3$ topological order
$\sGT_{S_3}^{n+2}$ with energy gap approaching $\infty$ .  The class of lattice
boundary Hamiltonians $\{H_{S_3}^\text{bndry}\}$, by definition, is said to
have $\sGT_{S_3}^{n+2}$ categorical symmetry.  We have argued that the class of
lattice Hamiltonians $\{H_{S_3}\}$, when restricted to the symmetric subspace,
is holo-equivalent to  the class of boundary Hamiltonians
$\{H_{S_3}^\text{bndry}\}$. For example, for each $S_3$-symmetric Hamiltonian
in the class $\{H_{S_3}\}$, we can find a boundary Hamiltonian in the class
$\{H_{S_3}^\text{bndry}\}$, such that the two Hamiltonians have the same low
energy properties.  In this sense, we say the $S_3$-symmetric lattice
Hamiltonians also has the $\sGT_{S_3}^{n+2}$ categorical symmetry.  In this
section, we like to ask, whether there are other algebraic higher symmetry
$\cR$, such that the $\cR$-symmetric lattice Hamiltonians also have the
$\sGT_{S_3}^{n+2}$ categorical symmetry.  In this case, we may say the
algebraic higher symmetry $\cR$ to be holo-equivalent to the $S_3$-symmetry
(\ie the $n\cRep S_3$-symmetry).

Certainly, the dual symmetry of the $n\cRep S_3$-symmetry, $n\cVec_{S_3}$ is
holo-equivalent to the $n\cRep S_3$-symmetry.  Do we have other algebraic
higher symmetry $\cR$ that is  holo-equivalent to the $n\cRep S_3$-symmetry?

So we first try to solve $\bulk(\cR) = \bulk(n\cRep S_3) = \sGT_{S_3}^{n+2}$,
in a physical way.  We consider a $(n+1)$d $S_3$ gauge theory
$\sGT_{S_3}^{n+2}$, whose excitations are described by $\Om^2\sGT_{S_3}^{n+2}$.
The excitations $\cR$ on a gapped boundary satisfy $\bulk(\cR) =
\sGT_{S_3}^{n+2}$.

Let us start with a gauge-flux-condensed boundary, whose excitations are
described by $n\cRep S_3$.  Such a boundary corresponds to a $S_3$-symmetric
phase.  Next, we try to obtain other boundaries by condensing the point $S_3$
charges on this boundary.

The $S_3$ charges are described by two bosonic fields: a real field $\si$ and a
complex field $\phi$.  Under the $Z_2$ transformation in $S_3$: $\si \to -\si$
and $\phi \to \phi^*$.  Under the $Z_3$ transformation in $S_3$: $\si \to \si$
and $\phi \to \ee^{\ii \frac{2\pi}{3}} \phi$.

In the first case,
we condense the $\si$ bosons but not the $\phi$ bosons on the  $n\cRep S_3$
boundary, by setting $\si=1,\ \phi=0$.  This condensation will change the
$n\cRep S_3$ boundary to another gapped boundary, denoted as $\cR_\si$.  This
new boundary $\cR_\si$ corresponds to a $S_3 \to Z_3$ spontaneous symmetry
breaking phase of $S_3$ symmetric systems.

In the second case,
we condense the $\phi$ bosons but not the $\si$ bosons on the  $n\cRep
S_3$ boundary, by setting $\phi=1,\ \si=0$.  This condensation will change the
$n\cRep S_3$ boundary to a gapped boundary, denoted as $\cR_\phi$.  The
boundary $\cR_\phi$ corresponds to a $S_3 \to Z_2$ spontaneous symmetry
breaking phase of $S_3$ symmetric systems.

Now we like to ask, does $\cR_\si$ describe the algebraic higher symmetry in
the $S_3 \to Z_3$ spontaneous symmetry breaking phase?  \ie can the $S_3 \to
Z_3$ spontaneous symmetry breaking phase be viewed as the trivial symmetric
phase of the $\cR_\si$-symmetry?  Also, does $\cR_\phi$ describe the algebraic
higher symmetry in the $S_3 \to Z_2$ spontaneous symmetry breaking phase?

Although $\cR_\si$ and $\cR_\phi$ satisfy $\bulk(\cR_\si) = \bulk(\cR_\phi) =
\sGT_{S_3}^{n+2}$, we still need to show they are local fusion higher
categories in order for them to describe algebraic higher symmetries.  For this
purpose, we start with the $n$d $S_3$-gauge theory constructed by stacking two
$n\cRep S_3$ via their common bulk $\sGT_{S_3}^{n+2}$ (a $(n+1)$d $S_3$-gauge
theory), as shown in Fig. \ref{RRgauge}.  The excitations in the $n$d
$S_3$-gauge theory are given by $n\cRep S_3 \ot{\sGT_{S_3}^{n+2}} (n\cRep
S_3)^\rev$.  Now we condense $\si$ on one boundary and condense $\phi$ on the
other boundary.  The resulting $n$d state has excitations given by $\cR_\si
\ot{\sGT_{S_3}^{n+2}} \cR_\phi^\rev$.  From our physical understanding, when
both $\si$ and $\phi$ condense, the ``$S_3$-gauge symmetry'' is completely
broken, and the $n$d topological order described by the $S_3$-gauge theory
becomes a trivial phase.  This implies that $\cR_\si \ot{\sGT_{S_3}^{n+2}}
\cR_\phi^\rev = n\cVec$. Then using Proposition~\ref{RBlocal}, we find that $\cR_\si$
and $\cR_\phi$ are both local fusion higher categories.  They describe a pair
of dual algebraic higher symmetries.

~

{\bf Acknowledgement}: 
LK and HZ are supported by the Science, Technology and Innovation Commission of Shenzhen Municipality (Grant Nos. ZDSYS20170303165926217) and by Guangdong Provincial Key Laboratory (Grant No.2019B121203002). LK is also supported by NSFC under Grant No. 11971219. XGW is partially supported by NSF DMS-1664412 and by the Simons Collaboration on
Ultra-Quantum Matter, which is a grant from the Simons Foundation (651440). HZ is also supported by NSFC under Grant No. 11871078.

%\bibliography{./all,./publst} 
\bibliography{../../bib/all,../../bib/allnew,../../bib/publst} 

\end{document}